\documentclass[11pt,a4paper]{article} 

\usepackage[german,english]{babel} 
\usepackage{jheppub}
\usepackage{booktabs} 
\usepackage{comment} 
\usepackage[utf8]{inputenc} 
\usepackage[T1]{fontenc} 
\usepackage{tcolorbox}
\usepackage{amsmath,amsfonts,amsthm}
\usepackage{amssymb}
\usepackage{tikz}
\usepackage{tikz-cd}
\usetikzlibrary{positioning,arrows} 
\usetikzlibrary{decorations.pathreplacing} 
\usepackage{tikzsymbols} 
\usepackage{blindtext} 
\usepackage{braket}
\usepackage{bm}
\newcommand{\sbra}[1]{\left(#1\right)}
\newcommand{\sqbra}[1]{\left[#1\right]}

\def\n3{\sqrt{3}}

\def\tr{\mathrm{Tr}}

\def\D{\mathrm{d}}
\usepackage{simplewick}
\usepackage{color}
\usepackage{framed}
\usepackage{titlesec}
\allowdisplaybreaks[3]

\usepackage[T1]{fontenc} 
\usepackage[utf8]{inputenc} 

\usepackage{natbib} 
\title{
The Markov Gap in the presence of islands
}
\author[a]{
Yizhou Lu,}
 \affiliation[a]{
 Department of Physics, Southern University of Science and Technology, Shenzhen 518055, China
 }
 \affiliation[b]{Interdisciplinary Center for Theoretical Study,
University of Science and Technology of China, Hefei, Anhui 230026, China}
\affiliation[c]{Peng Huanwu Center for Fundamental Theory, Hefei, Anhui 230026, China}
\author[b,c,1]{Jiong Lin\note{Corresponding author.}}
\emailAdd{luyz@sustech.edu.cn}

\emailAdd{jionglin@ustc.edu.cn}

\abstract{
    The Markov gap \cite{Hayden:2021gno}, namely the difference between reflected entropy and mutual information, is explicitly computed in the defect extremal surface model, JT gravity, and the generic 2d extremal black holes, in vacuum states.
    The phases that contain various island contributions are considered, and their existence is carefully checked.
    Moreover, we show explicitly how the Markov gap originates from the OPE coefficient of the boundary CFT.
And, as a generalization of \cite{Hayden:2021gno}, the lower bound of the Markov gap is given by $\frac{c}{3}\log 2$ times the number of EWCS boundaries on minimal surfaces.
    We propose a boundary way of counting the lower bound for the Markov gap, which states that the lower bound is given by $\frac{c}{3}\log 2$ times the number of gaps between two boundary regions in vacuum states.
    We discuss the limitation and possible generalization of the boundary counting, and its relation to tripartite entanglement.
} 

\begin{document}
 \hfill USTC-ICTS/PCFT-22-26\\
\maketitle

\section{Introduction}
The von Neumann entropy is an excellent measure of quantum entanglement between two subsystems in a pure state and thus is usually referred to as entanglement entropy.
Based on the developments in holographic entanglement entropy \cite{Ryu:2006bv,Hubeny:2007xt,Lewkowycz:2013nqa,Engelhardt:2014gca,Faulkner:2013ana}, the \emph{island} formula for entanglement entropy is proposed as \cite{Almheiri:2019hni,Penington:2019kki,Penington:2019npb,Almheiri:2019psf,Almheiri:2019qdq}
\begin{align}\label{island_EE}
    S(A)=\min \text{ext}_{\mathcal I_A}\sqbra{\frac{\text{Area}(\partial \mathcal I_A)}{4G_N}+S_{\rm bulk}(A\cup \mathcal I_A)
    },
\end{align}
where the region $\mathcal I_A$ is known as $A$'s island as it is separate from $A$, the second term is the quantum entanglement of bulk matter. 
\eqref{island_EE} stems from the QES formula for holographic entanglement entropy in AdS/CFT correspondence \cite{Engelhardt:2014gca}, and is derived via gravitational path integral in a specific JT gravity \cite{Almheiri:2019qdq,Penington:2019kki}.
With \eqref{island_EE}, the unitary Page curves for many black holes have been successfully recovered, making significant progress toward the information paradox.\footnote{
See \cite{Penington:2019kki,Penington:2019npb,Almheiri:2019psf,Almheiri:2019qdq,Almheiri:2019yqk,Hollowood:2020cou,Gautason:2020tmk,Goto:2020wnk,Hashimoto:2020cas,Wang:2021woy,Lu:2021gmv,Geng:2020qvw,Geng:2021hlu,Geng:2021mic,He:2021mst,Tian:2022pso,Chu:2021gdb,Alishahiha:2020qza,Hartman:2020khs,Balasubramanian:2020xqf,Chen:2020uac,Chen:2020hmv,Hernandez:2020nem,Grimaldi:2022suv,HosseiniMansoori:2022hok,Ageev:2022qxv,Goswami:2022ylc,Du:2022vvg,Yu:2021cgi,Yu:2021rfg,Yu:2022xlh,Bhattacharya:2021dnd,Bhattacharya:2021jrn,Bhattacharya:2021nqj,Yadav:2022fmo,Yadav:2022jib,Krishnan:2020fer,Krishnan:2020oun,Ghosh:2021axl,Uhlemann:2021nhu,Ahn:2021chg,Karch:2022rvr} and reference therein for a non-exhausted list of related researches and \cite{Almheiri:2020cfm} for a review.
}

However, the von Neumann entropy ceases to be a good measure of entanglement for tripartite systems or mixed states.
A measure of entanglement for mixed states is of significant importance, as it can be used to probe the entanglement structure of the state, and,
on the other hand, the states we encounter are not always pure.

Many quantities have been proposed to measure the bipartite correlation in mixed states and in tripartite systems, such as mutual information, \emph{entanglement of purification} $E_p$ \cite{Terhal2002TheEO,Takayanagi:2017knl,Nguyen:2017yqw},
\emph{balanced partial entanglement} (BPE) $s_{\rm BPE}$ \cite{Wen:2021qgx,Camargo:2022mme},
\emph{logarithmic entanglement negativity} $\mathcal E$ \cite{Vidal:2002zz,Calabrese:2012ew,Calabrese:2012nk,Rangamani:2014ywa,Chaturvedi:2016rcn,Kusuki:2019zsp}, odd entanglement entropy \cite{Tamaoka:2018ned}, and \emph{reflected entropy} $S_{R}$ \cite{Dutta:2019gen}.
Many quantities are conjectured to be dual to a geometric object called \emph{entanglement wedge cross-section} (EWCS) in the dual gravity side \cite{Takayanagi:2017knl,Nguyen:2017yqw,Akers:2019gcv,Dutta:2019gen}.
For instance, $\text{Area}[E_W]/4G_N\overset{{?}}{=}\frac23\mathcal E\overset{{?}}{=}E_p\overset{{?}}{=}\frac12S_{R}\overset{{?}}{=}s_{\rm BPE}$ for holographic 2D CFT in the ground state, where $\overset{{?}}{=}$ means the equality is a conjecture or based on some assumptions.

By subtracting the mutual information, one can define some UV-finite quantities, i.e., $g=2E_p-I$ and $h=S_R-I$ \cite{Zou:2020bly}, where $I(A:B)$ is the mutual information.
In particular, non-vanishing $g$ and $h$ imply non-trivial tripartite entanglement \cite{Akers:2019gcv,Zou:2020bly}.
For $g=0$, the state must be in a so-called \emph{triangle state} $\ket{\psi}_{ABC}$ up to local isometries \cite{Zou:2020bly}.
The triangle state is free of non-trivial tripartite entanglement as it is formed by bipartite-entangled states.
For $h=0$, the state must be in the sum of triangle states (SOTS) \cite{Zou:2020bly}.
In general, $g\geq h$, which means some types of tripartite entanglement cannot be seen by $h$.
In a holographic CFT at large-$c$ limit, it is conjectured that the two quantities coincide $g=h$.
Specifically, for a 1D tripartite holographic spin chain on a circle, the authors of \cite{Zou:2020bly} found that $g=h=\frac{c}{3}\log 2$.
A similar discovery was also made by Wen with balanced partial entanglement \cite{Wen:2021qgx}.
Later on, $h$ is shown to be related to the Markov recovery map, and a non-vanishing $h$ precludes a perfect Markov recovery map \cite{Hayden:2021gno}, because of which $h$ is termed as the \emph{Markov gap} by Hayden, Parrikar, and Sorce (HPS).
Using the geometric approach, they proved that for a static state in pure AdS$_3$, the lower bound of the Markov gap of boundary regions is related to the number of boundaries of EWCS:
\begin{align}\label{Hayden_ineq0}
    h(A:B)\geq \frac{\ell }{2G_N}\log 2\times (\text{\#~of~EWCS~boundaries}),
\end{align}
which is nice and neat.
$\ell$ is the AdS radius.
A BPE version of \eqref{Hayden_ineq0} was proposed in \cite{Camargo:2022mme}, and an interface CFT (ICFT) version was studied in \cite{Kusuki:2022bic}.

While \eqref{Hayden_ineq0} is proved to be valid for CFT$_2$ with a pure AdS$_3$ dual, it remains to be explored in the other cases, among which the presence of islands is of great interest.
Firstly, it is natural to consider the presence of an island as it arises after Page time during black hole evaporation.
Secondly, the island formula for reflected entropy has been proposed in \cite{Li:2020ceg, Chandrasekaran:2020qtn}. It is interesting to see if this island formula admits a lower bound for Markov gap like \eqref{Hayden_ineq0}.
We explicitly compute the Markov gaps for various phases in a model based on AdS/BCFT correspondence \cite{Takayanagi:2011zk}, called \emph{defect extremal surface} (DES) model \cite{Deng:2020ent,Li:2021dmf}.

In DES model, the RT formula is corrected by the quantum defect theory living on an end-of-the-world (EoW) brane in the bulk.
It is very exciting that by combining AdS/BCFT with braneworld holography \cite{Randall:1999ee,Randall:1999vf,Karch:2000ct}, the island formula emerges in an effective 2d description of DES model \cite{Deng:2020ent}.
The reflected entropy and entanglement negativity has been studied in this model \cite{Li:2021dmf,Basu:2022reu,Shao:2022gpg}.
Our results favor the HPS inequality even in the presence of islands,
if we do not take the boundary of EWCS on brane into account.
Even so, the geometric proof of \eqref{Hayden_ineq0} in our cases is not a trivial extension of HPS's, as generally the EoW brane in the bulk is neither necessarily along geodesics nor at asymptotic infinity.

The inequalities \eqref{Hayden_ineq0} are stated from a bulk point of view, as it incorporates EWCS.
We expect that, for a vacuum state, one can also read off some lower bound for the Markov gap from the boundary theory viewpoint.
This thought, together with our results, prompts us to conjecture that
\begin{align}\label{main_clain_boundary0}
    h(A:B)\geq \frac{c}{3}\log 2\times (\#\text{~of~gaps~between~}A\cup \mathcal I_{R,A}\text{~and~}B\cup \mathcal I_{R,B}),\quad \mathrm{for~} I(A:B)> 0,
\end{align}
where $c$ is the central charge, and $\mathcal I_R$ is the reflected island for the corresponding region at the asymptotic boundary \cite{Chandrasekaran:2020qtn,Deng:2020ent}.
We test the boundary proposal \eqref{main_clain_boundary0} in DES model, JT gravity, and generic 2d extremal black holes for various phases.
And the results satisfy \eqref{main_clain_boundary0} with the same lower bound. 
In addition, we show, using an explicit example, how the lower bound of the Markov gap originates from the OPE coefficient, which may kindle the general proof of \eqref{main_clain_boundary0} in future work.
However, in the most general situations where the boundary region contains multi-intervals, even though the inequality \eqref{main_clain_boundary0} is satisfied, the lower bound given by counting gaps could be underestimated.
We will discuss this in more detail and provide a generalization for multi-interval regions in Sec.\ref{sec:disc}.

This paper is organized as follows.
In Sec.\ref{sec2}, we first introduce reflected entropy, the Markov gap, and the HPS inequality.
Then we propose a DES version and a boundary version of HPS inequality.
In Sec.\ref{sec:des}, we calculate the Markov gap in the DES model for several phases, both disjoint and adjacent, and compare the results with our proposal.
In Sec.\ref{sec:jt}, we do a similar calculation in JT gravity, totally from a boundary point of view, using the island formula.
In Sec.\ref{sec:2dbh}, we calculate the lower bound for the Markov gap in general 2D extremal black hole setups.
In Sec.\ref{sec:disc}, we discuss our results and proposal.
Throughout this paper, we consider only the ground states of the field, and all the phases are assumed to be time-symmetric.
We will use ``$a \simeq b$'' to indicate that $a$ approaches $b$ but $a-b$ still has a relatively small value.

\section{The Markov gap and its bulk and boundary inequalities}\label{sec2}

\subsection{The Markov gap}
Reflected entropy \cite{Dutta:2019gen} was proposed as the von Neumann entropy in a canonically purified state $|\sqrt{\rho_{AB}}\rangle_{ABA^*B^*}$ in the doubled Hilbert space $\left(\mathcal{H}_{A} \otimes \mathcal{H}_{A}^{\star}\right) \otimes\left(\mathcal{H}_{B} \otimes \mathcal{H}_{B}^{\star}\right)$, i.e.
\begin{equation}
    S_R(A:B)=S(AA^*)_{|\sqrt{\rho_{AB}}\rangle},
\end{equation}
which serves as a measure of entanglement between $A$ and $B$.
The reflected entropy has been widely studied in various systems \cite{Akers:2022max,Akers:2022zxr,Bueno:2020vnx,Bueno:2020fle,Camargo:2021aiq,Berthiere:2020ihq}.

In \cite{Hayden:2021gno}, the difference between reflected entropy $S_{R}(A:B)$ and mutual information $I(A:B)=S(A)+S(B)-S(A\cup B)$ is called \emph{Markov gap} $h(A:B)\equiv S_{R}(A:B)-I(A:B)$, as a non-vanishing $h$ precludes a perfect Markov recovery map $\rho_{ABB^*}=\mathcal R_{B\rightarrow BB^*}(\rho _{AB})$.
Moreover, $h$ is considered as a smoking gun of certain tripartite entanglement, that is, a pure state $\ket{\psi}_{ABC}$ is a sum of triangle states iff $h(A:B)=0$ \cite{Akers:2019gcv,Zou:2020bly}.
In \cite{Dutta:2019gen}, the Markov gap is identified with conditional mutual information
\begin{align}
    h(A:B)=I(A:B^*|B)=I(B:A^*|A),
\end{align}
where the conditional mutual information is defined as
\begin{align}
    I(A:C|B)=I(A:BC)-I(A:B).
\end{align}
The Markov gap satisfies the following inequality in information-theoretic language
\begin{align}
    h\geq -\max_{\mathcal R_{B\rightarrow BB^*}}\log F\sbra{
    \rho_{ABB^*},\mathcal R_{B\rightarrow BB^*}(\rho_{AB})
    },
\end{align}
where $F$ is the quantum fidelity
\begin{align}
    F(\rho,\sigma)=\sqbra{\tr\sqrt{\sqrt{\rho}\sigma \sqrt{\rho}}}^2.
\end{align}

\subsection{HPS inequality and its bulk and boundary version in presence of island}
In AdS/CFT correspondence, Hayden, Parrikar and Sorce (HPS) show that the Markov gap satisfies the following inequality
\begin{align}
    h(A:B)&\geq \frac{\ell}{2G_N}\log 2\times (\text{\#~of~cross-section~boundaries})\notag\\
    &=\frac{c}{3}\log 2\times (\text{\#~of~cross-section~boundaries}) \label{Hayden_ineq}
\end{align}
to $\mathcal O(G_{N}^0)$ in pure AdS$_3$ space by using geometric argument \cite{Hayden:2021gno}.
$\ell$ is the AdS radius.
In the second line of Eq. \eqref{Hayden_ineq}, we used Brown-Henneaux formula $c=3\ell/2G_{N}$ \cite{Brown:1986nw}.

The island contribution naturally arises in many situations, for example,  an evaporating black hole.
So it should inevitably be taken into account.
Based on AdS/BCFT correspondence, the defect extremal surface (DES) model has been proposed as the holographic counterpart of the island formula \cite{Deng:2020ent,Li:2021dmf}, that is, the island formula emerges when we consider the effective 2D description of DES model by partial dimension reduction.
Our observation in Sec.\ref{sec:des} will indicate the HPS inequality \eqref{Hayden_ineq} is also obeyed if we do not take the EWCS boundary on brane into account.
Or one can instead make a little modification of HPS's statement
\begin{align}
    h(A:B)&\geq \frac{\ell}{2G_N}\log 2\times (\text{\#~of~cross-section~boundaries~on~the~minimal~surface~of~}AB)\notag\\
    &=\frac{c}{3}\log 2\times (\text{\#~of~cross-section~boundaries~on~the~minimal~surface~of~}AB) \label{main_clain_bulk}.
\end{align}
We also provide a geometric interpretation of our claim \eqref{main_clain_bulk} for DES model in Appendix.\ref{app:geointerpretation} in the case that the brane tension is zero, but in general, this claim remains to be proved.

\begin{figure}
\centering
\includegraphics[width=0.45\textwidth]{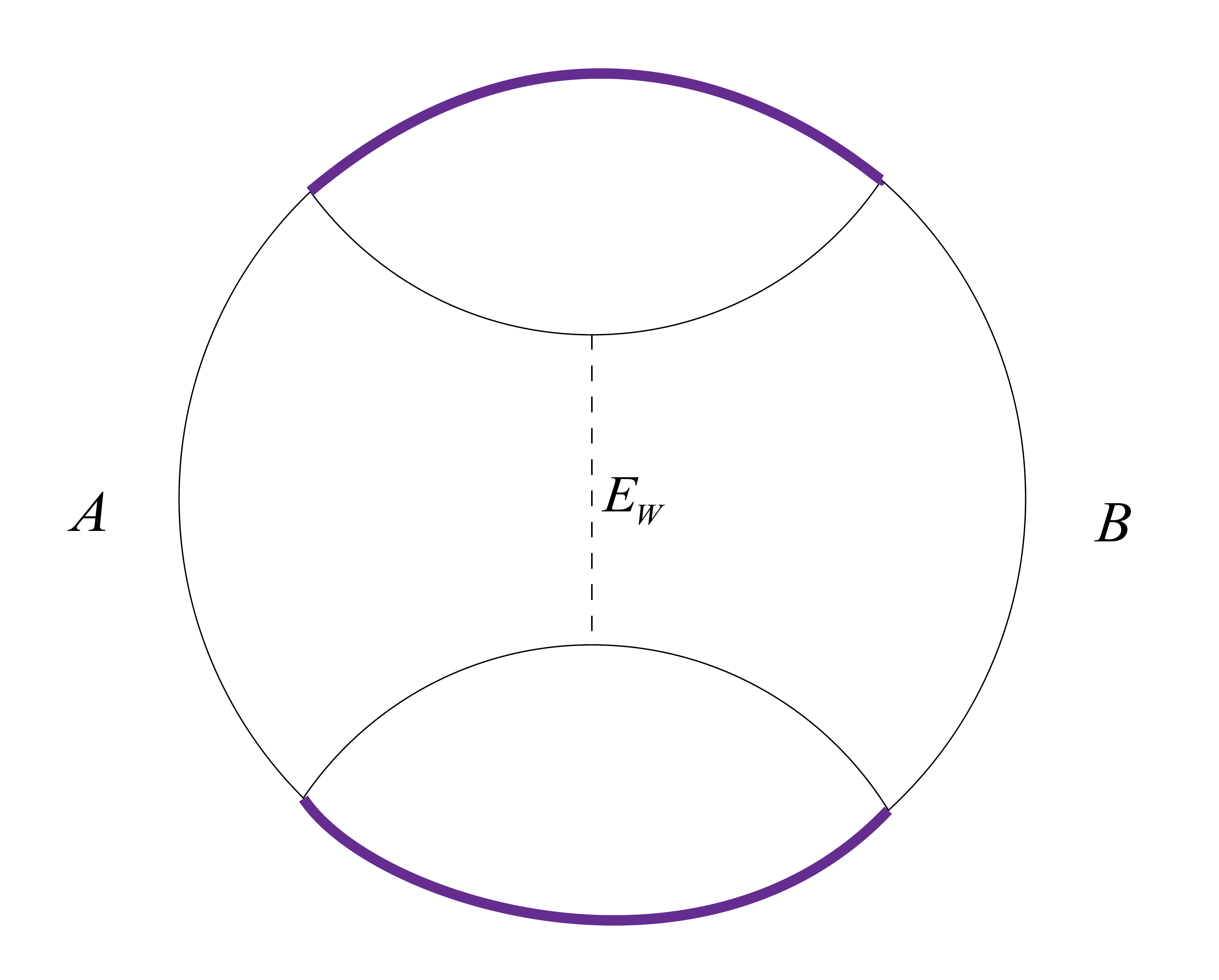}
\includegraphics[width=0.45\textwidth]{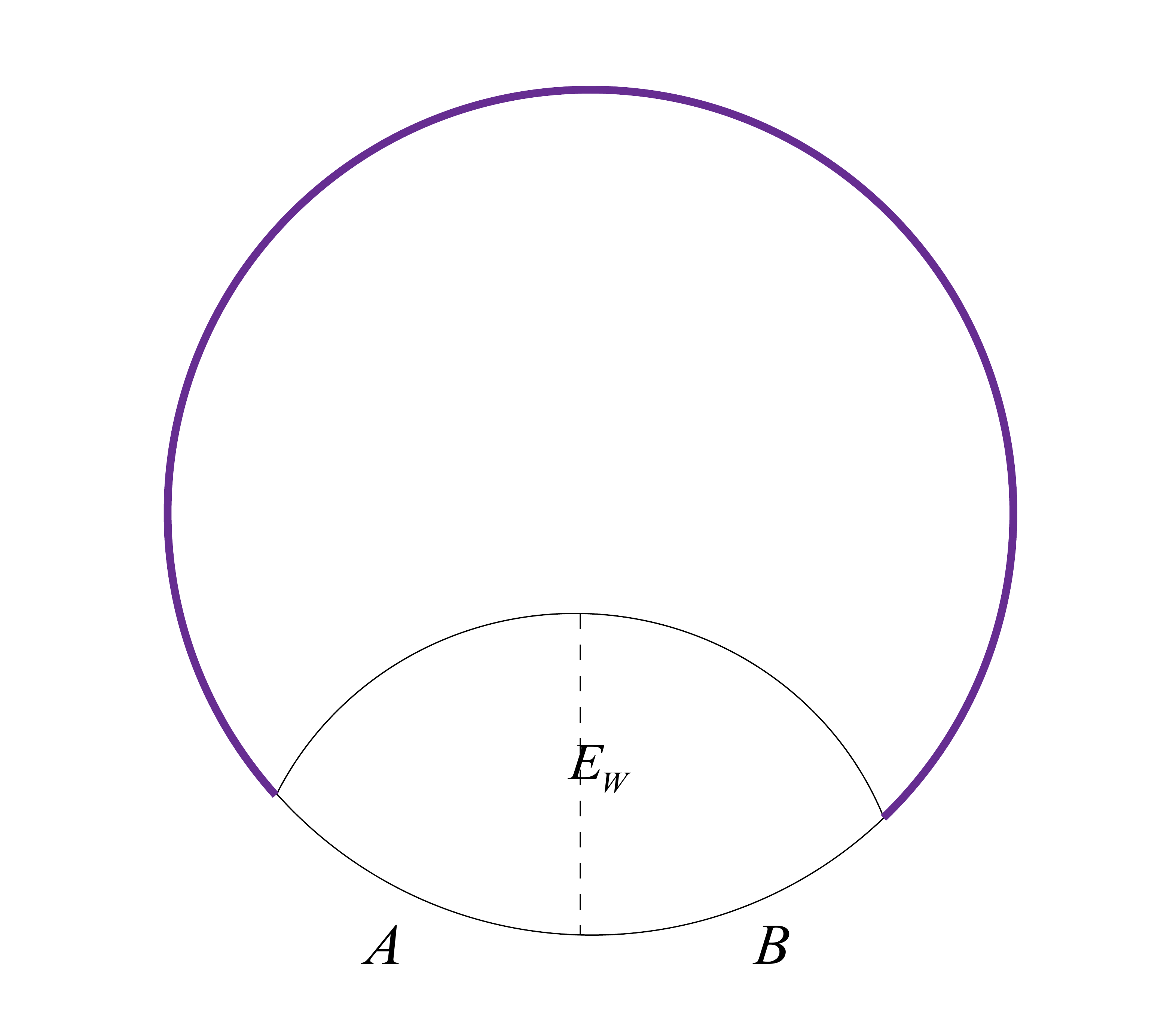}
\caption{The gap (denoted by purple curves) between two boundary regions $A$ and $B$ of a time slice of AdS$_3$ vacuum. 
{The lower bound of the Markov gap for the left phase is $\frac{2c}{3}\log 2$, and $\frac{c}{3}\log 2$ for the right as there is no boundary on asymptotic infinity.}
}
\label{fig:gapnumberbdy}
\end{figure}

On the other hand, \eqref{Hayden_ineq} and \eqref{main_clain_bulk} are counting the lower bound of the Markov gap from the bulk point of view.
In principle, this lower bound can be obtained from boundary theory.
Furthermore, we expect that one is also able to read off some lower bounds from the topology of the boundary regions.
Therefore, based on our observation, we propose that 
\begin{align}\label{main_clain_boundary}
    h(A:B)\geq \frac{c}{3}\log 2\times (\#\text{~of~gaps~between~}A\cup \mathcal I_{R,A}\text{~and~}B\cup \mathcal I_{R,B}),\quad \mathrm{for~} I(A:B)> 0,
\end{align}
where $\mathcal I_R$ is the island for reflected entropy.
In fact, our inequality \eqref{main_clain_boundary} also works without an island.
As shown in Fig.\ref{fig:gapnumberbdy}, for pure AdS$_3$, there are two gaps between two disjoint boundary intervals and one gap between two adjacent boundary intervals and thus according to our inequality \eqref{main_clain_boundary}, we have $h\geq \frac{2c}{3}\log 2$ for disjoint intervals and 
$h \geq \frac{c}{3}\log 2$ for adjacent intervals, which are consistent with the explicit calculation \cite{Takayanagi:2017knl,Zou:2020bly} and HPS inequality \eqref{Hayden_ineq}.
Mind that on a {time slice of vacuum CFT}, a region containing infinity $\infty$ is also regarded as a gap.

In Sec.\ref{sec:des} and Sec.\ref{sec:jt}, we will show that \eqref{main_clain_boundary} holds generally for DES model, JT gravity and generic 2D extremal black holes.
Phases with disjoint and adjacent intervals will be considered separately. 
Before going deep into the detailed calculations of the Markov gap for DES models, we will also  qualitatively analyze the recovery map of these phases, following the analysis in \cite{Hayden:2021gno}, which will enlighten the physical origin of the Markov gap of these phases.

\section{The Markov gap in defect extremal surface model}\label{sec:des}
We calculate the Markov gap in defect extremal surface (DES) model for both disjoint and adjacent phases.
To derive the lower bound of Markov gap, sometimes we must use the conditions for the phase to exist, which will be listed if necessary.

\subsection{Review of DES}
\begin{figure}
\centering
\includegraphics[width=0.45\textwidth]{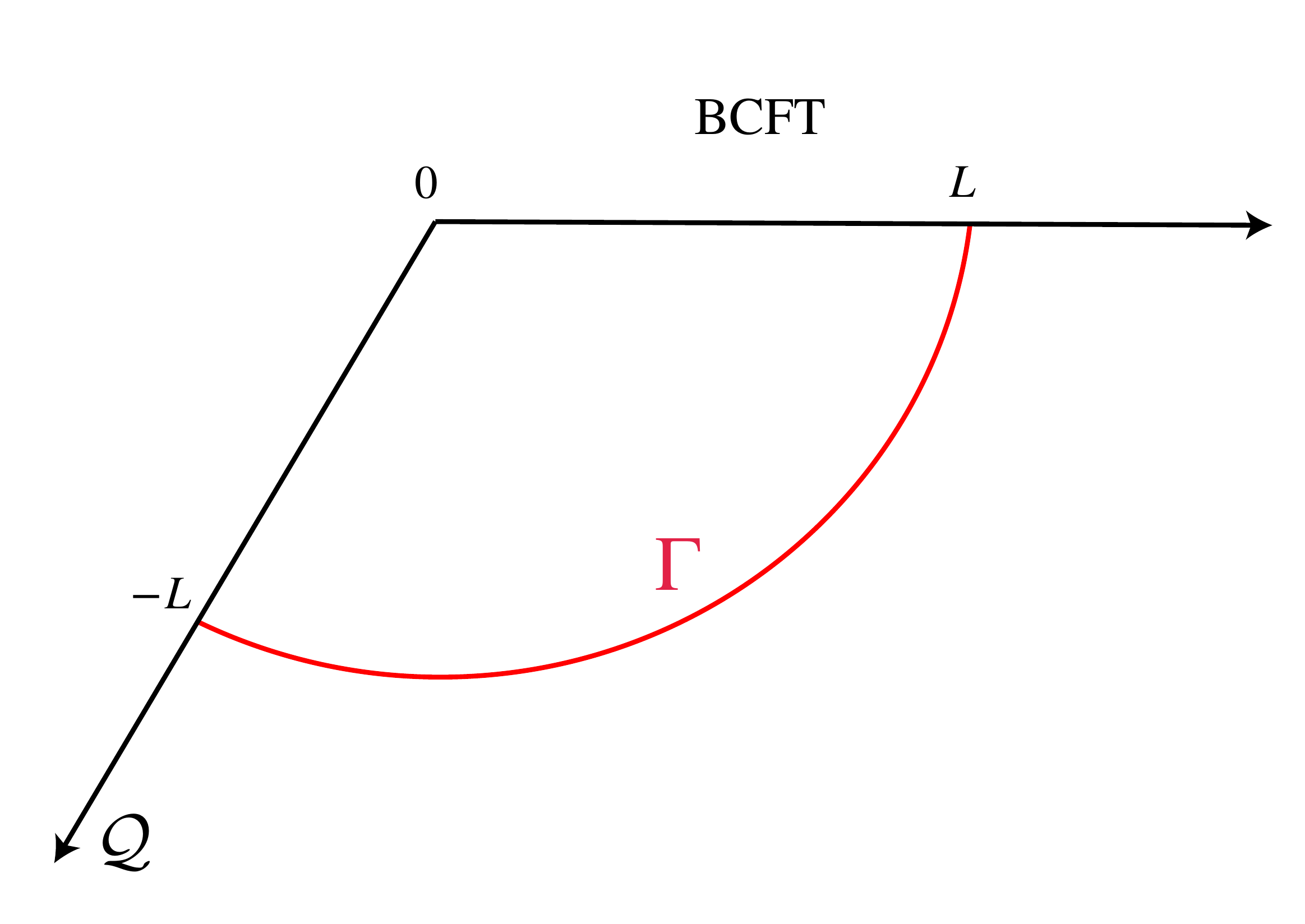}
\caption{The holographic dual  of BCFT on a half space in vacuum.}
\label{fig:adsbcft}
\end{figure}
{In this section, we will review the entanglement entropy and reflected entropy of the DES model \cite{Deng:2020ent,Li:2021dmf}.

\subsubsection{Entanglement entropy}
DES model is based on AdS/BCFT correspondence.
The holographic dual of a 2d BCFT on a half space in a vacuum state is known to be a part of Poincaré AdS$_3$ + end-of-the-world (EoW) brane $\mathcal Q$ where the Neumann boundary condition is imposed. 
The AdS$_3$ geometry is given by
\begin{align}
    \D s^2&=\D\rho^2+\ell^2\cosh^2\frac{\rho}{\ell}\frac{-\D t^2+\D w^2}{w^2}\\
   &= \frac{\ell^2}{y^2}\sbra{
    -\D t^2+\D x^2+\D y^2},
\end{align}
where these coordinates are related via
\begin{align}
    y=-\frac{w}{\cosh(\rho/\ell)},\quad x=w\tanh\frac{\rho}{\ell}.
\end{align}
The EoW brane $\mathcal Q$ lives at 
\begin{equation}
    y=\lambda x,
\end{equation}
where $\lambda$ is related to the brane tension $T$ by
\begin{equation}
    \lambda=\sqrt{\frac{1}{\ell^2T^2}-1}.
\end{equation}
The boundary BCFT lives at $x\geq0$.
The entanglement entropy for an interval $[0,L]$ on BCFT in the ground state is given by the area of the RT surface $\Gamma$, which connects the endpoint $L$ and EoW brane (see Fig.\ref{fig:adsbcft}),
\begin{align}
    S([0,L])=\frac{\text{Area}(\Gamma)}{4G_N}=\frac{c}{6}\log\frac{2L}{\epsilon}+\frac{c}{6}\text{arctanh}(\sin\theta_0),
\end{align}
where $\theta$ is defined as $(\cos\theta)^{-1}=\cosh(\rho/\ell)$ and the second term $\frac{c}{6}\text{arctanh}(\sin\theta_0)$ is the boundary entropy of BCFT.

If the brane has zero tension or no matter is on the brane, the brane is orthogonal to the BCFT at the boundary of CFT, which is our origin.
Now we add CFT matter on the brane and turn on the tension, and the brane will no longer be orthogonal to BCFT \cite{Deng:2020ent}.
This brane can be regarded as a defect in the bulk.
Holographically, the matter field on the brane also contributes to the entanglement entropy of a BCFT region and now the entanglement entropy on BCFT is given by defect extremal surface (DES) proposal
\begin{equation}
    S_{\rm DES}=\min_{\Gamma,X}\left\{
    \mathrm{ext}_{\Gamma,X}\sqbra{
    \frac{\mathrm{Area}(\Gamma)}{4G_{N}}+S_{\rm Defect}[D]
    }
    \right\},\quad X\equiv \Gamma\cap D,
\end{equation}
in which $\Gamma$ is the defect extremal surface (the corresponding minimal surface).
$S_{\rm Defect}[D]$ is the bulk semi-classical entanglement entropy and $D$ is a region on the brane where the bulk matter live on.

\subsubsection*{Intervals on the brane}
Let us first consider $S_{\rm Defect}[D]$.
For an brane interval $D=[x_1,0]$ with $x_1<0$ and touching the BCFT, we have
\begin{align}
    S_{\rm Defect}([x_1,0])=\frac{c'}{6}\log\sbra{\frac{2\ell}{\epsilon_w\cos\theta_0}}+\log g,
\end{align}
where $\log g$ corresponds to the boundary entropy of the brane \cite{Takayanagi:2011zk,Affleck:1991tk}, and $c'$ is the central charge for CFT on the brane.
In this paper, we take $c'=c$ and $\log g=0$.
We see that entanglement entropy does not depend on the length of the interval on the brane.
This nice fact makes the calculation in this model tractable with ease.

For an brane interval $D=[x_1,x_2]$ with $x_1<x_2<0$, $S_{\rm Defect}(D)$ possesses two phases that correspond to the dominance of the two channels: the bulk operator product expansion (OPE) and the boundary operator product expansion (BOE) \cite{Deng:2020ent,Sully:2020pza}
\begin{align}
    S_{\rm Defect}([x_1,x_2])=\frac{c}{3}\log\frac{2\ell}{\epsilon_w\cos\theta_0},\quad \eta>1
\end{align}
and 
\begin{align}
     S_{\rm Defect}([x_1,x_2])=\frac{c}{6}\log\sbra{\frac{\ell^2}{\epsilon_w^2}
     \frac{(x_1-x_2)^2}{x_1x_2\cos\theta_0}},\quad \eta<1,
\end{align}
where $\eta$ is given by
\begin{equation}
    \eta(x_1,x_2)=\frac{(x_1-x_2)^2}{4x_1x_2}.
\end{equation}

\subsubsection*{Intervals on BCFT}
\begin{figure}
\centering
\includegraphics[width=0.45\textwidth]{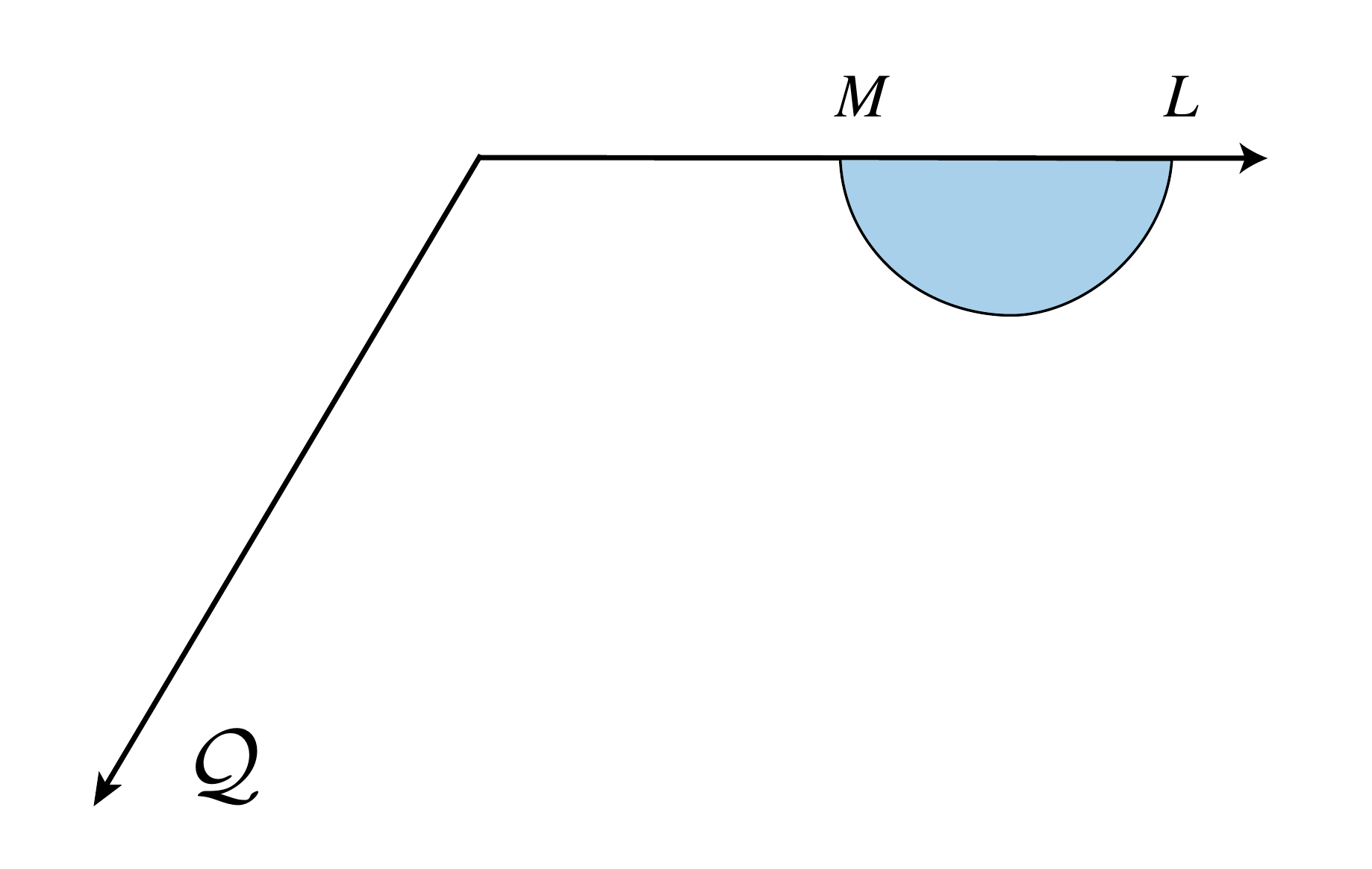}
\includegraphics[width=0.45\textwidth]{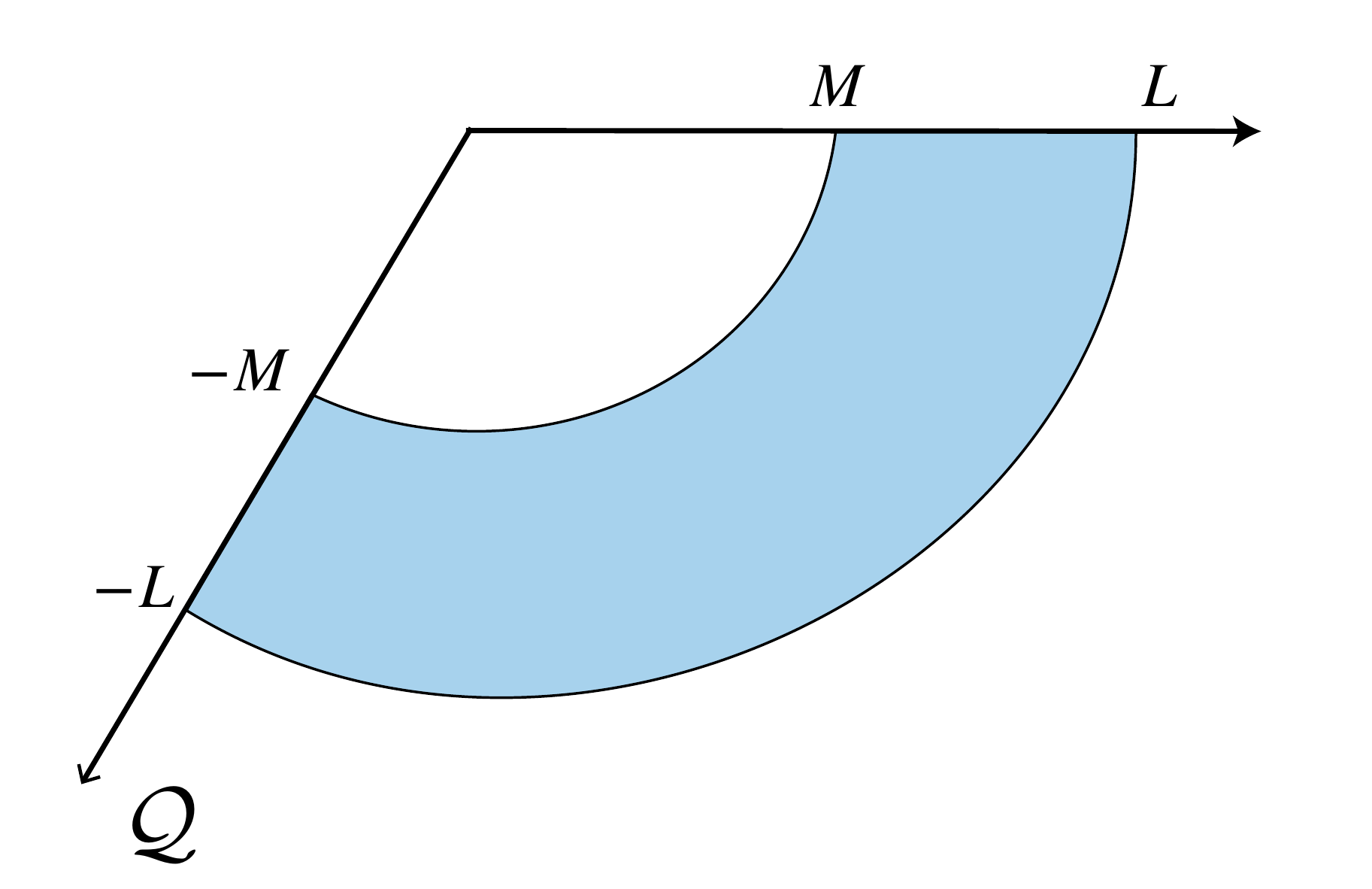}
\caption{The entanglement entropy phases for an interval $[M,L]$ in DES model.
Left: connected phase. 
Right: disconnected phase.}
\label{fig:des-two-phases}
\end{figure}
Now let us consider $S_{\text{DES}}$.
For an interval $[0,L]$ on BCFT, $S_{\rm Defect}$ is irrelevant to length or position on the brane, thus the extremization and minimization procedures reduce to finding a shortest geodesic distance between brane and a boundary point and the result is
\begin{align}\label{SDES}
    S_{\rm DES}([0,L])&=\frac{c}{6}\log\frac{2L}{\epsilon}+\frac{c}{6}\mathrm{arctanh}(\sin\theta_0)+\frac{c}{6}\log\sbra{\frac{2\ell}{\epsilon_w\cos\theta_0}}\notag\\
    &=\frac{c}{6}\log\frac{2L}{\epsilon}+\frac{c}{6}T(\theta_0)+\frac{c}{6}W(\theta_0),
\end{align}
where we defined 
\begin{equation}
T(\theta)\equiv\text{arctanh}(\sin\theta),\quad W(\theta)\equiv \log\frac{2\ell}{\epsilon_w\cos\theta}.
\end{equation}
For an interval $[M,L]$ on BCFT, there are two phases for $S_{\rm DES}$ (see Fig.\ref{fig:des-two-phases}).
The connected phase 
\begin{equation}
    S_{\rm DES}([M,L])=\frac{c}{3}\log\frac{L-M}{\epsilon},\quad \eta(M,L)<\eta_c(M,L),
\end{equation}
and the disconnected phase
\begin{equation}\label{S_ML}
    S_{\rm DES}([M,L])=\frac{c}{6}\sqbra{
    \log\frac{4LM}{\epsilon^2}+2\mathrm{arctanh}\sbra{\sin\theta_0}+2\log\frac{2\ell}{\epsilon_w\cos\theta_0}
    },\quad \eta(M,L)>\eta_c(M,L)
\end{equation}
where the critical point is
\begin{equation}\label{etac}
    \eta_c(M,L)=\exp\sqbra{2T(\theta_0)+2W(\theta_0)}>1.
\end{equation}
Note that there is an extremal value for $S_{\rm DES}$ only when $\eta>1$, which is automatically satisfied as $\eta_c>1$.

\subsubsection*{DES formula v.s. Island formula}
One can also seek the effective 2D boundary description for the DES model by taking partial dimension reduction, then the island formula emerges and gives the same result for entanglement entropy as the DES formula \cite{Deng:2020ent}.
To be specific, as shown in Fig.\ref{fig:des-island}, insert an imaginary boundary $\mathcal{Q}'$ with $(t,x,y)=(t,0,y)$ to decompose the bulk into two parts $W_1$ and $W_2$. 
For $W_1$, by performing  Randall-Sundrum reduction along $\rho$ direction, one could obtain a 2d gravity theory + CFT on the brane $\mathcal{Q}$ with the area term 
\begin{equation}
   S_{\text {area }}= \frac{1}{4 G_N^{\text{brane}}}=\frac{\rho_0}{4 G_N}=\frac{c}{6} \operatorname{arctanh}\left(\sin \theta_0\right).
\end{equation}
For $W_2$, we choose its dual BCFT description on the half-space boundary. 
Ultimately, we arrive at the 2D effective description of DES model.
By using boundary QES formula in this 2D boundary, the entanglement entropy for an interval $[0,L]$ agrees with the DES result, that is, 
\begin{align}
    S_{\rm QES}=&
    \text{Min}_a\{S_{\text {area }}(-a)+S_{\text {CFT }}([-a, L])\}\\\notag
    =&\frac{c}{6}\log\frac{2L}{\epsilon}+\frac{c}{6}\mathrm{arctanh}(\sin\theta_0)+\frac{c}{6}\log\sbra{\frac{2\ell}{\epsilon_w\cos\theta_0}}\\
    =&S_{\text{DES}},
\end{align}
where $S_{\text {CFT }}$ is the effective entropy of CFT in this 2D boundary and 
\begin{equation}
    S_{\text {CFT }}([w_1, w_2])=\frac{c}{6} \log \left(\frac{\left|w_1-w_2\right|^2}{\epsilon_{w_1} \epsilon_{w_2} \Omega\left(w_1, \bar{w}_1\right) \Omega\left(w_2, \bar{w}_2\right)}\right),\ \D s^2=\Omega^{-2} \D w \D \bar{w}.
\end{equation}

\begin{figure}
\centering
\includegraphics[width=0.9\textwidth]{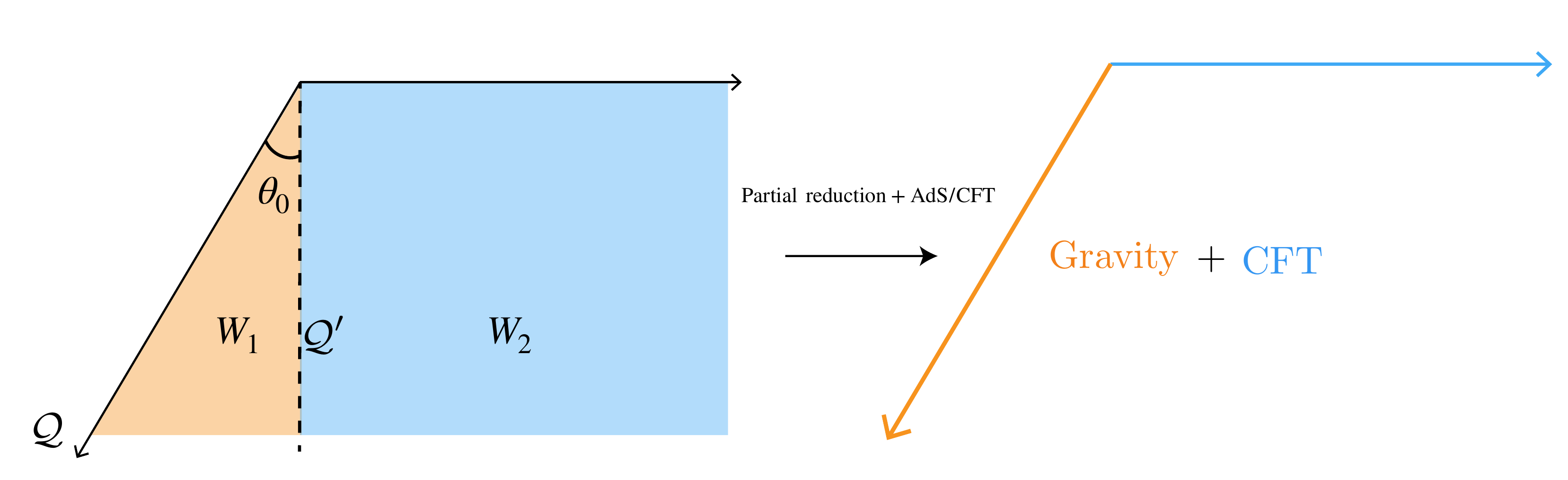}
\caption{2D effective description of DES model by performing partial reduction for $W_1$ and AdS/CFT for $W_2$.}
\label{fig:des-island}
\end{figure}

\subsubsection{Reflected entropy}
In DES model, the reflected entropy can be understood in both boundary theory and bulk theory viewpoints.
In boundary island point of view, the reflected entropy is proposed to be \cite{Li:2021dmf}
\begin{align}\label{DES_SR_proposal1}
    S^{\rm bdy}_R(A:B)=\min\text{ext}_{\gamma}\sqbra{
    S_{R}^{\rm eff}(A\cup \mathcal I_A:B\cup \mathcal I_B)+\frac{\text{Area}(\gamma)}{2G_N^{\rm brane}}
    },
\end{align}
where $\gamma=\partial \mathcal I_A\cap \partial \mathcal I_B$ is the intersection of entanglement wedge cross-section and brane $\mathcal Q$. 
In the bulk point of view, the reflected entropy is given by
\begin{align}\label{DES_SR_proposal2}
    S_R^{\text {bulk }}(\mathcal{A}: \mathcal{B})=\min \operatorname{ext}_{\mathrm{EW}}\left\{S_R^{(\mathrm{eff})}(\mathcal{A}: \mathcal{B})+\frac{\operatorname{Area}[\mathrm{EW}(A: B)]}{2 G_N}\right\}.
\end{align}
The two proposals are equivalent. 

For phase-D1 (Fig.\ref{fig:D1}) where the entanglement wedge\footnote{
    In this paper, we refer to the codimension-1 surface bounded by the boundary region and its codimension-2 minimal surface as the entanglement wedge because we assume time symmetry.
} of $AB$ is just the disconnected union of entanglement wedges of $A$ and $B$, the reflected entropy vanishes
\begin{equation}
S_{R}=0=E_W.
\end{equation}
Throughout this paper, {without loss of generality,} we assume that region $B$ is large enough so that it always receives the island contribution.

In phase-D2 (Fig.\ref{Fig:D2}), the entanglement wedge of $AB$ is connected ($B$ has an entanglement island while $A$ does not), and the entanglement wedge cross-section is the minimal geodesic with two endpoints on RT surfaces of $[0,b_1]$ and $[b_2,b_3]$. 
Given that the reflected entropy $S_R$ is dual to the area of EWCS, then we have
\begin{equation}\label{SR_phase2}
S_{R}(A:B)=\frac{c}{6}\log\sqbra{
\frac{b_2b_3-b_1^2+\sqrt{(b_2^2-b_1^2)(b_3^2-b_1^2)}}{b_2b_3-b_1^2-\sqrt{(b_2^2-b_1^2)(b_3^2-b_1^2)}}
}.
\end{equation}
\eqref{SR_phase2} is just twice the distance between two parallel geodesics in hyperbolic space $\D s^2=\ell^2(\D x^2+\D y^2)/y^2$ in unit of $\frac{1}{4G_N}$.
We leave the derivation of \eqref{SR_phase2} in Appendix.\ref{app1}. 
\eqref{SR_phase2} can be also obtained by employing the cross-section formula in \cite{Takayanagi:2017knl}
\begin{equation}\label{takayanagi_SR}
  E_W=\frac12 S_{R}=\frac{c}{6}\log\sbra{1+2z+2z\sqrt{1+1/z}},
\end{equation}
with the cross-ratio here 
\footnote{
The correct expression of the cross-ratio $z$ is vital for our calculation.
$z$ here is different from that in \cite{Li:2021dmf}.
The authors of \cite{Li:2021dmf} obtain the formula by using the result in \cite{Takayanagi:2017knl,Caputa:2018xuf}, but the situation here is slightly different.
In \cite{Takayanagi:2017knl}, the authors calculated the cross section between two intervals $[a_1,a_2]$ and $[b_1,b_2]$ and thus the cross ratio there is $z=\frac{(a_2-a_1)(b_2-b_1)}{(b_1-a_2)(b_2-a_1)}$, while here in phase-D2, we calculate the cross section between two intervals $[b_1,b_2]$ and $[b_3,-a_1]$ and thus the cross ratio here is $z=\frac{(b_3+a_1)(b_2-b_1)}{(b_1+a_1)(b_3-b_2)}$.
}
\begin{equation}
    z=\frac{(b_3+a_1)(b_2-b_1)}{(b_1+a_1)(b_3-b_2)},
\end{equation}
where $a_1$ is the island boundary for $[0,b_1]$, or simply $b_1$.

In phase-D2, on the other hand, one can also compute the reflected entropy using twist operators as was done in Refs.\cite{Dutta:2019gen,Chandrasekaran:2020qtn}
\begin{equation}\label{SR_Faulkner}
    S_{R}(A:B)=\frac{2c}{3}\log\sbra{
    \frac{1+\sqrt{1-x}}{\sqrt{x}}},\quad x=\frac{(b_1+a_1)(b_3-b_2)}{(a_1+b_2)(b_3-b_1)}.
\end{equation}
Again, the cross-ratio $x$ here should be properly chosen, and one easily checks that \eqref{SR_Faulkner} gives the same result as \eqref{SR_phase2} and \eqref{takayanagi_SR}.

For phase-D4 (Fig.\ref{fig:D4}) where $A$ and $B$ both have their islands and the entanglement wedge of $AB$ is connected, the reflected entropy is \cite{Li:2021dmf}
\begin{align}\label{SR_phase3}
S_{R}(A:B)&=\frac{c}{3}\log\frac{(b_3+a')(b_2+a')}{(b_3-b_2)a'}+\frac{c}{3}T(\theta_0)+\frac{c}{3}W(\theta_0),\quad a'=\sqrt{b_2b_3}
\end{align}
where $a'$ is the \emph{island cross-section}.

\subsection{The Markov gap}
In the following, we compute the Markov gap in several phases in DES model.
The goal of this section is to show that the Markov gap in DES model satisfies the inequalities \eqref{main_clain_bulk} and \eqref{main_clain_boundary}.
In some phases, we will present the conditions for the dominance of the phase.
These conditions are given by some inequalities that will be useful for later calculation.
\subsubsection{Disjoint intervals}
We first consider the two regions $A=[b_1,b_2]$ and $B=[b_3,b_4]$ are disjoint.
We assume that $b_4$ is large enough so that $B$ always admits the island.
\subsubsection*{Phase-D1}

\begin{figure}
\centering
\includegraphics[width=0.45\textwidth]{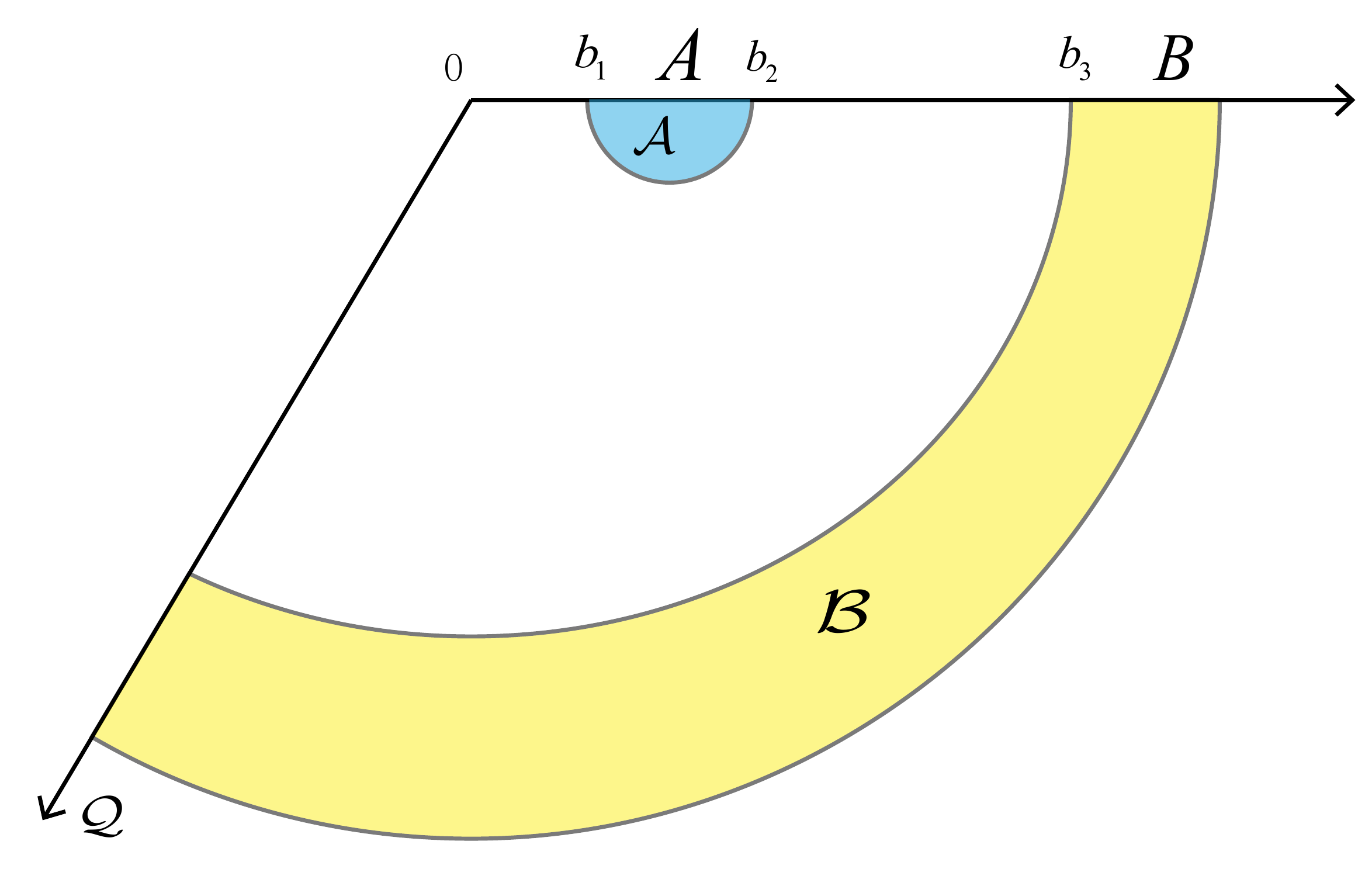}
\caption{Phase-D1. {The blue and yellow regions are the entanglement wedges of $A$ and $B$, respectively. }}
\label{fig:D1}
\end{figure}

\begin{figure}
\centering
\includegraphics[width=0.5\textwidth]{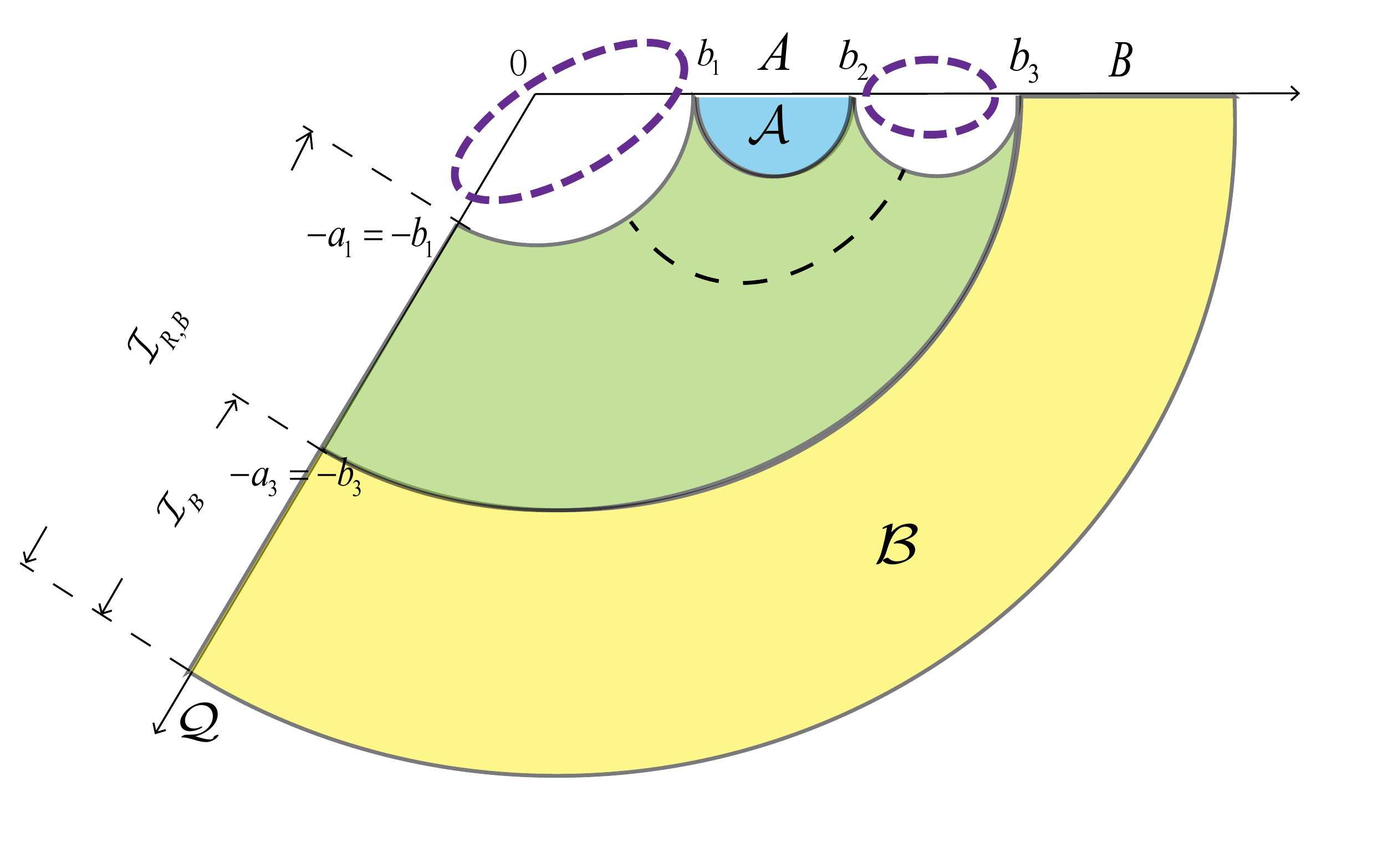}
\includegraphics[width=0.45\textwidth]{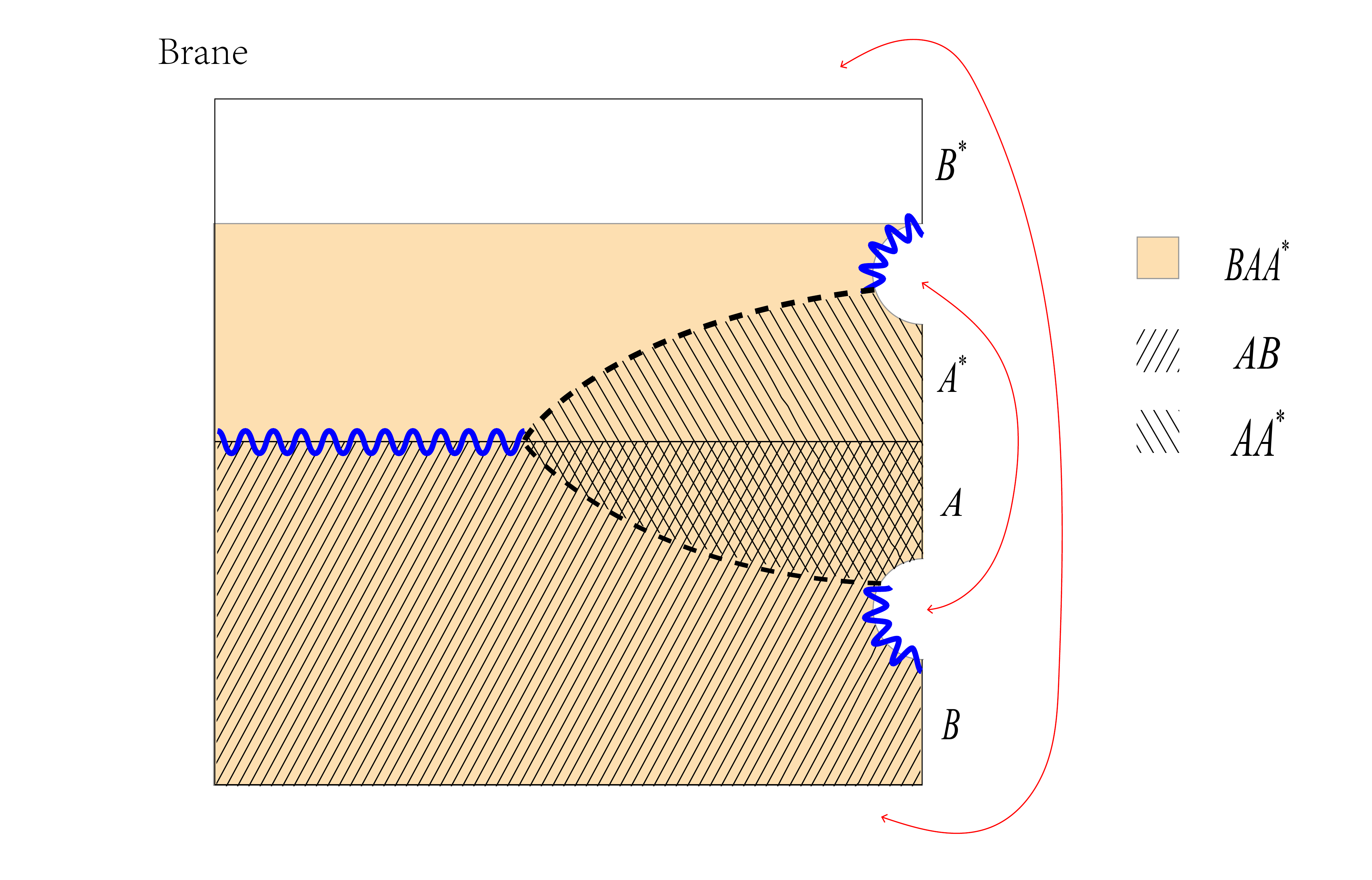}
\caption{Phase-D2.
{Left: The blue and yellow regions denote the entanglement wedges of $A$ and $B$, respectively.
{The green region denotes the shared entanglement wedge of $AB$. }
The black dashed line denotes the entanglement wedge cross-section between $A$ and $B$.
{From the 2d boundary viewpoint, there are islands on the brane.}
$\mathcal{I}$ and $\mathcal{I}_R$ denote the island for entanglement entropy and reflected entropy, respectively.
The purple dashed circles are the boundary gap between $A\cup \mathcal{I}_{R,A}$ and $B\cup \mathcal{I}_{R,B}$, and there are two gaps for phase-D2.
Right: The canonical purification of $\rho_{AB}$ and the entanglement wedges for $AB, AA^*, BAA^*$.
The red double-headed arrows denote the identification between two spacelike surfaces of the entanglement wedges.
The small tubular neighborhoods of two blue jagged lines here are completely visible to $BAA^*$, while not to $AB$ or $AA^*$.
The entanglement wedges satisfy $\mathrm W(AA^*B)\supset\mathrm W(AB)\cup \mathrm W(AA^*)$.
This signals the non-vanishing Markov gap for this phase.}}
\label{Fig:D2}
\end{figure}

In phase-D1, the entanglement wedge of $AB$ is disconnected, Fig.\ref{fig:D1}.
The reflected entropy and mutual information vanish
\begin{equation}
S_{R}(A:B)=I(A:B)=0.
\end{equation}
Thus the Markov gap is $h=S_R-I=0$.

\subsubsection*{Phase-D2}

In phase-D2, on one hand, there is no island for $A=[b_1,b_2]$ and $B=[b_3,b_4]$, which leads to the following inequalities
\begin{align}
	\log\frac{(b_2-b_1)^2}{4b_1b_2}&\leq 2T(\theta_0)+2W(\theta_0)\label{14},\\
\log\frac{(b_3-b_2)^2}{4b_2b_3}&\leq 2T(\theta_0)+2W(\theta_0)\label{15}.
\end{align}
On the other hand, we require that the entanglement wedge of $[-b_1,b_1]\cup [b_2,b_3]$ is disconnected, which is equivalent to a non-vanishing mutual information between $A$ and $B=[b_3,b_4]$: $I(A:B)\geq 0$.
This condition gives
\begin{align}
I(A:B)=&\frac{c}{6}\log\sqbra{\frac{(b_2-b_1)^2}{4b_1b_2}
\frac{4b_2b_3}{(b_3-b_2)^2}
}\label{17}
\\
=&\frac{c}{6}\log\frac{\eta(b_1,b_2)}{\eta(b_2,b_3)}> 0\notag\\
\Rightarrow \eta(b_1,b_2)&> \eta(b_2,b_3).
\end{align}
Before going deep into the computation of the Markov gap, let us analyze the Markov recovery $\rho_{BAA^*}=\mathcal{R}_{A\to AA^*}(\rho_{AB})$ first, using the same argument as in \cite{Hayden:2021gno}.
In Fig.\ref{Fig:D2}, we stretched the canonical purification of $\rho_{AB}$ and the entanglement  wedges of $AB,AA^*,BAA^*$.
For phase-D2, the entanglement wedge of $AB$ together with $AA^*$ cannot cover all the entanglement wedge of $BAA^*$
so that there are two jagged lines whose small tubular neighborhoods  are completely visible to $BAA^*$, but not to
$AA^*$ and $AB$.
Thus the Markov recovery $\rho_{BAA^*}=\mathcal{R}_{A\to AA^*}(\rho_{AB})$ must be precluded and we expect a non-vanishing Markov gap for phase-D2.

Now we compute the Markov gap.
The entanglement entropy for $A$ is
\begin{equation}
S(A)=\frac{c}{3}\log\frac{b_2-b_1}{\epsilon},
\end{equation}
and for $B$ with an island
\begin{equation}
S(B)=\frac{c}{6}\sqbra{\log\frac{4b_4b_3}{\epsilon^2}
+2T(\theta_0)+2W(\theta_0)
}.
\end{equation}
And the entanglement entropy for $AB$ is
\begin{equation}
S(AB)=\sum_iS_{\mathrm{RT}_i}+S_{\rm defect}.
\end{equation}
The areas of the RT and DES surfaces are given by 
\begin{align}
S_{\mathrm{RT}_1}&=\frac{c}{6}\log\frac{2b_4}{\epsilon}+\frac{c}{6}T(\theta_0),\\
S_{\mathrm{RT}_2}&=\frac{c}{6}\log\frac{2b_1}{\epsilon}+\frac{c}{6}T(\theta_0),\\
S_{\mathrm{RT}_3}&=\frac{c}{3}\log\frac{b_3-b_2}{\epsilon}.
\end{align}
Then 
\begin{align}
S(AB)=\frac{c}{6}\log\frac{4b_1b_4(b_3-b_2)^2}{\epsilon^4}+\frac{c}{3}T(\theta_0)+\frac{c}{3}W(\theta_0).
\end{align}
The mutual information is then
\begin{align}\label{phase2_mutual}
I(A:B)=\frac{c}{6}\log\sqbra{\frac{b_3}{b_1}\sbra{\frac{b_2-b_1}{b_3-b_2}}^2}.
\end{align}

The reflected entropy in this phase is given by \eqref{SR_phase2}, with which we get the Markov gap
\begin{align}
    h=\frac{c}{6}\log\sqbra{
\frac{b_2b_3-b_1^2+\sqrt{(b_2^2-b_1^2)(b_3^2-b_1^2)}}{b_2b_3-b_1^2-\sqrt{(b_2^2-b_1^2)(b_3^2-b_1^2)}}
\frac{b_1}{b_3}\frac{(b_3-b_2)^2}{(b_2-b_1)^2}
}.
\end{align}
It is direct to see that $\partial_{b_3}h >0$.
$h$ monotonically increase with $b_3$ with minimum at $b_3\simeq b_2$.
Hereafter by ``$b_3\simeq b_2$'' we mean we let $b_2\rightarrow b_3$ but assume the gap $[b_2,b_3]$ still exists so that the phase still makes sense.
In the limit $b_3\simeq b_2$, we obtain
\begin{align}
    h=\frac{c}{3}\log\sbra{\sqrt{\frac{b_1}{b_2}}+\sqrt{\frac{b_2}{b_1}}}+\frac{c}{3}\log 2\geq \frac{2c}{3}\log 2,
\end{align}
which is saturated at $b_1\simeq b_2\simeq b_3$.
Note that the reflected entropy $S_R$, the mutual information $I$ and thus $h$ are independent of $\theta_0$.
So one can always tune $\theta_0$ to make this phase happen, that is, to satisfy \eqref{14} and \eqref{15}.

{In Appendix.\ref{app:geointerpretation}, we also give a geometric interpretation of this lower bound in the case that the brane tension is zero.
For phase-D2, both boundaries of the cross-section are on the minimal surfaces of $AB$ and here both  inequalities \eqref{Hayden_ineq} and  \eqref{main_clain_bulk} give the same lower bound $\frac{2c}{3}\log 2$, which is consistent with our calculation above. 
On the other hand, from the island boundary viewpoint, there are two gaps (the purple dashed circles in Fig.\ref{Fig:D2}) between $A\cup \mathcal I_{R,A}$ and  $B\cup \mathcal I_{R,B}$, which also indicates the lower bound $\frac{2c}{3}\log 2$ according to our boundary inequality \eqref{main_clain_boundary}.
}

\begin{figure}
\label{fig:newphase2}
\centering
\includegraphics[width=0.5\textwidth]{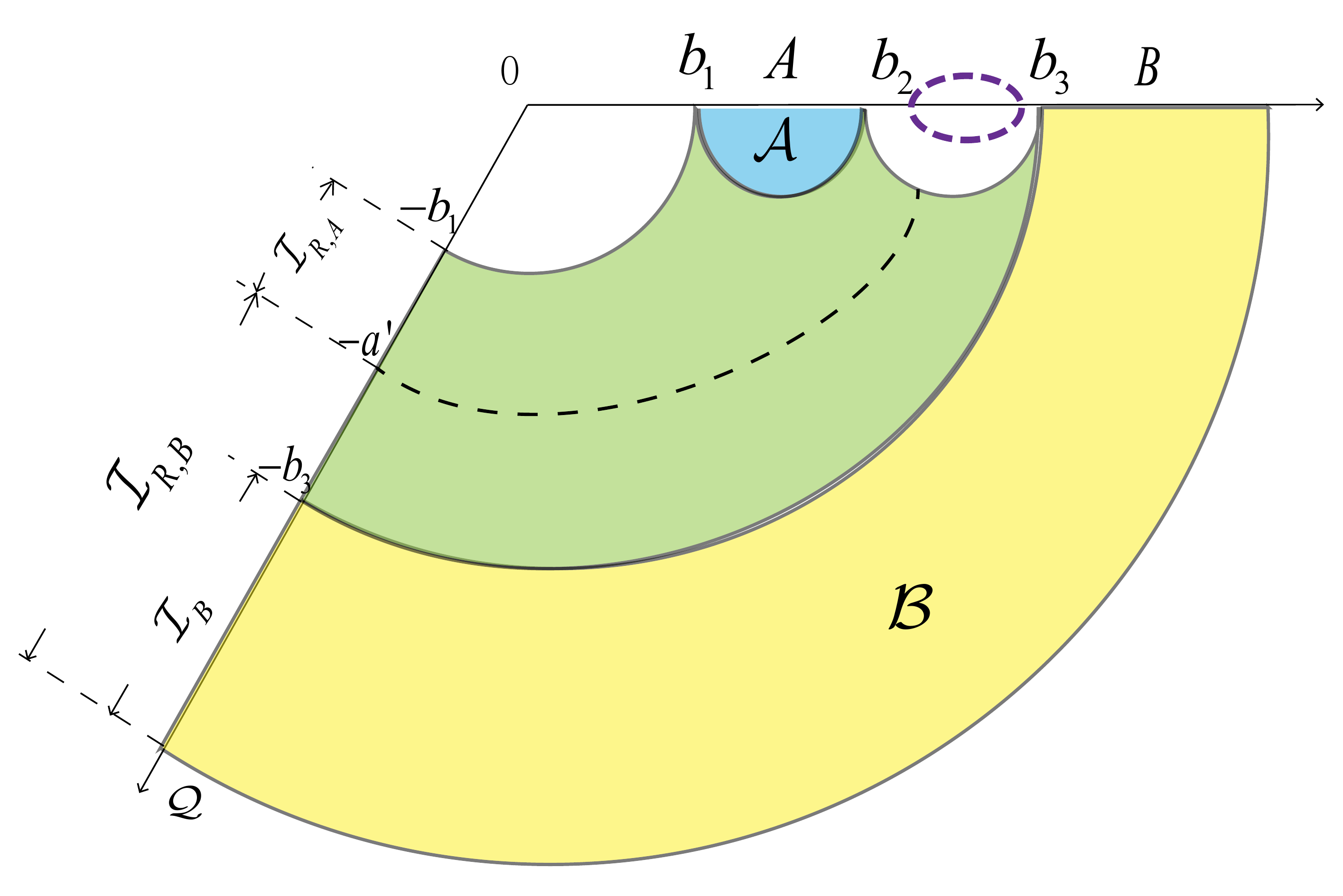}
\includegraphics[width=0.46\textwidth]{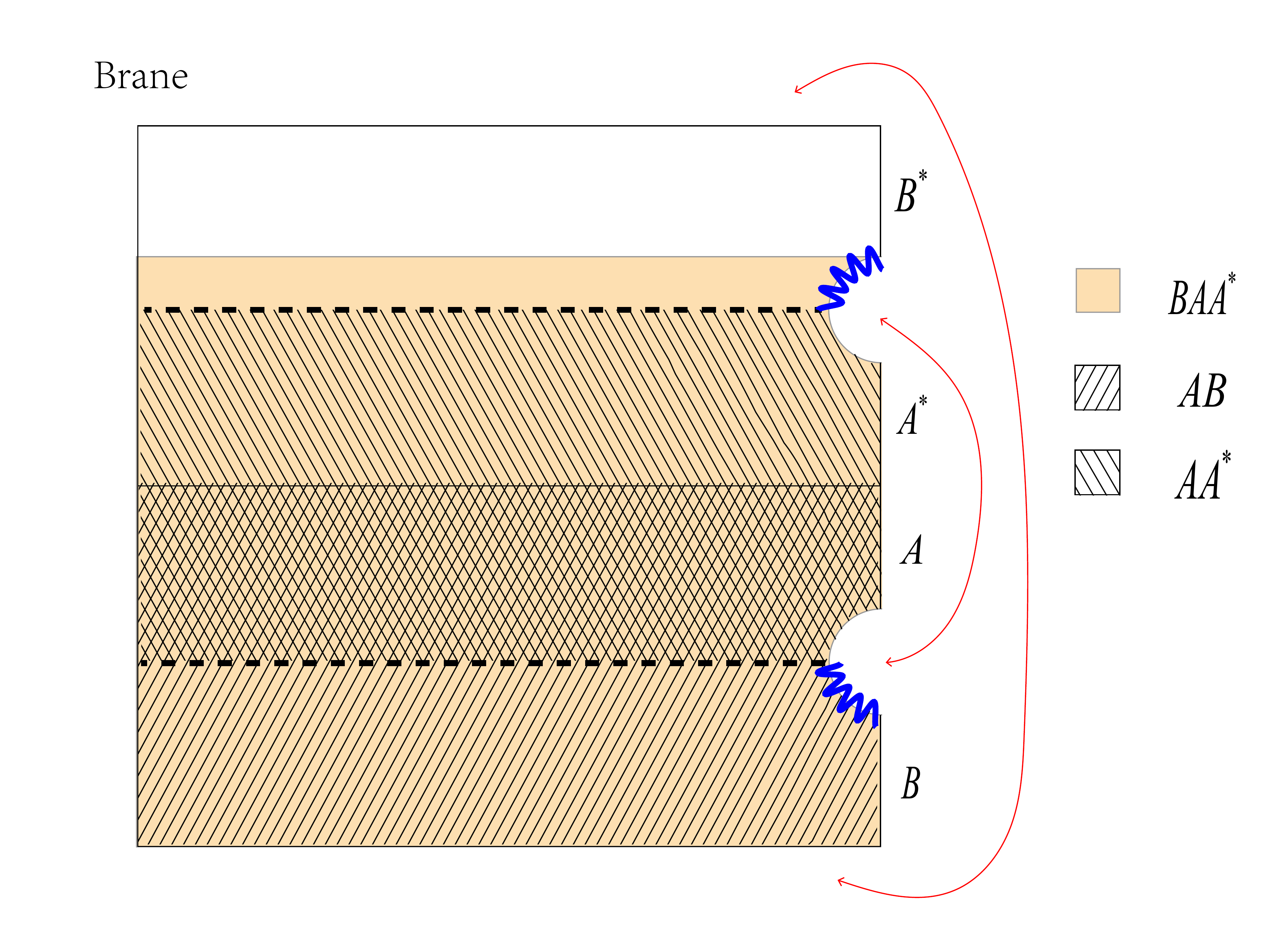}
\caption{Phase-D3.
$A$ and $B$ are disjoint.
$A$ admits no entanglement island but a reflected island.
There is one boundary of EWCS, denoted by a dashed curve, and one gap between $A\cup \mathcal I_{R,A}$ and $B\cup \mathcal I_{R,B}$, indicating $h\geq \frac{c}{3}\log 2$.
There is a jagged surface in this phase denoted in blue.
}
\label{fig:D3}
\end{figure}

\subsubsection*{Phase-D3}
{It is possible that the cross-section of phase-D2 is anchored at the brane, that is phase-D3 (Fig.\ref{fig:D3}).}
For this phase to exist, we require
\begin{align}
    \log&\frac{(b_2-b_1)^2}{4b_1b_2}\leq 2T(\theta_0)+2W(\theta_0)\label{422}\\
    \log&\frac{(b_3-b_2)^2}{4b_2b_3}\leq 2T(\theta_0)+2W(\theta_0)\label{423}\\
    S_{R}(A:B)&<\frac{c}{6}\log\sqbra{
\frac{b_2b_3-b_1^2+\sqrt{(b_2^2-b_1^2)(b_3^2-b_1^2)}}{b_2b_3-b_1^2-\sqrt{(b_2^2-b_1^2)(b_3^2-b_1^2)}}
}\label{424}.
\end{align}

In this case, the mutual information is the same as \eqref{phase2_mutual}.
And the reflected entropy is given by \eqref{SR_phase3}
\begin{align}
S_R(A:B)=\frac{c}{3}\log\frac{(b_3+a')(b_2+a')}{(b_3-b_2)a'}+\frac{c}{3}T(\theta_0)+\frac{c}{3}W(\theta_0),
\end{align}
where the island cross section $a'=\sqrt{b_2b_3}$.
The condition \eqref{424} becomes
\begin{align}\label{ineqf}
    f\equiv \log\sqbra{
\frac{b_2b_3-b_1^2+\sqrt{(b_2^2-b_1^2)(b_3^2-b_1^2)}}{b_2b_3-b_1^2-\sqrt{(b_2^2-b_1^2)(b_3^2-b_1^2)}}\frac{(b_3-b_2)^2}{(\sqrt{b_2}+\sqrt{b_3})^4}
}-2T(\theta_0)-2W(\theta_0)>0.
\end{align}
Apply \eqref{423} and \eqref{422}, we have
\begin{align}
    f\leq \log \sqbra{
\frac{b_2b_3-b_1^2+\sqrt{(b_2^2-b_1^2)(b_3^2-b_1^2)}}{b_2b_3-b_1^2-\sqrt{(b_2^2-b_1^2)(b_3^2-b_1^2)}}\frac{(b_3-b_2)^2}{(\sqrt{b_2}+\sqrt{b_3})^4}\frac{4b_1b_2}{(b_2-b_1)^2}
}\label{DES_phase3_f1},\\
f\leq \log \sqbra{
\frac{b_2b_3-b_1^2+\sqrt{(b_2^2-b_1^2)(b_3^2-b_1^2)}}{b_2b_3-b_1^2-\sqrt{(b_2^2-b_1^2)(b_3^2-b_1^2)}}\frac{(b_3-b_2)^2}{(\sqrt{b_2}+\sqrt{b_3})^4}\frac{4b_2b_3}{(b_3-b_2)^2}
}\label{DES_phase3_f2}.
\end{align}
We are only interested in the existence of this phase, then it is sufficient to pick a special case.
 Set $b_1=1$, $b_2=2\times10^3$ and $b_3=2\times10^4$. Then the R.H.S of \eqref{DES_phase3_f1} and \eqref{DES_phase3_f2} become roughly $\log 10672$ and $\log (2.6\times10^6)$, which allow a positive $f$ to exist.
Taking $\theta_0=\pi/6$, $\ell=1$ and $\epsilon_w=0.01$, we then have $f\approx\log 33>0$ and thus conditions  \eqref{ineqf}, \eqref{DES_phase3_f1} and \eqref{DES_phase3_f2} (or equivalently \eqref{422}, \eqref{423} and \eqref{424} ) for phase-D3 to exist are satisfied.

{As shown in Fig.\ref{fig:D3},
the entanglement wedge of $AB$ together with $AA^*$ does not cover all $BAA^*$ and there is a jagged line for phase-D3. 
Thus a non-vanishing Markov gap is expected.
}
Now let us compute the Markov gap. 
Use \eqref{phase2_mutual} and \eqref{SR_phase3}, and we obtain
\begin{align}
h(A:B)&=\frac{c}{6}\log\sqbra{\frac{(\sqrt{b_2}+\sqrt{b_3})^4}{(b_3-b_2)^2}
\frac{b_1}{b_3}\sbra{\frac{b_3-b_2}{b_2-b_1}}^2
}+\frac{c}{3}T(\theta_0)+\frac{c}{3}W(\theta_0)\notag\\
&\geq \frac{c}{6}\log\sqbra{
\frac{(b_3+b_2+2\sqrt{b_2b_3})^2}{4b_2b_3}
}
=\frac{c}{6}\log\sqbra{
\frac14\sbra{
\sqrt{\frac{b_3}{b_2}}+\sqrt{\frac{b_2}{b_3}}+2
}^2
}\notag\\
&\geq \frac{c}{3}\log 2 \label{newphase2_h}
\end{align}
where we have used \eqref{422} in the second line.
The equality in second line is taken at critical point $S(A)_{\rm RT}=S(A)_{\rm island}$ for $A$, which is dependent of $\theta_0$ that is in turn related to the brane tension.
So \eqref{newphase2_h} is saturated when $b_2\simeq b_3$ and near the critical point of $A$.
To sum up, $h=\frac{c}{3}\log 2$ iff
\begin{align}
    &\log\frac{(b_2-b_1)^2}{4b_1b_2}= 2T(\theta_0)+2W(\theta_0),\\
    &b_2\simeq b_3.
\end{align}

In fact, it is possible that we cannot take the equality, as if $f$ is negative, phase-D2 takes over the reflected entropy.
Then it is necessary to check this.
First, at the critical point, \eqref{DES_phase3_f1} takes equality.
Then it is easy to see that in the limit $b_2\simeq b_3$, we have
\begin{align}\label{DES_phase3_f3}
    f=\log \sqbra{
    \frac{(b_1 + b_2)^2}{b_1b_2}
    }\geq 2\log 2.
\end{align}
Therefore, we deduce that $f\geq 2\log2$, implying there is no problem taking this limit.
On the other hand, \eqref{DES_phase3_f2} must be compatible with \eqref{DES_phase3_f3}, and this can be seen explicitly by inserting $b_3=b_2+\epsilon$ in \eqref{DES_phase3_f2}
\begin{align}
    f\leq 2\log \frac{(b_2-b_1)^2}{b_1\epsilon},
\end{align}
the R.H.S of which is divergent as $\epsilon\rightarrow 0$.

{In fact, for phase-D3, only one boundary of the cross-section is anchored at the minimal surface of $A\cup B$.  
Then the inequality \eqref{main_clain_bulk} also implies  $h\geq\frac{c}{3}\log 2$.
Besides, one can also obtain the same lower bound using our boundary inequality \eqref{main_clain_boundary}. 
Note that although there is no entanglement island for $A$, the cross-section for phase-D3 is anchored at the brane so that there is an island of reflected entropy for $A$ and thus only one gap between $A\cup \mathcal I_{R,A}$ and  $B\cup \mathcal I_{R,B}$.}

\subsubsection*{Phase-D4}

\begin{figure}
\centering
\includegraphics[width=0.45\textwidth]{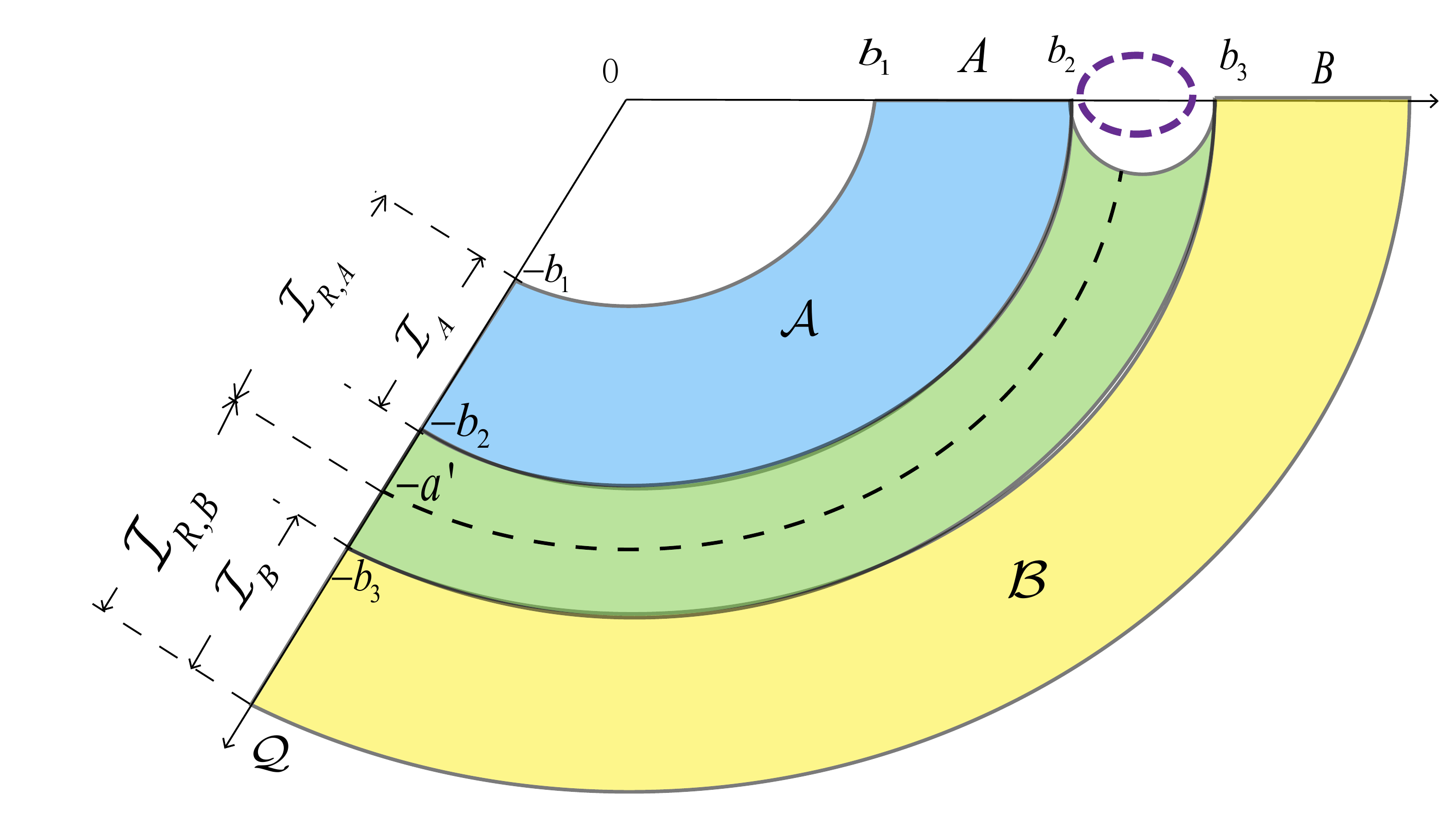}
\includegraphics[width=0.45\textwidth]{figure/jagD4.pdf}
\caption{Phase-D4.
    In this phase, the entanglement islands for $A$ and $B$ are different from their reflected island.
    One boundary of EWCS and one gap between $A\cup \mathcal I_{R,A}$ and $B\cup \mathcal I_{R,B}$ give the lower bound $h\geq \frac{c}{3}\log 2$.
    There is one jagged surface, with $\mathrm W(AA^*B)\supset \mathrm W(AA^*)\cup \mathrm{W}(AB)$.
}
\label{fig:D4}
\end{figure}
{
For phase-D4 (Fig.\ref{fig:D4}) where both intervals contain islands, the entanglement wedge of $AB$ together with that of $AA^*$ cannot cover all the entanglement wedge of $BAA^*$ so that there is a jagged line and thus we also expect the non-vanishing  Markov gap for phase-D4.}

The entanglement entropies for $A$ and $B$ in phase-D4 are
\begin{align}
S(A)&=\frac{c}{6}\log\frac{2b_1}{\epsilon}+\frac{c}{6}\log\frac{2b_2}{\epsilon}+\frac{c}{3}T(\theta_0)+\frac{c}{3}W(\theta_0),\\
S(B)&=\frac{c}{6}\log\frac{2b_3}{\epsilon}+\frac{c}{6}\log\frac{2b_4}{\epsilon}+\frac{c}{3}T(\theta_0)+\frac{c}{3}W(\theta_0).
\end{align}
And the entanglement entropy for $A\cup B$ is
\begin{align}
S(AB)=\frac{c}{6}\log\frac{4b_1b_4}{\epsilon^2}+\frac{c}{3}\log\frac{b_3-b_2}{\epsilon}+\frac{c}{3}T(\theta_0)+\frac{c}{3}W(\theta_0).
\end{align}
Then the mutual information is given by
\begin{align}
I(A:B)&=S(A)+S(B)-S(AB)\notag\\
&=\frac{c}{6}\log\sqbra{
\frac{4b_2b_3}{(b_3-b_2)^2}
}+\frac{c}{3}T(\theta_0)+\frac{c}{3}W(\theta_0).\label{DES_phase3_mutual}
\end{align}
The reflected entropy for phase-D4 is given by \eqref{SR_phase3}
\begin{align}
\label{srD4}
S_R(A:B)=\frac{c}{3}\log\frac{(b_3+a)(b_2+a)}{(b_3-b_2)a}+\frac{c}{3}T(\theta_0)+\frac{c}{3}W(\theta_0),
\end{align}
where $a=\sqrt{b_2b_3}$.
Then, the Markov gap is given by
\begin{align}
h=\frac{c}{3}\log\sqbra{
\frac12\sbra{\sqrt{\frac{b_3}{b_2}}+\sqrt{\frac{b_2}{b_3}}+2
}
}\geq \frac{c}{3}\log 2,
\end{align}
with the equality taken at $b_2\simeq b_3$.

For phase-D4, the analysis of the lower bound in terms of our inequalities \eqref{main_clain_bulk} and\eqref{main_clain_boundary} is similar to phase-D3 and they also give the lower bound $\frac{c}{3}\log2$.

\subsubsection{Adjacent intervals}\label{sec.adjacent}

\subsubsection*{Phase-A1}

In phase-A1 (Fig.\ref{fig:A1}), the two intervals $A$ and $B$ are adjacent and both contain island.

\begin{figure}
\centering
\includegraphics[width=0.45\textwidth]{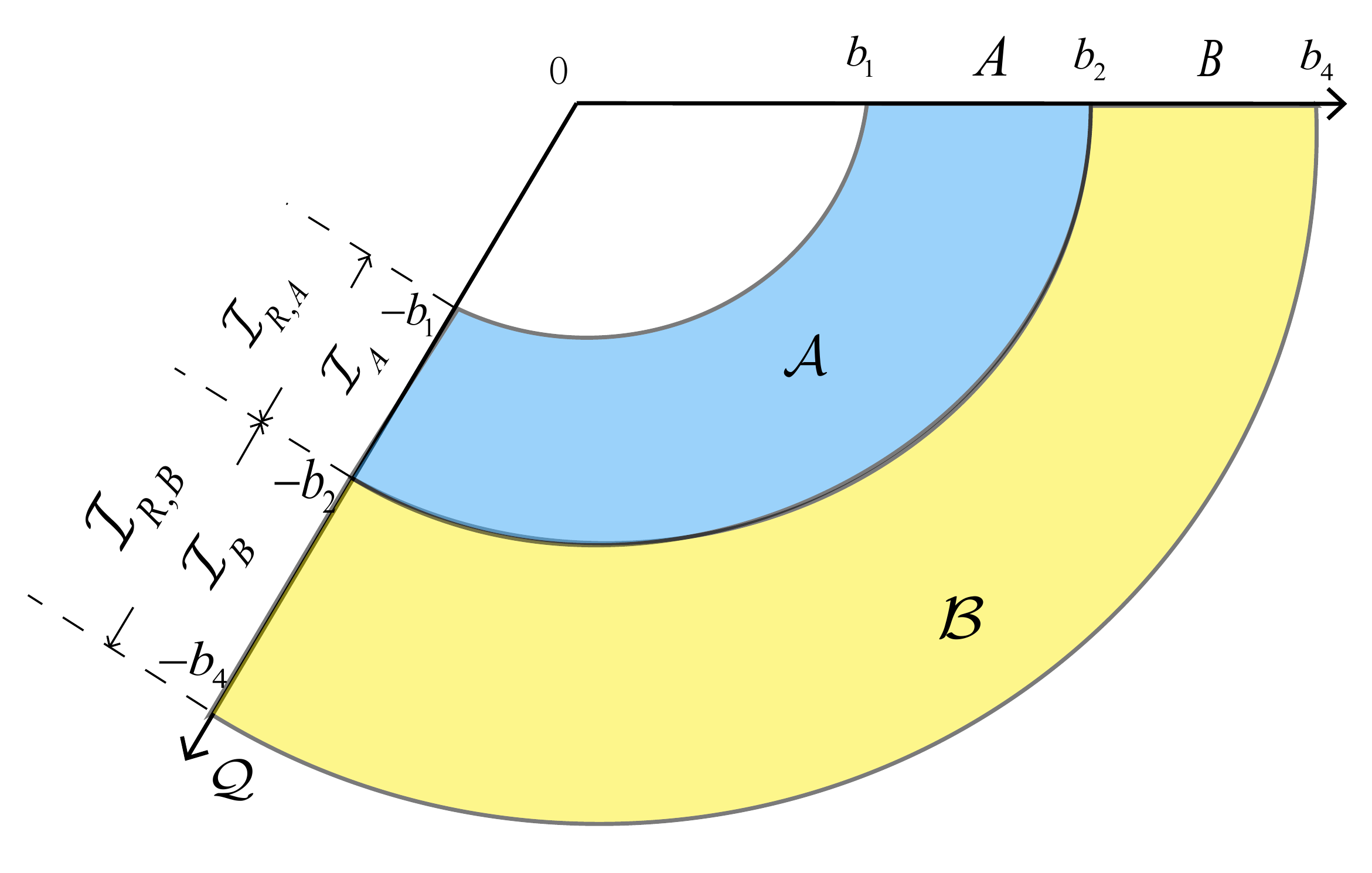}
\includegraphics[width=0.45\textwidth]{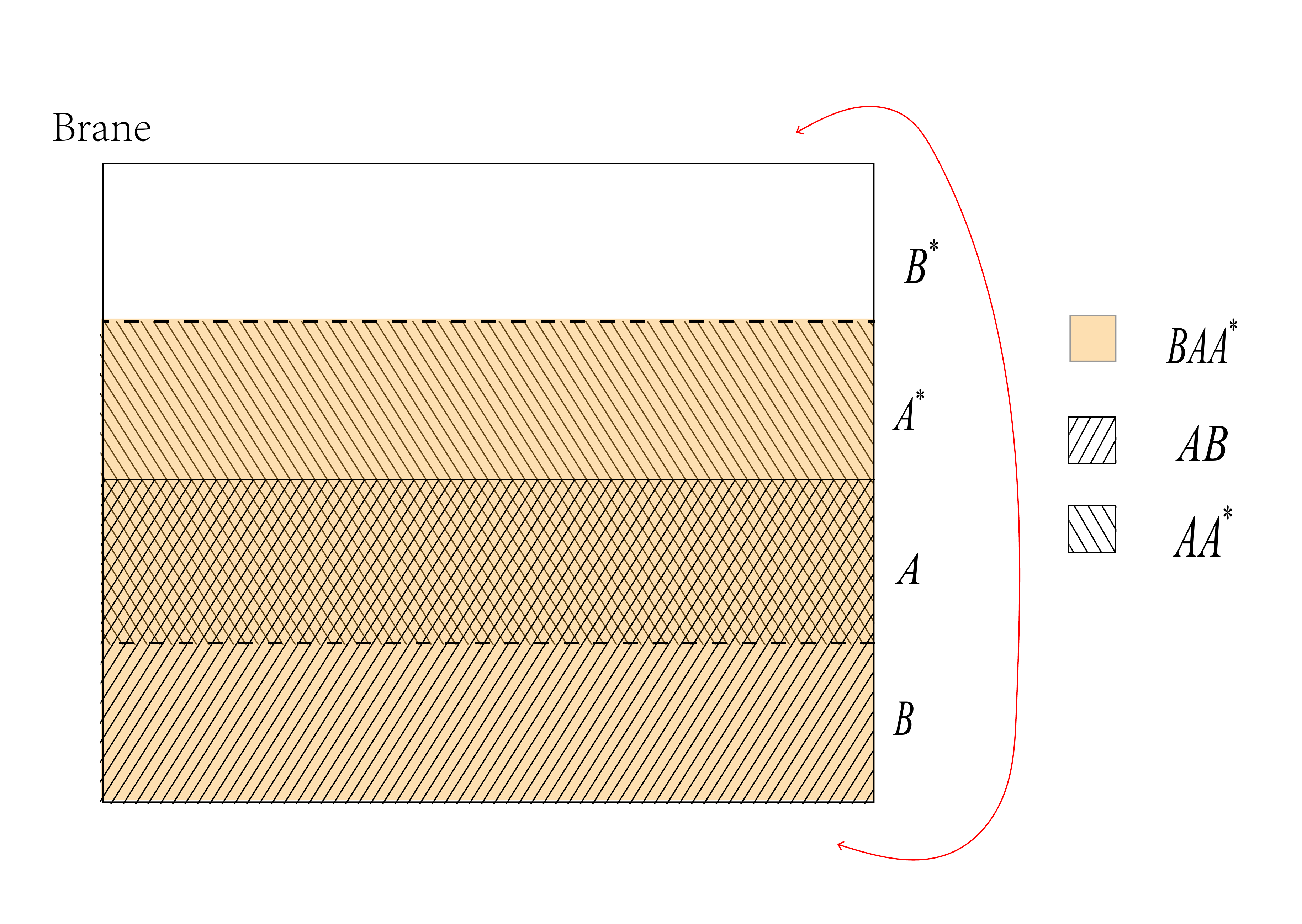}
\includegraphics[width=0.45\textwidth]{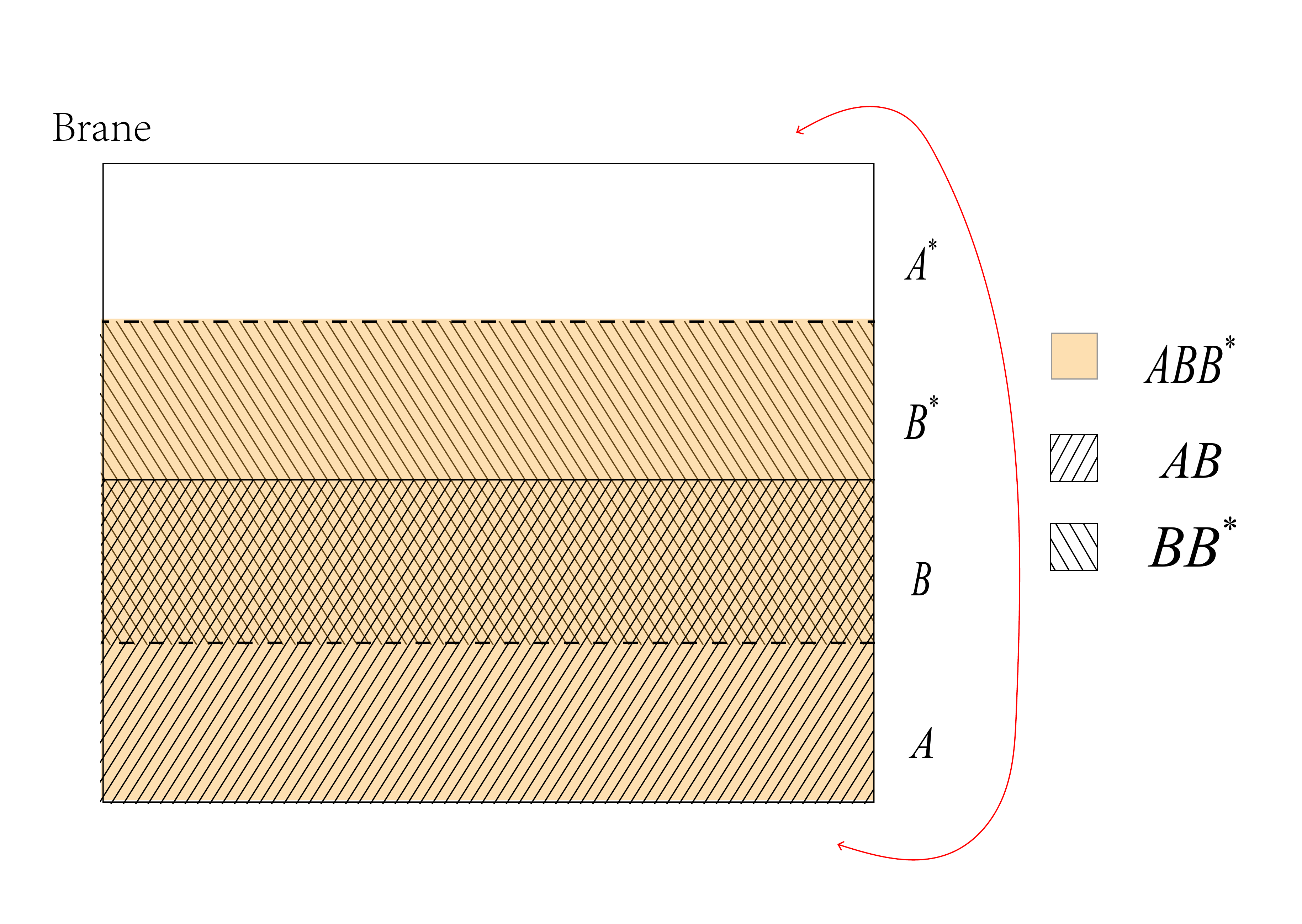}
\caption{Phase-A1.
    In this phase, the entanglement islands of $A$ and $B$ are identically their reflected island.
}
\label{fig:A1}
\end{figure}

The entanglement entropies for $A$ and $B$ in phase-A1 read
\begin{align}
    S(A)&=\frac{c}{6}\log\frac{2b_1}{\epsilon}+\frac{c}{6}\log\frac{2b_2}{\epsilon}+\frac{c}{3}T(\theta_0)+\frac{c}{3}W(\theta_0),\\
S(B)&=\frac{c}{6}\log\frac{2b_2}{\epsilon}+\frac{c}{6}\log\frac{2b_4}{\epsilon}+\frac{c}{3}T(\theta_0)+\frac{c}{3}W(\theta_0).
\end{align}
The entanglement entropy for $AB$ is
\begin{align}
    S(AB)=\frac{c}{6}\log \frac{2b_1}{\epsilon}+\frac{c}{6}\log \frac{2b_4}{\epsilon}+\frac{c}{3}T(\theta_0)+\frac{c}{3}W(\theta_0).
\end{align}
Then the mutual information is given by
\begin{align}\label{DES_phase4_mutual}
    I(A:B)=\frac{c}{3}\log \frac{2b_2}{\epsilon}+\frac{c}{3}T(\theta_0)+\frac{c}{3}W(\theta_0).
\end{align}

Now let us  derive the reflected entropy. 
According to proposal \eqref{DES_SR_proposal2}, we have in large-$c$ limit
\begin{align}\label{DES_phase4_SR1}
    S_R(A:B)=\min\operatorname{ext}_{\rm EW}\left\{
    S_{R}^{\rm eff}(I_A:I_B)+\frac{\operatorname{Area}[\operatorname{EW}(A:B)]}{2 G_N}
    \right\}.
\end{align}
The first term can be computed via correlation functions of twist operators.
We refer to \cite{Sully:2020pza,Li:2021dmf} for details, and just quote the result here
\begin{equation}
    S_R^{(\mathrm{eff})}\left(I_A: I_B\right)=
    \begin{cases}
    &\frac{c}{3} \log \frac{2 \ell}{\xi \epsilon_w \cos \theta_0}, \quad \xi>1\\
    &\frac{c}{3}\log \frac{2\ell}{\epsilon_w\cos\theta_0},\quad \xi<1
    \end{cases}
\end{equation}
where $\xi=\frac{2 a b_1}{a^2-b_1^2}$.
The second term in \eqref{DES_phase4_SR1} is just twice the length of a geodesic connecting $b_2$ and $-a$, which is given by \cite{Deng:2020ent,Takayanagi:2011zk}
\begin{align}
    \frac{c}{3}\log\frac{2L'}{\epsilon}+\frac{c}{3}T(\theta_0').
\end{align}
The quantities $L'$ and $\theta_0'$ satisfy
\begin{align}\label{448}
    L'&=\frac{a^2+b_2^2+2ab_2\sin\theta_0}{2(b_2+a\sin\theta_0)}\\
    \theta'&=\arcsin
    \frac{b_2^2+2ab_2\sin\theta_0-a^2\cos 2\theta_0}{b_2^2+2ab_2\sin\theta_0+a^2}\label{449}
\end{align}
Then the reflected entropy is given by
\begin{equation}
    S_R(A:B)=\min\operatorname{ext}_{\rm EW}    
    \begin{cases}
    &\frac{c}{3} \log \frac{2 \ell}{\xi \epsilon_w \cos \theta_0}+\frac{c}{3}\log\frac{2L'}{\epsilon}+\frac{c}{3}T(\theta'),\quad \xi>1\\
    &\frac{c}{3}\log \frac{2\ell}{\epsilon_w\cos\theta_0}+\frac{c}{3}\log\frac{2L'}{\epsilon}+\frac{c}{3}T(\theta'),\quad  \xi<1
    \end{cases}
\end{equation}
For $\xi>1$, we find no real solution to $\partial_a S_R=0$.
For $\xi<1$, the minimization process reduces to finding the entanglement island for $[0,b_2]$, which is $a=b_2$.
Substitute $a=b_2$ in \eqref{448} and \eqref{449}, and we get 
\begin{align}
  L'=b_2,\quad \theta'= \theta_0.
\end{align}
Now the reflected entropy is given by
\begin{align}\label{DES_phase4_SR}
    S_R(A:B)=\frac{c}{3}\log \frac{2b_2}{\epsilon}+\frac{c}{3}T(\theta_0)+\frac{c}{3}W(\theta_0).
\end{align}
Thus the correct Markov gap is
\begin{align}
    h=0.
\end{align}
A vanishing Markov gap implies the existence of a perfect recovery map.

One may notice that \eqref{DES_phase4_mutual} can be obtained by the following replacement in \eqref{DES_phase3_mutual}
\begin{align}
 b_3-b_2\rightarrow \epsilon.
\end{align}
Naively, if we take the same replacement in reflected entropy \eqref{srD4}, we get
\begin{align}\label{DES_phase4_wrongSR}
    S_{R}(A:B)=\frac{c}{3}\log
    \frac{4b_2}{\epsilon
    }+\frac{c}{3}T(\theta_0)+\frac{c}{3}W(\theta_0),
\end{align}
which leads to 
\begin{align}\label{DES_phaseA1_wrongh}
    h=\frac{c}{3}\log 2\quad (\text{incorrect}).
\end{align}
This is owing to the fact that when we evaluate the reflected entropy, or equivalently the entanglement wedge cross-section, we cannot take $b_3=b_2+\epsilon$, otherwise the cutoff of $y$ coordinate would become $y_{\rm UV}\sim\epsilon/2$.
We demonstrate this in Appendix.\ref{app.adjacent}.
The factor $1/2$ in $y_{\rm UV}$ contributes the term $\frac{c}{3}\log 2$ in \eqref{DES_phase4_wrongSR} and \eqref{DES_phaseA1_wrongh}.
Recall that all formulae should use the standard cutoff $y_{\rm UV}=\epsilon$.
In this sense, we should really set $b_3-b_2=2\epsilon$, and this gives us the correct result:
\begin{align}
    S_R(A:B)&=\frac{c}{3}\log \frac{2b_2}{\epsilon}+\frac{c}{3}T(\theta_0)+\frac{c}{3}W(\theta_0),\\
    h(A:B)&=0.
\end{align}
Based on above analysis, we conclude that when evaluating reflected entropy in the adjacent limit, the $x$-axis gap between two intervals should be $2\epsilon$.

 For phase-A1, there are no boundaries anchored at the minimal surfaces of $A\cup B$ and no boundary gaps between $A\cup \mathcal I_{R,A}$ and  $B\cup \mathcal I_{R,B}$, thus both the inequalities \eqref{main_clain_bulk} and \eqref{main_clain_boundary} imply that the lower bound is zero.

\subsubsection*{Phase-A2}

\begin{figure}
\centering
\includegraphics[width=0.5\textwidth]{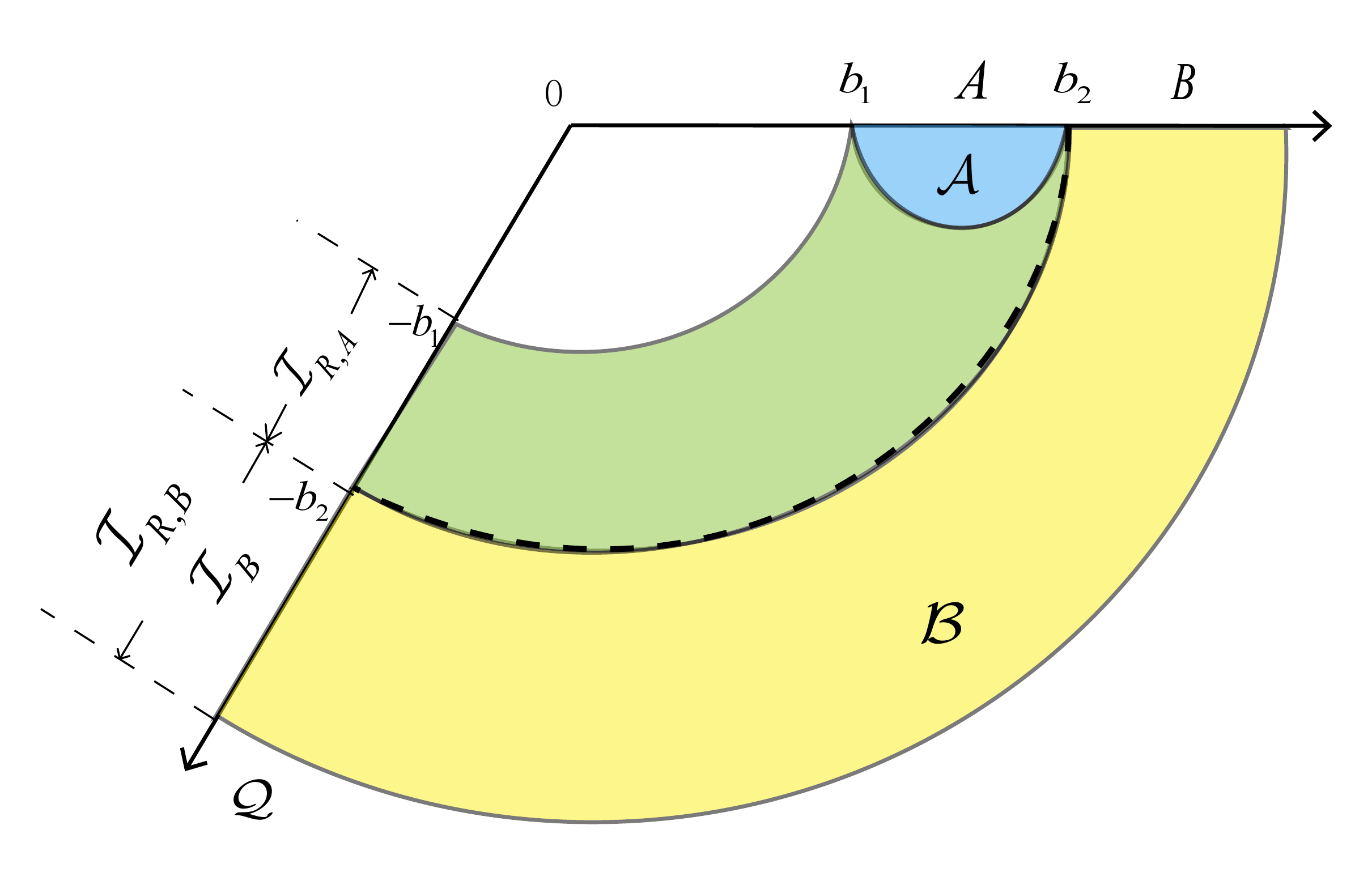}
\includegraphics[width=0.45\textwidth]{figure/jagA1.pdf}
\includegraphics[width=0.5\textwidth]{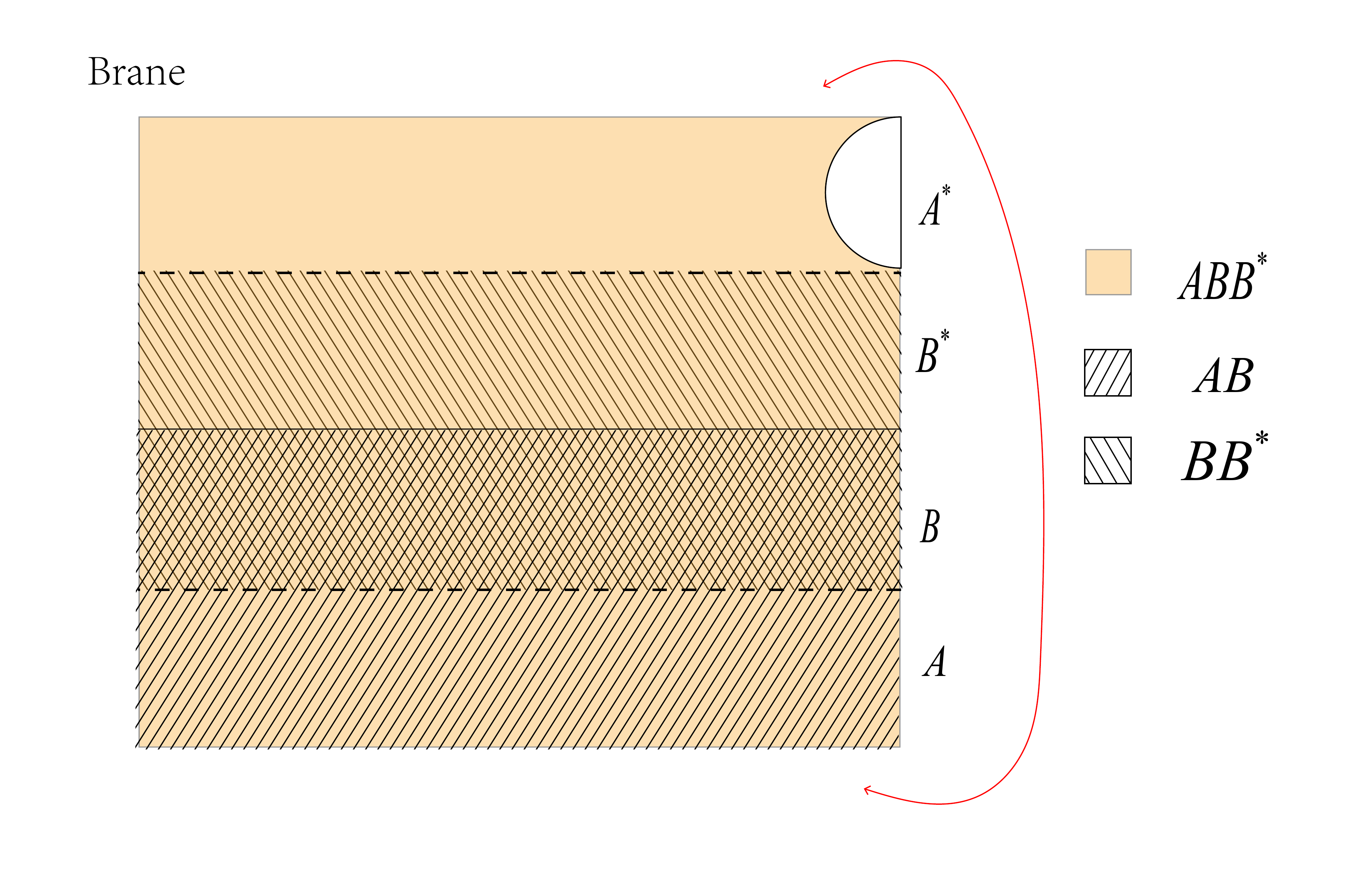}
\caption{Phase-A2.
    The region $A$ admits no entanglement island but a reflected island.
    The entanglement wedges satisfy $\mathrm{W}(AA^*B)=\mathrm{W}(AA^*)\cup \mathrm{W}(AB)$, but $\mathrm{W}(ABB^*)\supset \mathrm{W}(AB)\cup \mathrm{W}(BB^*)$.
}
\label{fig:A2}
\end{figure}

Phase-A2 (Fig.\ref{fig:A2}) is just like phase-D3 except that $A$ and $B$ are now adjacent. 
On one hand, we require $A$ has no island.
On the other hand, the area of entanglement wedge cross-section should be less than that of phase-A3, which is given in next phase by \eqref{DES_phaseA3_SR}
The above conditions lead to the following inequalities
\begin{align}\label{DES_phase-5_cond1}
    &\frac{c}{6}\log\frac{(b_2-b_1)^2}{4b_1b_2}\leq \frac{c}{3}T(\theta_0)+\frac{c}{3}W(\theta_0),\\
    &\frac{c}{3}\log\frac{2b_1b_2}{b_2^2-b_1^2}+\frac{c}{3}T(\theta_0)+\frac{c}{3}W(\theta_0)\leq 0.\label{DES_phase-5_cond2}
\end{align}

Though, this phase is easy to calculate, we still show how to obtain this phase by taking adjacent limit from phase-D3.
First, let $b_3-b_2=\epsilon$, and we obtain the mutual information from \eqref{phase2_mutual}
\begin{align}\label{DES_phaseA2_mutual}
    I(A:B)=\frac{c}{6}\log\sqbra{\frac{b_2}{b_1}\sbra{\frac{b_2-b_1}{\epsilon}}^2}.
\end{align}
Then let $b_3-b_2=2\epsilon$, and we have the reflected entropy from \eqref{SR_phase3}
\begin{align}\label{DES_phaseA2_SR}
    S_R(A:B)=\frac{c}{3}\log\frac{2b_2}{\epsilon}+\frac{c}{3}T(\theta_0)+\frac{c}{3}W(\theta_0).
\end{align}

It is easy to confirm that the mutual information is indeed given by \eqref{DES_phaseA2_mutual}.
The reflected entropy for $A:B$ is given by \eqref{DES_phase4_SR}, which is exactly \eqref{DES_phaseA2_SR}.
Then the Markov gap is given by
\begin{align}\label{DES_phaseA2_h}
    h&=\frac{c}{6}\log\frac{4b_1b_2}{(b_2-b_1)^2}+\frac{c}{3}T(\theta_0)+\frac{c}{3}W(\theta_0)
    \geq 0,
\end{align}
which is just the condition \eqref{DES_phase-5_cond1}.
Notably, in this case, the Markov gap between $A$ and $B$ is just the difference between two different phases of $S(A)$, and is guaranteed to be non-negative.
{The equality in \eqref{DES_phaseA2_h} is taken when the phase transition between phase-A1 and phase-A2 happens.
Away from the phase transition, we have $h>0$ and thus an imperfect Markov recovery for phase-A2.
}

We should also apply the second condition \eqref{DES_phase-5_cond2}, and this leads to
\begin{align}\label{DES_phaseA2_h_upper}
    h\leq\frac{c}{3}\log\sbra{
    \sqrt{\frac{b_1}{b_2}}+\sqrt{\frac{b_2}{b_1}}
    }.
\end{align}
Combine \eqref{DES_phaseA2_h} and \eqref{DES_phaseA2_h_upper}, and we have
\begin{align}
    0\leq h\leq \frac{c}{3}\log\sbra{\sqrt{\frac{b_1}{b_2}}+\sqrt{\frac{b_2}{b_1}}}.
\end{align}
That is, the Markov gap is not only lower-bounded but also upper-bounded in this phase.

The analysis of Markov recovery for phase-A1 and phase-A2 is as follows.
For both phases, as shown in Fig.\ref{fig:A1} and Fig.\ref{fig:A2}, the entanglement wedge of $AB$ together with $AA^*$ covers all the entanglement wedge of $BAA^*$.
However, this information is not enough to tell us whether there is a perfect Markov recovery or not.
In this case, one should resort to the direct calculation of $h$, which informs us that there is a perfect Markov recovery ($h=0$) for phase-A1 while no perfect Markov recovery ($h>0$) for phase-A2 away from the phase transition.  
In fact, as shown in Fig.\ref{fig:A2}, for another Markov recovery map $\rho_{ABB^*}=\mathcal{R}_{B\to BB^*}(\rho_{AB})$, 
the entanglement wedge of $AB$ together with that of $BB^*$ cannot cover all the entanglement wedge of $ABB^*$, which obviously signals an imperfect Markov recovery for phase-A2.


\subsubsection*{Phase-A3}

\begin{figure}
\centering
\includegraphics[width=0.5\textwidth]{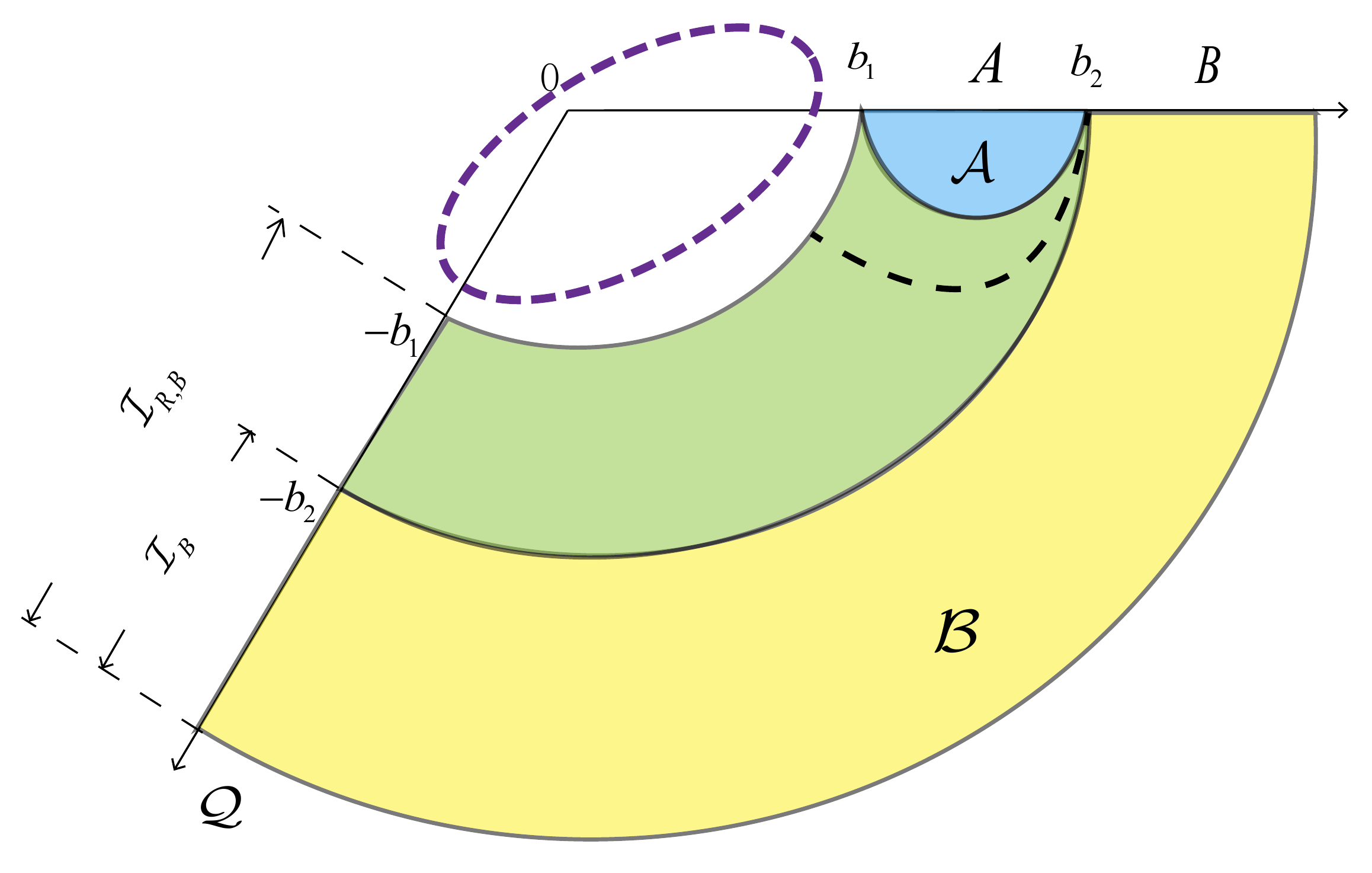}
\includegraphics[width=0.45\textwidth]{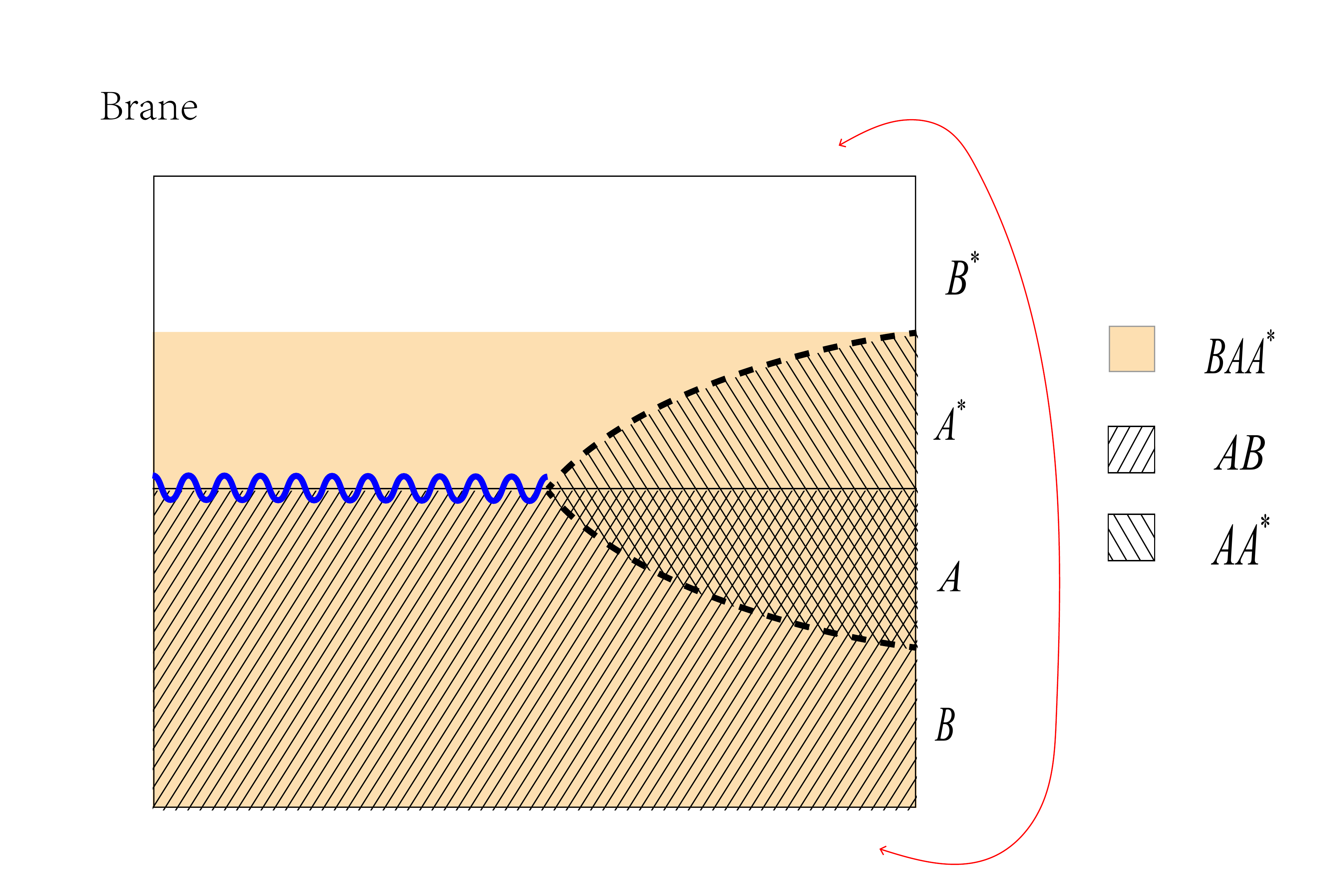}
\caption{Phase-A3.
    In this phase, $A$ and $B$ are adjacent, and $A$ admits no island.
}
\label{fig:A3}
\end{figure}

Phase-A3 (Fig.\ref{fig:A3}) is like phase-D2 except that now $A$ and $B$ are adjacent.
{Unlike phase-D2, there is only one jagged surface due to the vanishing spacing between $A$ and $B$.
The Markov recovery is precluded and a non-vanishing Markov gap is expected.}
The condition for this phase to dominate is
\begin{align}
   & \label{DES_phaseA3_cond1}\log\frac{(b_2-b_1)^2}{4b_1b_2}\leq 2T(\theta_0)+2W(\theta_0)\\
&S_R(A:B)<\frac{c}{3}\log\frac{2b_2}{\epsilon}+\frac{c}{3}T(\theta_0)+\frac{c}{3}W(\theta_0)\label{DES_phaseA3_cond2},
\end{align}
where the first inequality follows from that $A$ has no entanglement island, while the second inequality follows from that its reflected entropy should be smaller than \eqref{DES_phase4_SR}.

We can do adjacent limit from phase-D2 to get the result of this phase.
First, let $b_3=b_2+\epsilon$ to get the mutual information from \eqref{phase2_mutual}
\begin{align}
    I(A:B)=\frac{c}{6}\log\sqbra{
    \frac{b_2}{b_1}\sbra{\frac{b_2-b_1}{\epsilon}}^2
    }.
\end{align}
And then we let $b_3=b_2+2\epsilon$ to get the reflected entropy
\begin{align}
    S_R(A:B)=\frac{c}{3}\log\frac{b_2^2-b_1^2}{b_1\epsilon}.
\end{align}

We can also obtain these from direct calculation.
In this phase, the mutual information is given by \eqref{DES_phaseA2_mutual}.
And the reflected entropy equals twice the minimum length of geodesics that connect $b_2$ and the RT surface of $[0,b_1]$.
The minimum length can be derived with simple geometric relation
\begin{align}\label{DES_phase5_SR0}
    S_R(A:B)=\frac{c}{3}\log\frac{2(b_2-L)}{\epsilon}+\frac{c}{3}\operatorname{arctanh}\sbra{\frac{b_2-L}{L}},
\end{align}
where $L$ reads
\begin{equation}\label{DES_phase5_L}
    L=\frac{b_1^2+b_2^2}{2b_2}.
\end{equation}
We leave the derivation of the above result in Appendix.\ref{app1}.
Inserting \eqref{DES_phase5_L} into \eqref{DES_phase5_SR0}, we obtain
\begin{align}\label{DES_phaseA3_SR}
    S_{R}(A:B)=\frac{c}{3}\log\frac{b_2^2-b_1^2}{\epsilon b_1}.
\end{align}
So we obtain the same result as from the adjacent limit.

Now we consider if this phase could exist.
Rewrite \eqref{DES_phaseA3_cond2} as
\begin{align}
    \frac{c}{3}\log\frac{2b_1b_2}{b_2^2-b_1^2}+\frac{c}{3}T(\theta_0)+\frac{c}{3}W(\theta_0)>0.
\end{align}
Obviously, this can be satisfied for $b_2$ very close to $b_1$, and \eqref{DES_phaseA3_cond1} can be fulfilled by tuning $\theta_0$.

Subtract the mutual information from reflected entropy, and we have the Markov gap
\begin{align}
    h=\frac{c}{6}\log\sqbra{\sbra{\sqrt{\frac{b_2}{b_1}}+
    \sqrt{\frac{b_1}{b_2}}
    }^2}\geq \frac{c}{3}\log 2,\quad \lim_{b_2\rightarrow b_1}h=\frac{c}{3}\log 2.
\end{align}
As we have said, if $b_1$ is close to $b_2$, the requirement \eqref{DES_phaseA3_cond2} can be satisfied.

\section{The Markov gap in JT gravity}\label{sec:jt}
In this section, we consider the JT gravity model in \cite{Almheiri:2019hni,Almheiri:2019qdq}, where the AdS$_2$ JT gravity, coupled with CFT matter, is glued with a flat CFT.
We do not apply the double holography description \cite{Chandrasekaran:2020qtn}.
Instead, we work in a pure boundary way to test \eqref{main_clain_boundary}.

\subsection{Entanglement entropy for extremal JT black holes coupled to a bath}
Consider a system where a 2D  Jackiw-Teitelboim (JT) gravity with CFT matter is glued with a flat 2D CFT bath along its boundary, $x=0$, at which the transparent boundary condition is imposed.
The action is
\begin{align}
    S=\frac{1}{4 \pi} \int d^2 x \sqrt{-g}\left[\phi R+2\left(\phi-\phi_0\right)\right]+I_{\mathrm{CFT}}
\end{align}
For extremal JT black holes, the metric and the dilaton in the gravity region are given by
\begin{align}
\D s^2 &=-4 \frac{\D x^{+} \D x^{-}}{\left(x^{-}-x^{+}\right)^2},\ 
\phi =\phi_0+2 \frac{\phi_r}{x^{-}-x^{+}},
\end{align}
where $x^{\pm} =t \pm x, x \in(-\infty, 0]$ and $\phi_r$ is the renormalized boundary field \cite{Maldacena:2016upp}.
$x=0$ is the interface of the JT gravity with the 2D flat CFT.

\begin{figure}
    \centering
    \includegraphics[width=0.6\textwidth]{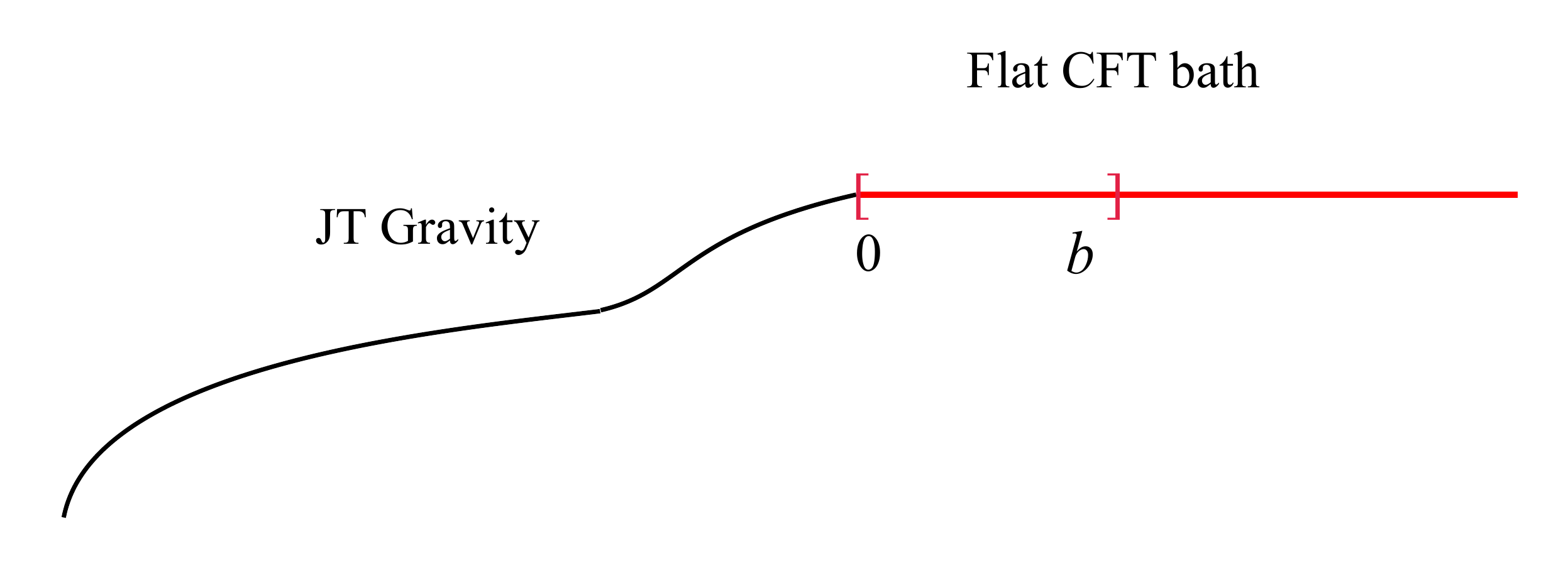}
    \caption{JT gravity plus CFT matters is jointed with a flat CFT bath. 
    }
    \label{fig:jt}
\end{figure}

According to the quantum extremal surface prescription \cite{Engelhardt:2014gca}, the entanglement entropy for an interval $[0,b]$ on the flat CFT bath (see Fig.\ref{fig:jt}) is given by the generalized entropy $S_{[0,b]}^{\rm gen}(a)$ with the island by
\begin{align}\label{Sgen}
    S([0,b])=S^{\rm gen}_{[0,b]}(a)=\phi_0+\frac{\phi_r}{a}+S_{\rm eff}([-a,b]),\quad S_{\rm eff}([-a,b])=\frac{c}{6}\log\frac{(a+b)^2}{a},
\end{align}
where the boundary of island $a$, which is on the gravity side, is given by extremization and minimization of the generalized entropy
\begin{align}
    \tilde a=\frac12\sbra{
    1+\tilde b+\sqrt{1+6\tilde b+\tilde b^2}
    },
\end{align}
where $\tilde a=a/q$, $\tilde b=b/q$ and $q=6\phi_r/c$.

The entanglement entropy for an interval $[b_1,b_2]$ with $b_1,b_2>0$ is given by the minimum entanglement entropy among possible saddles. And this is the key point to recovering the Page curve \cite{Almheiri:2019qdq,Penington:2019kki}. So we have
\begin{equation}
    \begin{aligned}\label{minimalS}
    S([b_1,b_2])=&
    \min\left\{S_{\rm no-island},S_{\rm island}\right\}\\=&
    \min\left\{\frac{c}{3}\log(b_2-b_1),S^{\rm gen}_{[b_1,b_2]}(a(b_1),a(b_2))\right\},
\end{aligned}
\end{equation}
where the generalized entropy
\begin{equation}
   S^{\rm gen}_{[b_1,b_2]}(a(b_1),a(b_2))
   =2\phi_0+\frac{\phi_r}{a(b_1)}+\frac{\phi_r}{a(b_2)}+S_{\rm eff}([-a(b_2),-a(b_1)]\cup [b_1,b_2]).
\end{equation}
Note in calculating $S_{\rm eff}([-a(b_2),-a(b_1)]\cup [b_1,b_2])$ at large-$c$ limit, $t$-channel will dominate and the 4-point correlation function of twist operators can be factorized into two 2-point functions \cite{Hartman:2013mia,Chandrasekaran:2020qtn}
\begin{equation}
    \begin{aligned}
    \braket{\sigma(b_1)\tilde\sigma(-a(b_1))\tilde\sigma(b_2)\sigma(-a(b_2))}\to 
    \braket{\sigma(b_1)\tilde\sigma(-a(b_1))}\braket{\tilde\sigma(b_2)\sigma(-a(b_2))},
    \end{aligned}
\end{equation}
where $\sigma$ and $\tilde\sigma$ are twist operators.
Thus at large-$c$ limit, we have
\begin{equation}
\begin{split}
   S^{\rm gen}_{[b_1,b_2]}(a(b_1),a(b_2))
  \approx&2\phi_0+\frac{\phi_r}{a(b_1)}+\frac{\phi_r}{a(b_2)}+S_{\rm eff}([-a(b_1),b_1])+S_{\rm eff}([-a(b_2),b_2])\\
  =&S^{\rm gen}_{[0,b_1]}(a(b_1))+S^{\rm gen}_{[0,b_2]}(a(b_2)).
  \end{split}
\end{equation}

\subsection{The Markov gap}\label{sec:42}
We consider only the phases in which $A$ and $B$ are disjoint.
The adjacent cases can be obtained by taking the adjacent limit, that is, set $b_3-b_2=\epsilon$ in mutual information and $b_3-b_2=2\epsilon$ in reflected entropy.
\subsubsection*{Phase-D2}
\begin{figure}
    \centering
    \includegraphics[width=0.8\textwidth]{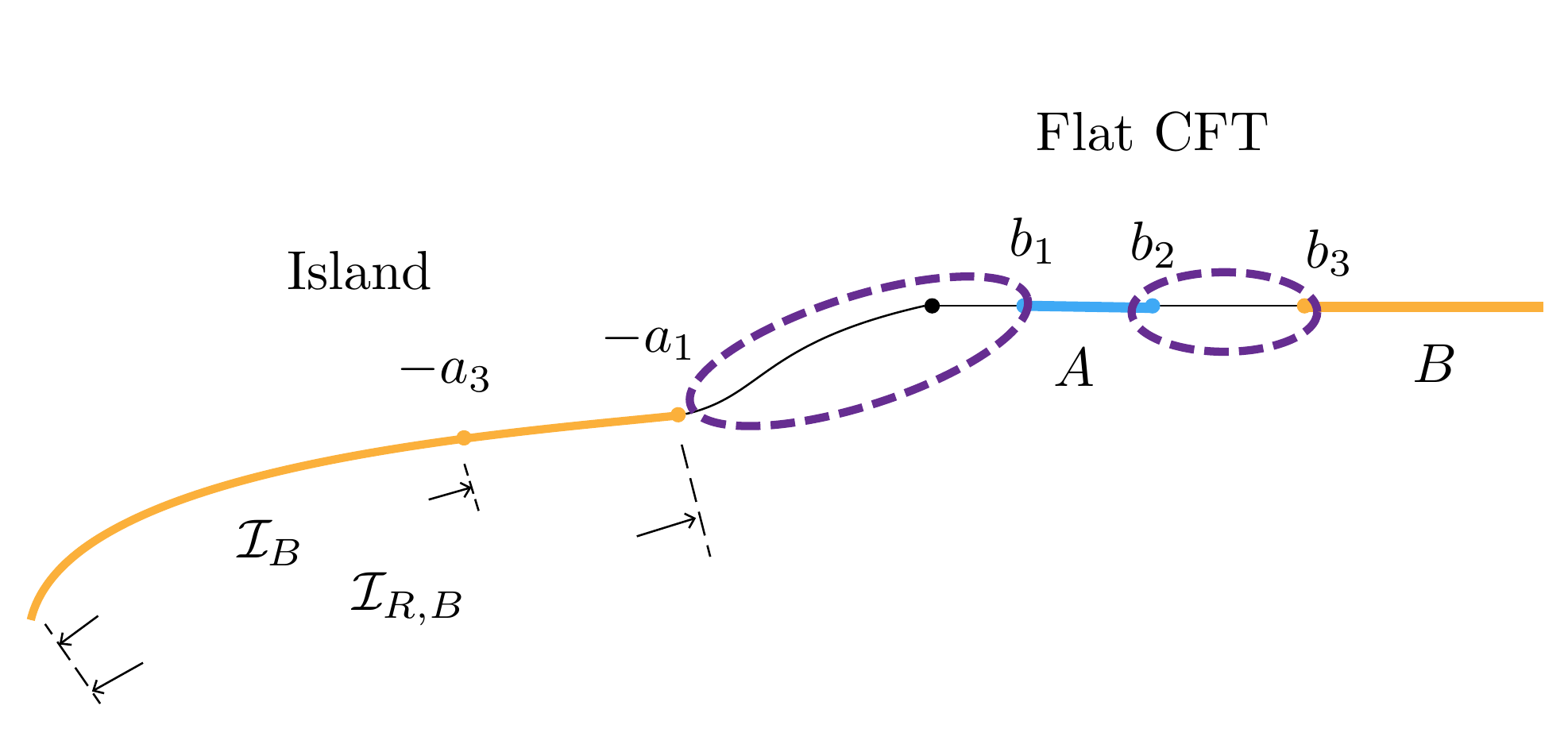}
    \caption{Phase-D2 for JT gravity.
    The purple dashed circles denote the boundary gaps between  $A\cup \mathcal I_{R,A}$ and  $B\cup \mathcal I_{R,B}$ }
    \label{fig:jtD2}
\end{figure}

Consider phase-D2 (Fig.\ref{fig:jtD2}) where $A=[b_1,b_2]$, $B=[b_3,\infty]$ and $C=[b_2,b_3]$ have no entanglement island and reflected island, but $B$ admits both entanglement and reflected island.
The conditions are
\begin{align}
    &\frac{c}{6}\log\frac{(b_2-b_1)^2a_1a_2}{(b_2+a_2)^2(b_1+a_1)^2}\leq 2\phi_0+\phi_r\left(\frac{1}{a_1}+\frac{1}{a_2}\right),\label{jt_phase2_cond1}\\
    &\frac{c}{6}\log\frac{(b_3-b_2)^2a_3a_2}{(b_2+a_2)^2(b_3+a_3)^2}\leq 2\phi_0+\phi_r\left(\frac{1}{a_3}+\frac{1}{a_2}\right).\label{jt_phase2_cond2}
\end{align}
For the entanglement wedge of $A\cup B$ to be connected, we require the mutual information $I(A:B)> 0$.

The entanglement entropies for $A$, $B$ and $A\cup B$ are
\begin{align}
    S(A)&=S([b_1,b_2])=\frac{c}{3}\log\frac{b_2-b_1}{\epsilon},\\
    S(B)&=S^{\rm gen}_{[0,b_3]}(a_3),\\
    S(AB)&=S^{\rm gen}_{[0,b_1]}(a_1)+\frac{c}{3}\log\frac{b_3-b_2}{\epsilon}.
\end{align}
Then the mutual information is given by
\begin{align}\label{jt_phase2_mutual}
    I(A:B)&=\frac{c}{3}\log\frac{b_2-b_1}{b_3-b_2}+S^{\rm gen}_{[0,b_3]}(a_3)-S^{\rm gen}_{[0,b_1]}(a_1)\notag\\
    &=\frac{c}{6}\log\sqbra{\frac{(b_2-b_1)^2}{(b_3-b_2)^2}\frac{(b_3+a_3)^2}{(b_1+a_1)^2}\frac{a_1}{a_3}}+\phi_r\sbra{\frac{1}{a_3}-\frac{1}{a_1}}> 0
\end{align}
In this case, the reflected entropy is given by \eqref{SR_Faulkner}.
Physically speaking, since we are working in a field theory manner, we should use \eqref{SR_Faulkner}, even though it is mathematically equivalent to \eqref{SR_phase2} and \eqref{takayanagi_SR}.
Then the Markov gap is given by
\begin{align}\label{jt_phase2_h}
    h=\frac{c}{3}\log\sqbra{
    \sbra{\frac{1+\sqrt{1-x}}{\sqrt{x}}}^2\frac{(b_3-b_2)(b_1+a_1)}{(b_2-b_1)(b_3+a_3)}\sqrt{\frac{a_3}{a_1}}
    }+\phi_r\sbra{\frac{1}{a_1}-\frac{1}{a_3}}
\end{align}
Notice that $\phi_0$ does not appear in \eqref{jt_phase2_h} and mutual information \eqref{jt_phase2_mutual}, so we can always tune $\phi_0$ to satisfy \eqref{jt_phase2_cond1} and \eqref{jt_phase2_cond2}.
It is not hard to find that $\partial _{b_3} h>0$, so that $h$ monotonically increases with $b_3$, the minimum is at $b_3\rightarrow b_2$.
In this limit, the reflected entropy reads
\begin{align}\label{jt_phase2_lim_SR}
    S_{R}(A:B)\approx\frac{c}{3}\log\frac{(b_2-b_1)(a_1+b_2)}{(b_1+a_1)(b_3-b_2)}+\frac{2c}{3}\log 2.
\end{align}
We write the reflected entropy in a suggestive way, in which the term $\frac{2c}{3}\log 2$ is isolated,
and we explicitly include the divergent factor $-(c/3)\log(b_3-b_2)$.
But the divergence in $S_{R}$ is doomed to be cancelled by mutual information.

The Markov gap is then
\begin{align}\label{jt_phase2_lim_h}
    h=\frac{c}{3}\log\sqbra{
    \frac{b_2+a_1}{b_2+a_2}\sqrt{\frac{a_2}{a_1}}
    }+\phi_r\sbra{
    \frac{1}{a_1}-\frac{1}{a_2}
    }+\frac{2c}{3}\log 2.
\end{align}
Again, we find $h$ monotonically increases with $b_2$, and the minimum now is taken at $b_2\simeq b_1$.
It is obvious that $b_1=b_2$ is a solution to $h=\frac{2c}{3}\log 2$, which turns out to be the only acceptable solution.
The other solution is either imaginary or excluded by $b_2>b_1$.
Therefore, we conclude that
\begin{equation}
    h\geq \frac{2c}{3}\log 2,
\end{equation}
with $h\rightarrow \frac{2c}{3}\log 2$ when $b_1\simeq b_2\simeq b_3$.

Notice that there is a subtlety.
When taking the limit $b_2\rightarrow b_3$, we let $b_3-b_2\ll b_2-b_1$, and then we take $b_2\rightarrow b_1$.
Therefore, in the limit $b_3\rightarrow b_2\rightarrow b_1$, we still require that
$b_3-b_2\ll b_2-b_1\ll b_1$.
We can see this behavior in numerical computation.
In Fig.\ref{fig:JT_phase2_num}, we show the Markov gap against $b_2-b_1$ with a specific parameter setting, that is $b_1=1$, $\phi_r=100$ and $c=12000$, the same as that used in \cite{Chandrasekaran:2020qtn}.
For fixed $b_1$, the minima are located at positions that satisfy $b_3-b_2\ll b_2-b_1$.
As $b_3-b_2$ gets smaller, the minimum approaches $h=\frac{2c}{3}\log 2$ at $b_2-b_1=0$.

\begin{figure}
    \centering
    \includegraphics[width=0.8\textwidth]{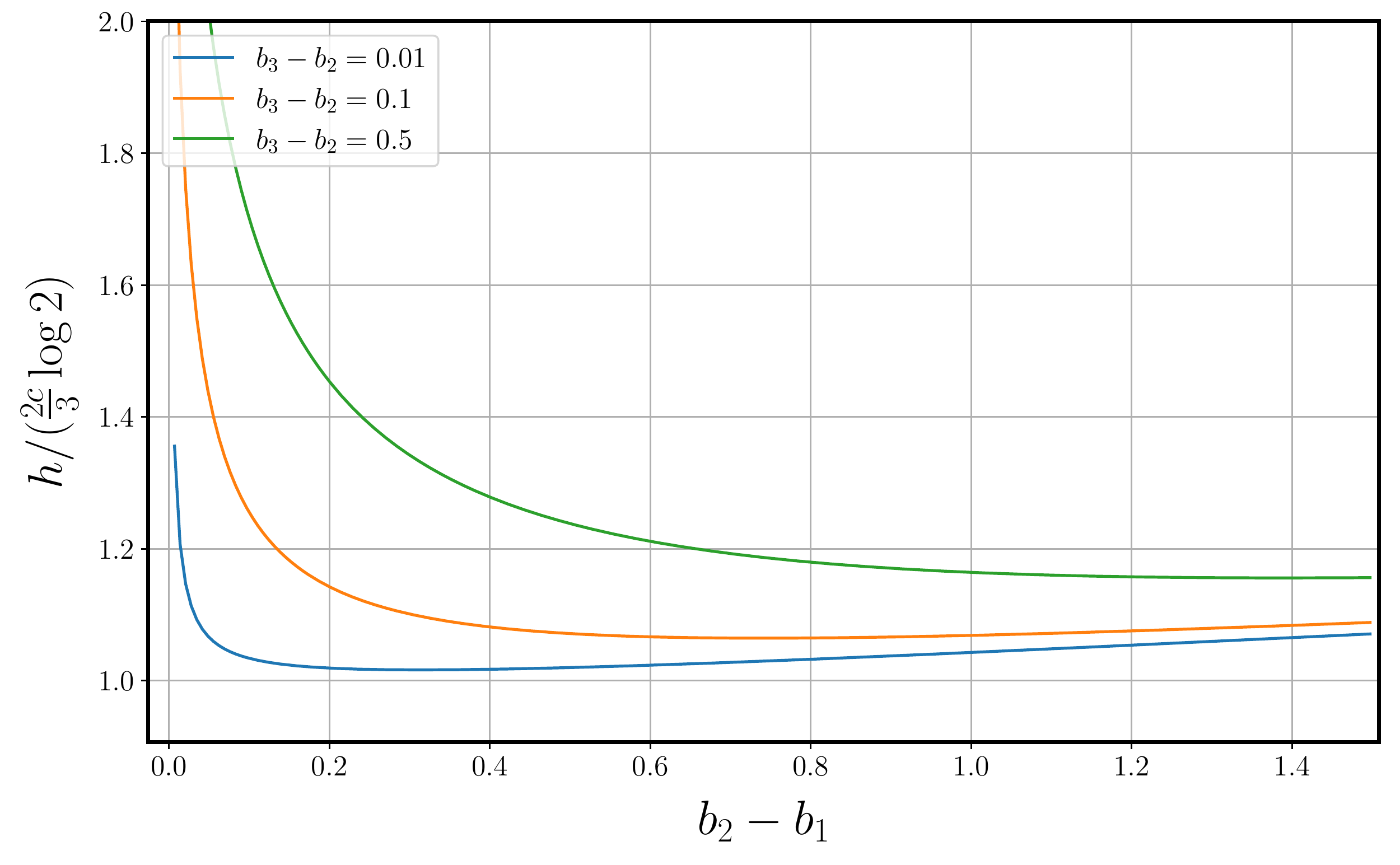}
    \caption{The Markov gap $h$ \eqref{jt_phase2_h} for phase-D2 of JT model.
    We set $\phi_r=100$, $c=12000$ and $b_1=1$. 
    The horizontal axis is $b_2-b_1$, while the vertical axis is $h/(\frac{2c}{3}\log 2)$.
    Every curve corresponds to a constant difference $b_3-b_2$.
    At minima, we have $b_3-b_2\ll b_2-b_1$.
    }
    \label{fig:JT_phase2_num}
\end{figure}


\subsubsection*{Phase-D3}

\begin{figure}
    \centering
    \includegraphics[width=0.8\textwidth]{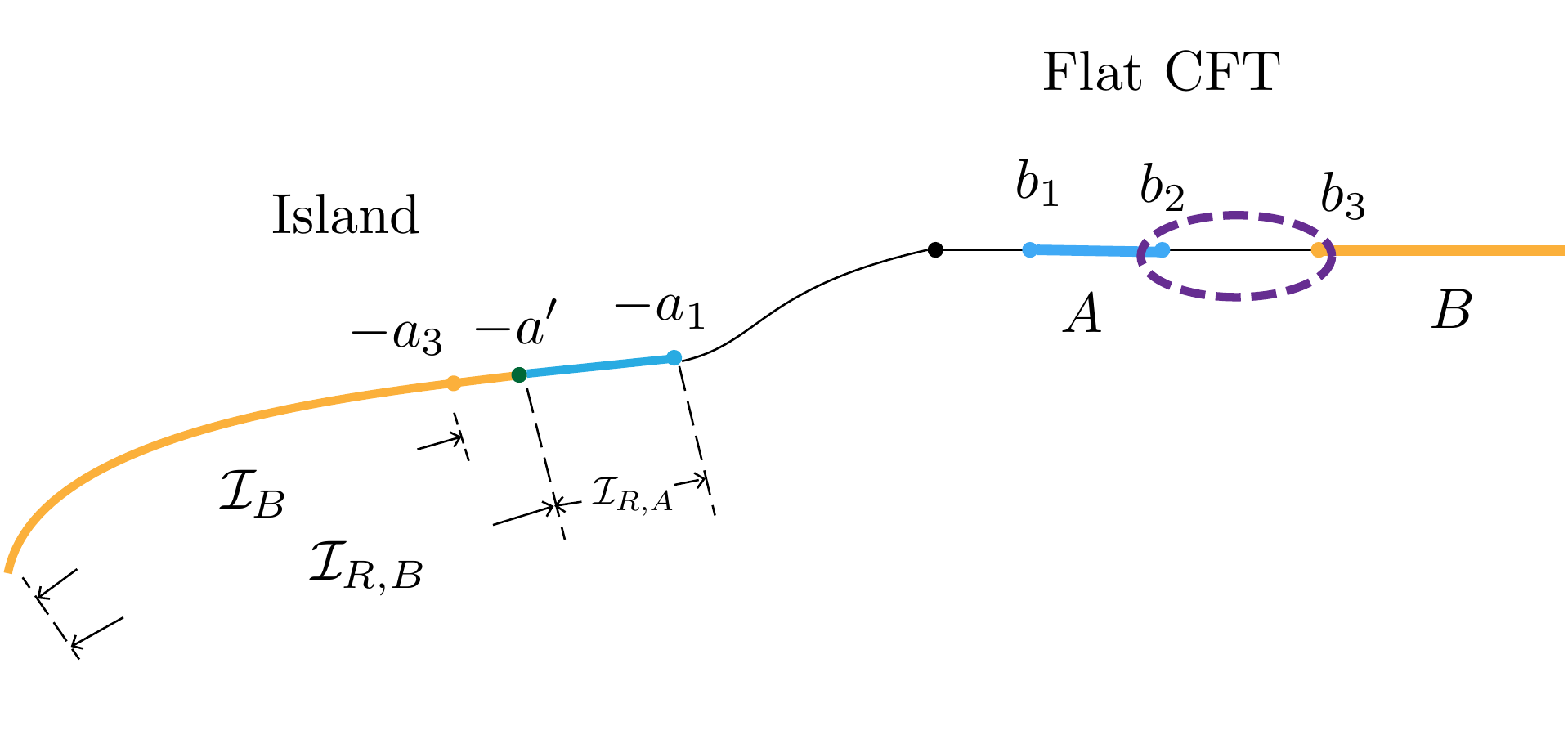}
    \caption{Phase-D3 for JT gravity.
    The purple dashed circle denotes the boundary gap between  $A\cup \mathcal I_{R,A}$ and  $B\cup \mathcal I_{R,B}$. 
    $a'$ denotes the island cross-section.}
    \label{fig:jtD3}
\end{figure}

In this phase (Fig.\ref{fig:jtD3}), we require $A$ to have no entanglement island, but admits the reflected island, which gives the conditions for this phase
\begin{align}
    S([b_1,b_2])&<S^{\rm gen}_{[0,b_1]}(a_1)+S^{\rm gen}_{[0,b_2]}(a_2),\label{JT_D3_cond1}\\
    S_R(A:B)&<\frac{2c}{3}\log\sbra{\frac{1+\sqrt{1-x}}{\sqrt{x}}}.
\end{align}
The entanglement entropies are
\begin{align}
    S(A)=&S([b_1,b_2]),\\
    S(B)=&S^{\rm gen}_{[0,b_3]}(a_3),\\
    S(AB)=&S^{\rm gen}_{[0,b_1]}(a_1)+S([b_2,b_3]).
\end{align}
Then the mutual information is given by
\begin{align}
    I(A:B)=\frac{c}{3}\log\sqbra{
    \frac{(b_3+a_3)(b_2-b_1)}{(b_1+a_1)(b_3-b_2)} \sqrt{\frac{a_1}{a_3}}
    }+\phi_r\sbra{
    \frac{1}{a_3}-\frac{1}{a_1}
    }.
\end{align}
The reflected entropy can be derived via replica trick by the correlation functions of twist operators, and the result is \cite{Chandrasekaran:2020qtn}
\begin{align}
    S_R(A:B)=\frac{c}{3}\log\frac{(b_3+a')(b_2+a')}{a'(b_3-b_2)}+2\frac{\phi_r}{a'}+2\phi_0+\frac{c}{3}\log2,
\end{align}
in which $a'$ is island cross-section $\partial \mathcal I_{R,A}\cap \partial\mathcal I_{R,B}$, given by the following equation
\begin{equation}\label{jta'}
       \partial_{a'}S_R=0\Rightarrow \frac{1}{b_2+a'}+\frac{1}{b_3+a'}-\frac{1}{a'}-\frac{q}{a'^2}=0.
\end{equation}

We arrive at the Markov gap
\begin{align}
    h&=\frac{c}{3}\log\sqbra{
    \frac{(b_3+a')(b_2+a')(b_1+a_1)}{a'(b_3+a_3)(b_2-b_1)}\sqrt{\frac{a_3}{a_1}}
    }+\phi_r\sbra{
    \frac{2}{a'}-\frac{1}{a_3}+\frac{1}{a_1}
    }+2\phi_0+\frac{c}{3}\log 2\notag\\
    &\geq\frac{c}{3}\log\sqbra{ \frac{\sqrt{a_2a_3}(b_3+a')(b_2+a')}{a'(b_3+a_3)(b_2+a_2)}}+\phi_r\sbra{
    \frac{2}{a'}-\frac{1}{a_3}-\frac{1}{a_2}
    }+\frac{c}{3}\log 2,\label{JT_D3_h}
\end{align}
where we have used \eqref{JT_D3_cond1}.

If $q\ll 1$, we get a simple solution to island cross section \eqref{jta'}, that is
\begin{equation}\label{jta'1}
    a'\approx \sqrt{b_2b_3}.
\end{equation}
This is reminiscent of the DES result of $a'$ in \eqref{SR_phase3}.
In the limit $q\rightarrow0$, the second term $\phi_r/a$ in generalized entropy \eqref{Sgen} can be ignored, so that the generalized entropy is given by an effective term plus a constant area term $\phi_0$.
This is indeed similar to the boundary QES description of the DES model \cite{Deng:2020ent}.
In addition, we also have
\begin{equation}\label{jta2a3}
    a_2\approx b_2,\ 
    a_3\approx b_3
\end{equation}
Inserting \eqref{jta'1} and \eqref{jta2a3} into \eqref{JT_D3_h}, we get
\begin{align}
    h\geq&\frac{c}{3}\log\frac{(\sqrt{b_2}+\sqrt{b_3})^2}{4\sqrt{b_2b_3}}+
    \frac{cq}{6}\left(\frac{2}{\sqrt{b_2b_3}}-\frac{1}{b_3}-\frac{1}{b_2} \right)+\frac{c}{3}\log 2\notag\\
    \gtrsim& \frac{c}{3}\log\frac{(\sqrt{b_2}+\sqrt{b_3})^2}{4\sqrt{b_2b_3}}+\frac{c}{3}\log 2\notag\\
    \geq & \frac{c}{3}\log 2.
\end{align}

For general $q$, it is hard to analytically solve the lower bound of the Markov gap. 
Numerically we find that the R.H.S of \eqref{JT_D3_h} grows with $b_3$, and clearly, if $b_3=b_2$ the second line of \eqref{JT_D3_h} equals $\frac{c}{3}\log 2$, as $a'$ in this case reduces to $a_2$.
Therefore, we conclude that
\begin{align}
    h>\frac{c}{3}\log 2,\quad \lim_{b_3\rightarrow b_2}h=\frac{c}{3}\log 2.
\end{align}

\subsubsection*{Phase-D4}

\begin{figure}
    \centering
    \includegraphics[width=0.8\textwidth]{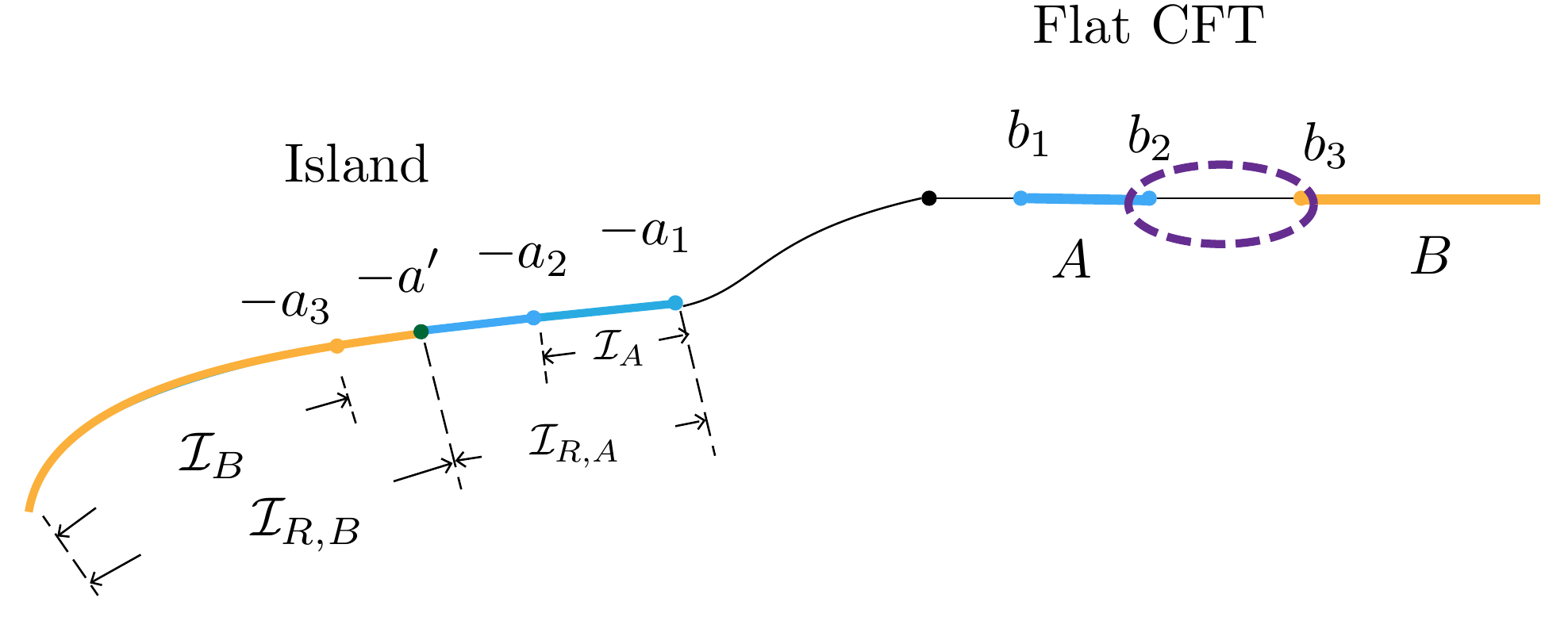}
    \caption{Phase-D4 for JT gravity.
    The purple dashed circle denotes the boundary gap between  $A\cup \mathcal I_R(A)$ and  $B\cup \mathcal I_R(B)$.
    $a'$ denotes the island cross-section.}
    \label{fig:jtD4}
\end{figure}

In phase-D4 (Fig.\ref{fig:jtD4}), both $A$ and $B$ have their islands, and the mutual information $I(A:B)>0$.
The entanglement entropies for $A$, $B$ and $AB$ are 
\begin{align}
    S(A)=&S^{\rm gen}_{[0,b_1]}(a_1)+S^{\rm gen}_{[0,b_2]}(a_2),\\
    S(B)=&S^{\rm gen}_{[0,b_3]}(a_3),\\
    S(AB)=&S^{\rm gen}_{[0,b_1]}(a_1)+S([b_2,b_3]).
\end{align}
Then the mutual information is given by
\begin{align}
    I(A:B)&=S^{\rm gen}_{[0,b_2]}(a_2)+S^{\rm gen}_{[0,b_3]}(a_3)-S([b_2,b_3])\notag\\
    &=2\phi_0+\phi_r\sbra{\frac{1}{a_2}+\frac{1}{a_3}}+\frac{c}{6}\log
    \frac{(b_2+a_2)^2(b_3+a_3)^2}{a_2a_3(b_3-b_2)^2}.
\end{align}
The reflected entropy for this phase is given by
\begin{align}
    S_R(A:B)=\frac{c}{3}\log\frac{(b_3+a')(b_2+a')}{a'(b_3-b_2)}+2\frac{\phi_r}{a'}+2\phi_0+\frac{c}{3}\log2,
\end{align}

Then the Markov gap is
\begin{align}
    h&=S_R-I\notag\\
    &=\phi_r\sbra{
    \frac{2}{a'}-\frac{1}{a_2}-\frac{1}{a_3}
    }
    +\frac{c}{3}\log\sqbra{
    \frac{\sqrt{a_2a_3}}{a'}\frac{(b_3+a')(b_2+a')}{(b_2+a_2)(b_3+a_3)}
    }+\frac{c}{3}\log 2\notag\\
    &> \frac{c}{3}\log2,
\end{align}
where the second line is just the second line in \eqref{JT_D3_h}.
So we conclude that
\begin{align}
    h>\frac{c}{3}\log 2,\quad \lim_{b_3\rightarrow b_2}h=\frac{c}{3}\log 2.
\end{align}

\section{The Markov gap for generic 2D extremal black holes}\label{sec:2dbh}

\begin{figure}
    \centering
    \includegraphics[width=0.8\textwidth]{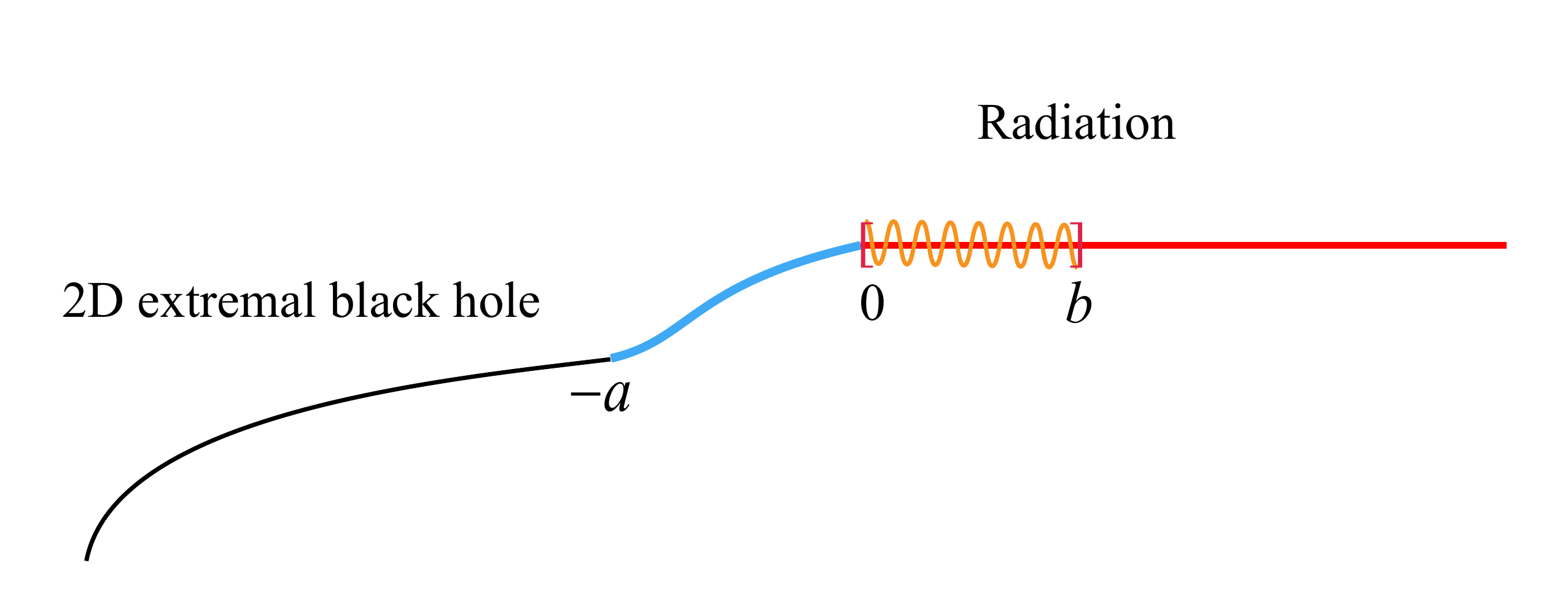}
    \caption{An illustration for the interval of 2D extremal black holes.}
    \label{fig:extbh}
\end{figure}

In this section, we derive the Markov gap in a rather generic 2D extremal black hole coupled to CFT at the large central charge limit.
The computation is performed using the correlation functions of twist operators.
\subsection{Setups}
The metric of generic 2d extremal black holes can be written in a conformally-flat vacuum coordinate\footnote{Note that for AdS black holes, to make black holes evaporate, we glue the original AdS spacetime with a flat spacetime along the boundary and impose the transparent boundary condition.}, that is,
\begin{equation}
    \mathrm{d}s^2=-\frac{1}{\Omega^2}\mathrm{d}x^+\mathrm{d}x^-,
\end{equation}
where $x^{\pm} =t \pm x$.
At the static time slice, the entanglement entropy for an interval $[0,b]$ on radiation region (see Fig.\ref{fig:extbh}) is given by minimization of the generalized entropy, that is, 
\begin{equation}\label{sgen}
   S([0,b])=\text{Min}\{\text{Ext}_{a}\{S^{\text{gen}}_{[0,b]}(a)\}\},\ S^{\text{gen}}_{[0,b]}(a)=\frac{\text{Area}(-a)}{4G^{(2)}_N}+S_{\text{eff}}([-a,b]),
\end{equation}
where {after minimization, $-a$ represents the boundary of the island of the interval $[0,b]$. $\text{Area}(-a)$ is the area term at the boundary of island $-a$, and the effective entropy for 2D CFT is
\begin{equation}\label{seff}
    S_{\text{eff}}([-a,b])
   =\frac{c}{6}\log \frac{d_{-ab}^{2}}{\Omega_{-a}\Omega_{b}},
\end{equation}
where $d_{-ab}$ is the distance between two points $(t_{-a},x_{-a})$ and $(t_b,x_{b})$ on the time slice $t_{-a}=t_b$ of the flat spacetimes $\D s^2=-\D x^+\D x^-$.

For an interval $[b_1,b_2]$ which does not admit an entanglement island, the entanglement entropy is 
\begin{equation}\label{snoisland}
    S([b_1,b_2])=S_{\text{eff}}([b_1,b_2])=\frac{c}{6}\log \frac{d_{b_1b_2}^{2}}{\Omega_{b_1}\Omega_{b_2}}.
\end{equation}
This reduces to $\frac{c}{3}\log d_{b_1b_2}$ for flat CFT with $\Omega=1$.

\subsection{The Markov gap}

Without loss of generality, we consider only phase-D2 and phase-D4.

\subsection*{Phase-D2}
With the same procedure in Sec.\ref{sec:42}, the Markov gap for phase-D2 (Fig.\ref{fig:jtD2}) is given by
\begin{align}\label{srd2}
    h=\frac{c}{3}\log\sqbra{
    \sbra{\frac{1+\sqrt{1-x}}{\sqrt{x}}}^2\frac{d_{b_3b_2}d_{-a_1b_1}}{d_{b_2b_1}d_{-a_3b_3}}\sqrt{\frac{\Omega_{-a_3}}{\Omega_{-a_1}}}
    }+\frac{\mathcal{A}(-a_1)}{4G_N}-\frac{\mathcal{A}(-a_3)}{4G_N},
\end{align}
where  the cross-ratio is given by
\begin{equation}
x=\frac{d_{-a_1b_1}d_{b_3b_2}}{d_{-a_1b_2}d_{b_3b_1}}.
\end{equation}
It is easy to find that as $b_2\to b_3$, the Markov gap \eqref{srd2} decreases. 
Thus it is sufficient to prove our boundary inequality \eqref{main_clain_boundary} in the limit of $b_2\to b_3$.
The reflected entropy in this limit is reduced to
\begin{equation}
\begin{split}
    S_R(A:B)=& \frac{2c}{3}\log \frac{2}{\sqrt{x}}
    =\frac{c}{3}\log\frac{d_{b_3b_1}d_{-a_1b_2}}{d_{-a_1b_1}d_{b_3b_2}}+\frac{2c}{3}\log2, 
    \end{split}
\end{equation}
which can be further written as
\begin{equation}\label{srd21}
    S_R(A:B)=\frac{c}{6}\left(\log\frac{d^2_{b_3b_1}}{\Omega_{b_3}\Omega_{b_1}}
    +\log\frac{d^2_{-a_1b_2}}{\Omega_{b_2}\Omega_{-a_1}}
    -\log\frac{d^2_{-a_1b_1}}{\Omega_{b_1}\Omega_{-a_1}}
    -\log\frac{d^2_{b_3b_2}}{\Omega_{b_3}\Omega_{b_2}}\right)+\frac{2c}{3}\log2.
\end{equation}
Using \eqref{sgen}, \eqref{seff} and \eqref{snoisland}, \eqref{srd21} can be written as
\begin{equation}
    S_R\overset{\text{math}}{=}S([b_1,b_3])+S^{\text{gen}}_{[0,b_2]}(a_1)-S^{\text{gen}}_{[0,b_1]}(a_1)
    -S([b_2,b_3])+\frac{2c}{3}\log2,
\end{equation}
where $\overset{\text{math}}{=}$ means that the equal sign should be understood from the mathematical aspect rather than the physical aspect.
Notice that we have
\begin{align}
    &S([b_1,b_3])>S([b_1,b_2])=S(A),\\
    &S([b_2,b_3])+S^{\text{gen}}_{[0,b_1]}(a_1)=S(AB).
\end{align}
Then the reflected entropy should satisfy
\begin{align}\label{srd22}
   S_R
    &\geq S(A)+S^{\text{gen}}_{[0,b_2]}(a_1)-S(AB)+\frac{2c}{3}\log 2\notag\\
    &\geq S(A)+S(B)-S(AB)+\frac{2c}{3}\log 2\notag\\
    &=I(A:B)+\frac{2c}{3}\log 2.
\end{align}
where the second line is from the fact that $S^{\text{gen}}_{[0,b_2]}(a_2)$ is the minimum in varying $a$, i.e. $S^{\text{gen}}_{[0,b_2]}(a_1)>S^{\text{gen}}_{[0,b_2]}(a_2)$ and $S^{\text{gen}}_{[0,b_2]}(a_2)\approx S(B)$ in the  limit of $b_2\to b_3$.
The result is just as expected from our boundary inequality \eqref{main_clain_boundary}.
And the equality $h=\frac{2c}{3}\log 2$ is taken at $b_3\simeq b_2\simeq b_1$.

\subsection*{Phase-D4}
For phase-D4 (Fig.\ref{fig:jtD4}), 
in calculating $S_{\rm eff}(A\cup B\cup \mathcal{I}_A\cup \mathcal{I}_B)$ at large-$c$ limit, the multi-point correlation function of twist operators can be factorized into \cite{Hartman:2013mia,Chandrasekaran:2020qtn}
\begin{equation}
\begin{split}
    &\left\langle\sigma_{g_{A}}\left(b_{1}\right) \sigma_{g_{A}^{-1}}(a_1) \sigma_{g_{A}^{-1}}\left(b_{2}\right) \sigma_{g_{B}}\left(b_{3}\right) \sigma_{g_{A} g_{B}^{-1}}\left(a^{\prime}\right)\right\rangle\\
    \to&
    \left\langle\sigma_{g_{A}}\left(b_{1}\right) \sigma_{g_{A}^{-1}}(a)\right\rangle
    \left\langle \sigma_{g_{A}^{-1}}\left(b_{2}\right) \sigma_{g_{B}}\left(b_{3}\right) \sigma_{g_{A} g_{B}^{-1}}\left(a^{\prime}\right)\right\rangle,
    \end{split}
\end{equation}
\begin{equation}
\begin{split}
    \left\langle\sigma_{g_{m}}\left(b_{1}\right) \sigma_{g_{m}^{-1}}(a) \sigma_{g_{m}^{-1}}\left(b_{2}\right) \sigma_{g_{m}}\left(b_{3}\right)\right\rangle
    \to
    \left\langle\sigma_{g_{m}}\left(b_{1}\right) \sigma_{g_{m}^{-1}}(a)\right\rangle
    \left\langle \sigma_{g_{m}^{-1}}\left(b_{2}\right) \sigma_{g_{m}}\left(b_{3}\right) \right\rangle,
    \end{split}
\end{equation}
and the reflected entropy at large $c$ limit is 
\begin{equation}\label{srd41}
\begin{split}
&S_{R}(A:B)\\
=&2\frac{\mathcal{A}(-a')}{4G_N^{(2)}}+\lim_{n,m\to 1}\frac{1}{1-n} \log \frac{\left\langle\sigma_{g_{A}}\left(b_{1}\right) \sigma_{g_{A}^{-1}}(a_1)
\right\rangle\left\langle
\sigma_{g_{A}^{-1}}\left(b_{2}\right) \sigma_{g_{B}}\left(b_{3}\right) \sigma_{g_{A} g_{B}^{-1}}\left(-a^{\prime}\right)  \right\rangle_{\text{CFT}^{\otimes mn}}}{\left(\left\langle\sigma_{g_{m}}\left(b_{1}\right) \sigma_{g_{m}^{-1}}(a_1)
\right\rangle\left\langle
\sigma_{g_{m}^{-1}}\left(b_{2}\right) \sigma_{g_{m}}\left(b_{3}\right)\right\rangle_{\text{CFT}^{\otimes m}}\right)^{n}},
\end{split}
\end{equation}
where $\sigma_{g_A},\sigma_{g_Ag_B^{-1}},\sigma_{g_m}$ are twist operators living at the endpoints of the intervals (branch points in the replica manifold) with the scaling dimensions \cite{Dutta:2019gen}
\begin{equation}
    \Delta_{g_Ag_B^{-1}}=\frac{c}{12n}(n-1)(n+1)=2\Delta_n,\
\Delta_{g_A}=\Delta_{g_A^{-1}}=n\Delta_m=\frac{cn(m^2-1)}{24m}.
\end{equation}
Note that due to $\Delta_{g_{A(B)}}=n\Delta_m$, 2-point function of twist operators at $b_1$ and $a_1$ in \eqref{srd41} will be canceled and \eqref{srd41} is  reduced to
\begin{equation}\label{srd42}
\begin{split}
S_{R}(A:B)
=&2\frac{\mathcal{A}(-a')}{4G_N^{(2)}}+\lim_{n,m\to1}\frac{1}{1-n} \log \frac{\left\langle
\sigma_{g_{A}^{-1}}\left(b_{2}\right) \sigma_{g_{B}}\left(b_{3}\right) \sigma_{g_{A} g_{B}^{-1}}\left(-a^{\prime}\right)  \right\rangle}{\left\langle
\sigma_{g_{m}^{-1}}\left(b_{2}\right) \sigma_{g_{m}}\left(b_{3}\right)\right\rangle^n}\\
=&2\frac{\mathcal{A}(-a')}{4G_N^{(2)}}+\frac{c}{6}\log\left(
\frac{d_{-a'b_3}^2}{\Omega_{-a'}\Omega_{b_3}}
\frac{d_{-a'b_2}^2}{\Omega_{-a'}\Omega_{b_2}}
\frac{\Omega_{b_2}\Omega_{b_3}}{d_{b_2b_3}^2}\right)+C'_{n\to1,m\to1},
\end{split}
\end{equation}
where \begin{equation}
    C'_{n,m}\equiv\frac{1}{1-n}\log C_{n,m}
\end{equation}
and $C_{n,m}=(2m)^{-4\Delta_n}$ is the structure constant of 3-point correlation function and
\begin{equation}
    C'_{n\to1,m\to1}=\frac{c}{3}\log2.
\end{equation}
Using \eqref{sgen}, \eqref{seff} and \eqref{snoisland}, \eqref{srd42} can be written as
\begin{equation}\label{srd43}
    S_R\overset{\text{math}}{=}S^{\text{gen}}_{[0,b_3]}(a')+S^{\text{gen}}_{[0,b_2]}(a')
    -S([b_2,b_3])+\frac{c}{3}\log2
\end{equation}
Again, the equal sign here should be understood from the mathematical aspect rather than the physical aspect. 
With the same argument as in the previous phase, we can arrive at
\begin{align}\label{srd44}
    S_{R}\geq I(A:B)+\frac{c}{3}\log 2,
\end{align}
where we used
\begin{align}
    &S^{\text{gen}}_{[0,b_3]}(a')\geq S^{\text{gen}}_{[0,b_3]}(a_3)=S(B),\\
    &S^{\text{gen}}_{[0,b_2]}(a')+S^{\text{gen}}_{[0,b_1]}(a_1)\geq S(A),\\
    &S^{\text{gen}}_{[0,b_1]}(a_1)+S([b_2,b_3])=S(AB).
\end{align}
The equality in \eqref{srd44} is taken at $b_2\simeq b_3$.
As expected from \eqref{main_clain_boundary}, the lower bound is $\frac{c}{3}\log 2$ as there is only one gap between $A\cup \mathcal{I}_{R,A}$ and $B\cup \mathcal{I}_{R,B}$.

From the above analysis for phase-D4, it is insightful to see that the lower bound of Markov gap $\frac{c}{3}\log 2$ stems from the 3-point structure constant from the boundary viewpoint. 

\section{Discussion}\label{sec:disc}
We have studied the Markov gap in the DES model, JT gravity and generic 2d extremal black holes in the presence of islands for different phases.
Some of these phases are not considered in the literature. For example, phase-D3, where $A$ has no entanglement island but admits {a} reflected island.
In doing this, we correct some little errors in literature as by-products.
Then, all the results respect the bulk inequality \eqref{main_clain_bulk} and the boundary inequality \eqref{main_clain_boundary}.
However, the rigorous proof {remains} unknown, either from the bulk gravity side or the boundary theory side.
We point out the obstacle.
In \cite{Hayden:2021gno}, this inequality is proved by {using} a property of the right-angled pentagon.
That is, for a right-angled pentagon in hyperbolic space, the lengths of its three sides satisfy $\alpha+\beta-\sigma\geq \log 2$, where $\alpha$ and $\beta$ are adjacent, and $\sigma$ is non-adjacent to $\alpha$ and $\beta$.
The right-angled pentagon is enclosed by geodesics and degenerate sides at infinity.
In the DES model, the EoW brane, {which locates along $\theta_0$ in bulk,} is neither a geodesic nor asymptotic infinity in Poincar\'e half-plane\footnote{
The only exception is when the brane has no tension.
In this case, the brane is located at $\theta_0=0$, a geodesic.
We show the geometric proof for this case in Appdendix.\ref{app:geointerpretation}.
}.

While all the results respect \eqref{main_clain_boundary}, there are some points we would like to stress.
For two single intervals $A$ and $B$, whereas $AB$ admits entanglement island $\mathcal I_{AB}$, there could be no island cross-section $a'$\footnote{
By no island cross-section, we mean it would not give a minimum reflected entropy.
}, like in phase-D2 and phase-A3.
Since $\mathcal I_{R,A}\cup \mathcal I_{R,B}=\mathcal I_{AB}$, we have either $\mathcal I_{R,A}=\mathcal I_{AB},\mathcal I_{R,B}=\emptyset$, or $\mathcal I_{R,B}=\mathcal I_{AB},\mathcal I_{R,A}=\emptyset\textbf{}$.
This can be determined from bulk using the entanglement wedge cross-section, which divides the entanglement wedge of $AB$ into two parts.
Nevertheless, from the boundary {topology}, this is subtle.
If only one of them admits an entanglement island, it is natural to assign the reflected island to this one.
If both $A$ and $B$ have their entanglement island, there is always an island cross-section that will divide the reflected island into their corresponding parts.
To show this in DES model, we just change the ``$\leq$'' into ``$\geq$'' in \eqref{DES_phase3_f1},
\begin{align}
    f\geq \log \sqbra{
\frac{b_2b_3-b_1^2+\sqrt{(b_2^2-b_1^2)(b_3^2-b_1^2)}}{b_2b_3-b_1^2-\sqrt{(b_2^2-b_1^2)(b_3^2-b_1^2)}}\frac{(b_3-b_2)^2}{(\sqrt{b_2}+\sqrt{b_3})^4}\frac{4b_1b_2}{(b_2-b_1)^2}
}\label{disc_1}.
\end{align}
If we prove the R.H.S is positive, then $f$ is also positive, indicating the existence of the island cross-section.
Note that the R.H.S  monotonically increases with $b_3$ as long as $b_3>b_2$.
Thus the R.H.S is always larger than its value at $b_3=b_2$, which leads to
\begin{align}
    f\geq 2\log\sbra{\sqrt{\frac{b_2}{b_1}}+\sqrt{\frac{b_1}{b_2}}}\geq 2\log 2>0. 
\end{align}
So if both $A$ and $B$ have entanglement islands, they have reflected islands.
Furthermore, if none of them has an entanglement island, then we cannot tell whether $\mathcal I_{R,A}=\mathcal I_{R}$ or $\mathcal I_{R,B}=\mathcal I_{R}$ from the simple topology of boundary regions.
Fortunately, we do not need to bother as they both have one gap. 
See Fig.\ref{fig:disc2} for an illustration.
\begin{figure}
    \centering
    \includegraphics[width=0.6\textwidth]{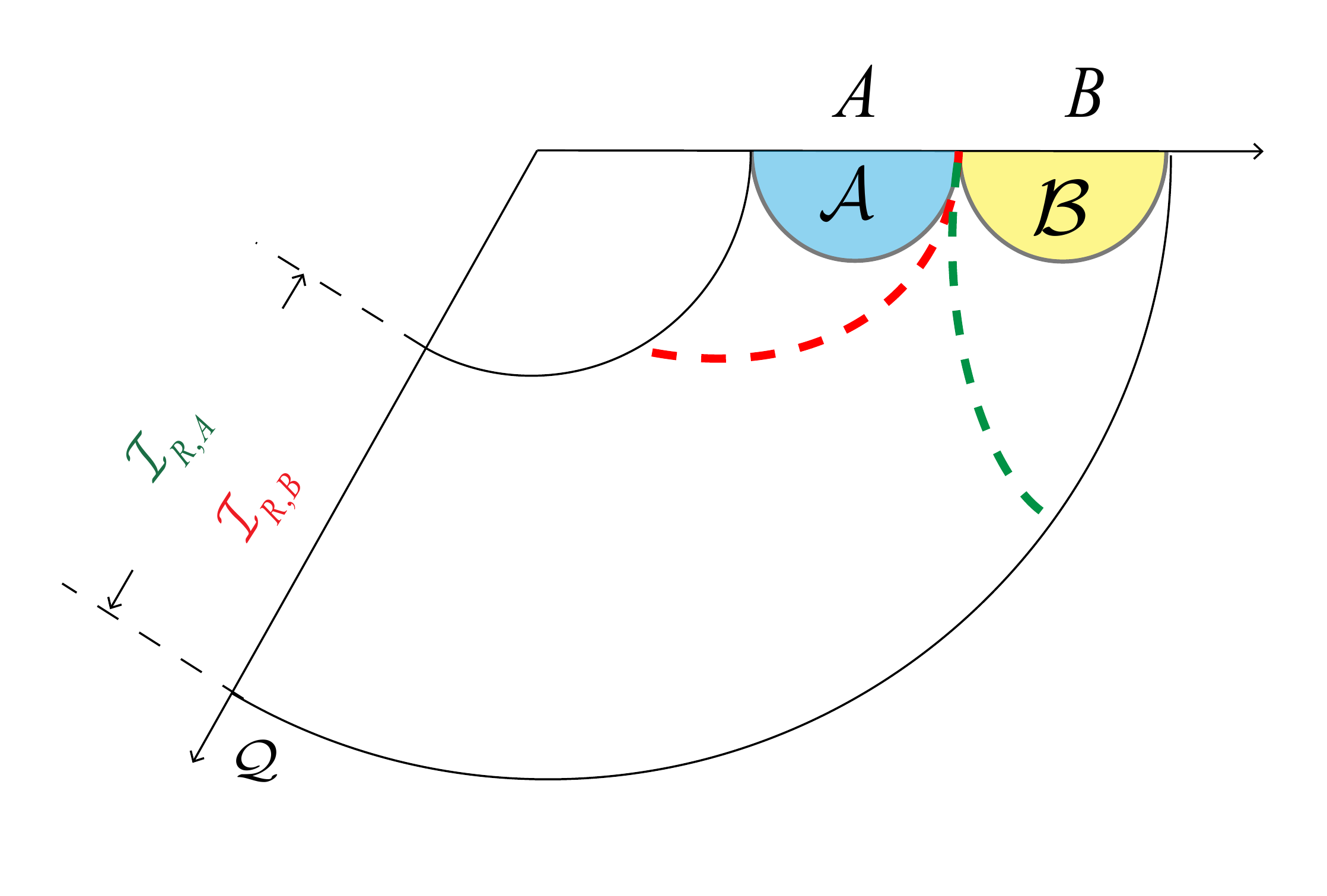}
    \caption{The phase where $AB$ has an island, but they do not admit an individual.
        The two possible EWCSs are shown with dashed curves, the corresponding reflected islands are denoted with the same color.
        These two give the same lower bound for the Markov gap $h\geq \frac{c}{3}\log 2$.
    }
    \label{fig:disc2}
\end{figure}

\begin{figure}
    \centering
    \includegraphics[width=0.45\textwidth]{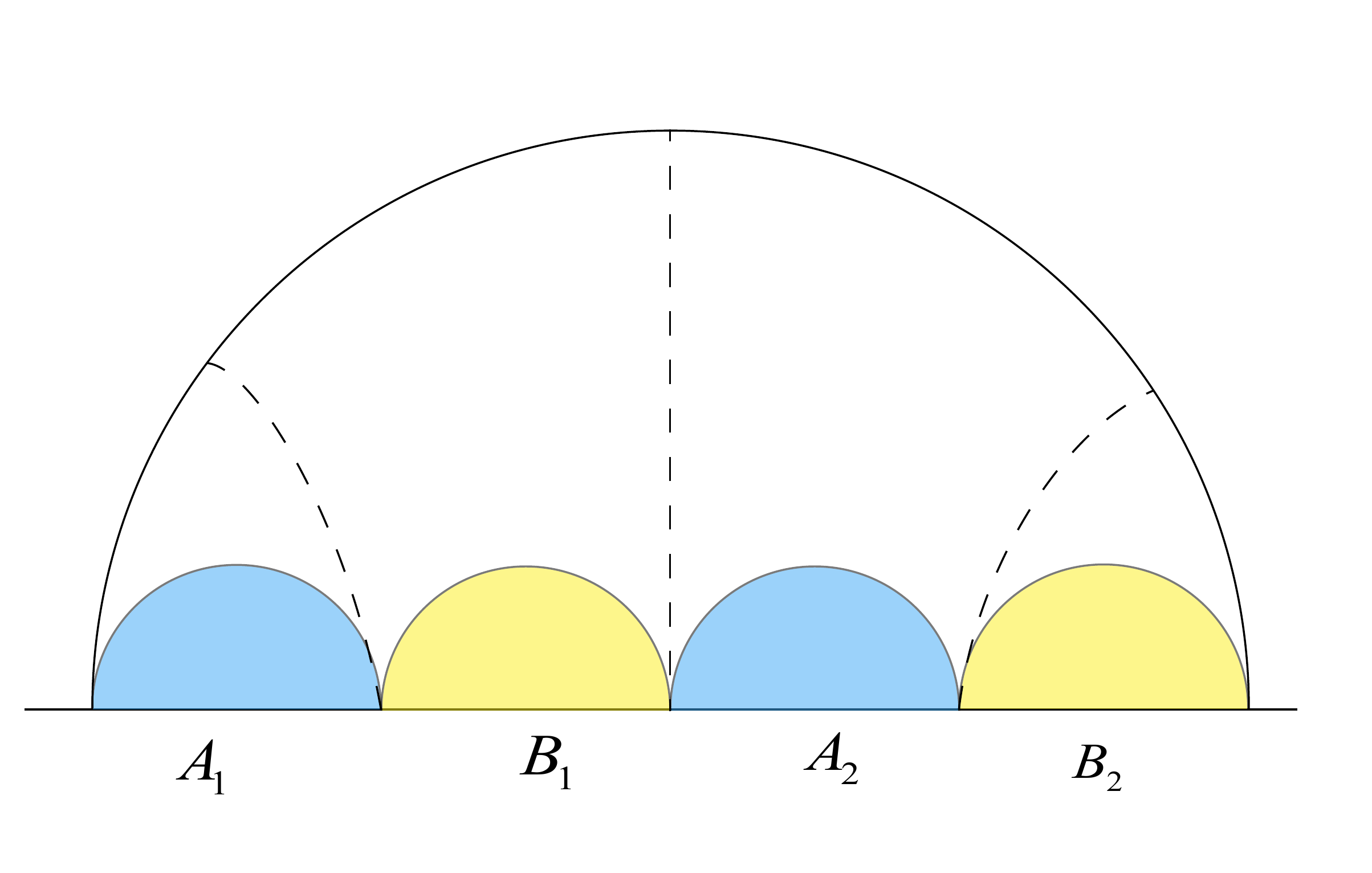}
    \includegraphics[width=0.45\textwidth]{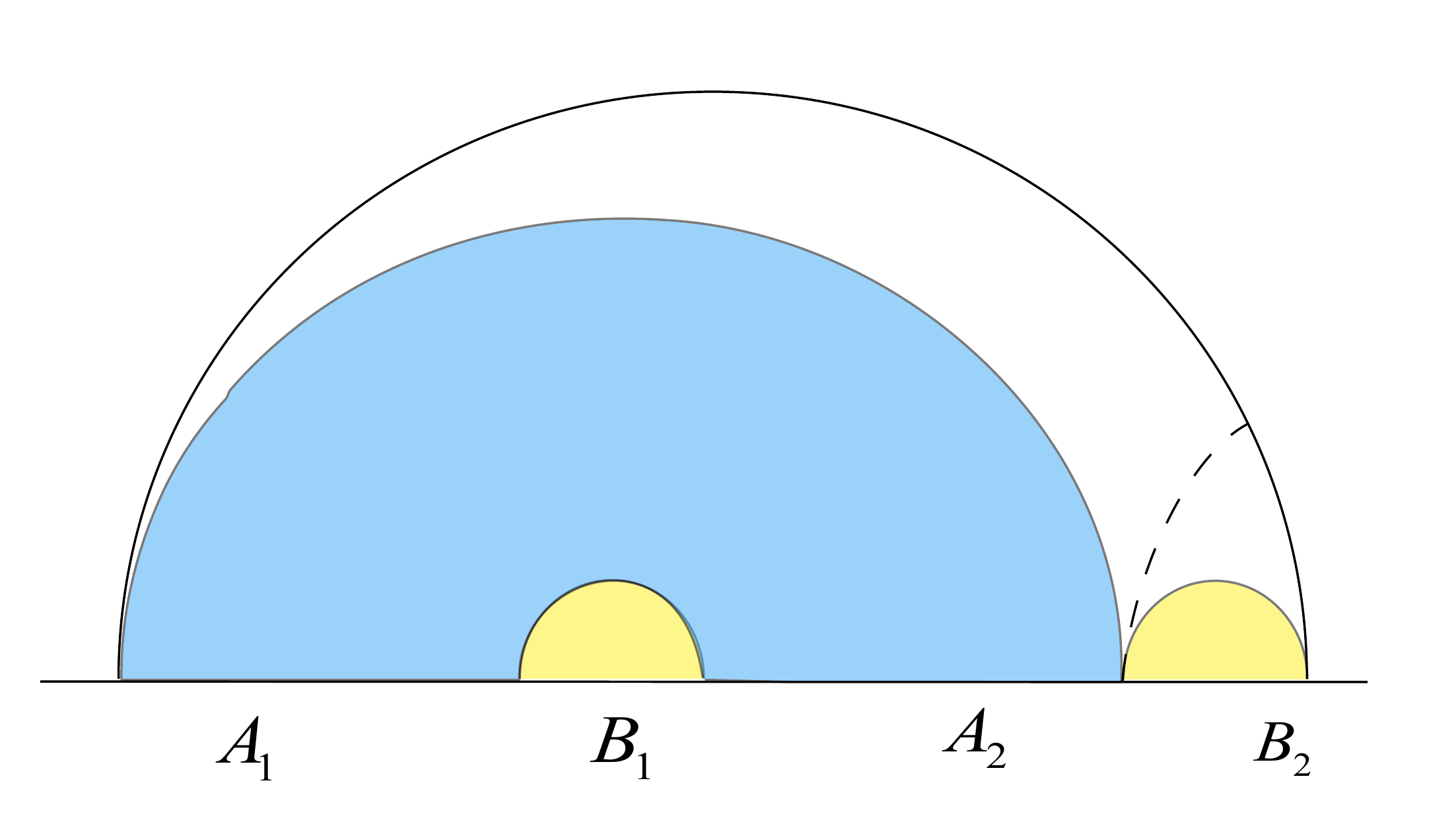}
    \caption{
   Left: Both entanglement wedges of $A$ and $B$ are disconnected.
   Right: The entanglement wedge of $A$ is connected while $B$ is disconnected. 
   The black dashed line denotes the cross-section.}
    \label{fig:disc}
\end{figure}

Nevertheless, the boundary statement \eqref{main_clain_boundary0} is not valid generally for multi-interval regions with disconnected EWCS.
An example is shown in Fig.\ref{fig:disc} on the left.
The reason is that we miss some information here.
The HPS proposal \eqref{Hayden_ineq} relies on the bulk object, namely the entanglement wedge cross-section, which is determined after we know the states of the boundary regions.
In contrast, the boundary statement considers only the topological information of the two boundary regions with a non-vanishing mutual information.
Even if the inequality is satisfied, the lower bound might be underestimated due to missing information.
In order to get a more accurate lower bound, more information about the regions is required to incorporate.
That would make it more challenging.

However, we conjecture that the lower bound can also be obtained by counting gaps, but in a more complicated way, as we should input the information about the lengths of intervals.
The basic idea is to decompose the disconnected region $A=\cup_iA_i$ into different multi-interval subregions $\mathbb{A}_j$ each with a closed contraction, like in \cite{Hartman:2013mia} using the monodromy method.
In bulk, every subregion should correspond to its individual single connected entanglement wedge \cite{Hartman:2013mia, Faulkner:2013yia}, which can be determined by their lengths in the vacuum state of CFT.
We denote the number of gaps between a subregion $\mathbb A_j$ and a subregion $\mathbb B_k$ in $B$ by $N_{jk}$.
Then the total number of gaps between $A$ and $B$ is given by $N=\sum _{j,k}N_{jk}$.
Note that only gaps between two subregions with non-vanishing mutual information are counted.
Although this statement seems much more elaborate than the HPS inequality, determining EWCS for generic two multi-interval regions is not direct.
One should determine the whole entanglement wedge first.
In doing this, the decomposition into subregions $\mathbb{A}_j$ and $\mathbb{B}_k$ is already done.
See Fig.\ref{fig:disc} for two examples to demonstrate the above observation.
On the left, the EWCS is the three disconnected dashed geodesics with 3 boundaries.
The entanglement wedges of $A=A_1\cup A_2$ and $B=B_1\cup B_2$ are disconnected.
So $\mathbb{A}_1=A_1,\mathbb{A}_2=A_2$ and $\mathbb{B}_1=B_1,\mathbb{B}_2=B_2$.
Then we count the gaps
\begin{align}
    N=N_{11}+N_{21}+N_{22}=1+1+1=3,
\end{align}
notice that $N_{12}=0$ {due to $I(A_1:B_2)=0$ }.
On the right, the entanglement wedge of $A$ is connected, while that of $B$ is disconnected.
So $\mathbb{A}_1=A_1\cup A_2$ and $\mathbb{B}_1=B_1,\mathbb{B}_2=B_2$.
Then the number of gaps is given by
\begin{align}
    N=N_{12}=1.
\end{align}
These are precisely the number of boundaries of EWCS.
As far as the phases in this paper are concerned, the decomposition is trivial: $\mathbb A=A\cup \mathcal I_{R,A}$ and $\mathbb B=B\cup \mathcal I_{R,B}$.


In holographic CFT, a non-vanishing Markov gap $h$ indicates the existence of non-trivial tripartite entanglement.
Since $h$ quantifies the deviation from having a perfect Markov recovery map, this suggests that tripartite entanglement prevents a perfect Markov recovery map.
Moreover, in the spirit of \cite{VanRaamsdonk:2010pw}, tripartite entanglement serves to assign boundaries to EWCS in the dual spacetime \cite{Hayden:2021gno}.
On the boundary, this is realized by adding gaps between the two regions, which can be rephrased as a physical gap leads to a gap {in} quantum recovery.
To us, the boundary inequality \eqref{main_clain_boundary} seems comprehensible, as the tripartite entanglement can be interpreted as entanglement among $A$, $B$, and the gap.

However, the simple relation \eqref{main_clain_boundary} should be considered {as} a property of the vacuum state because we did not input much information about the state.
For the CFT in a mixed state, there must be further tripartite entanglement between $A$, $B$, and a generic purification.
In this sense, the HPS inequality \eqref{Hayden_ineq} has more promising validity in general states, as the information about the state is embedded in its gravity dual.
But if the lower bound of the Markov gap will change or not requires further investigation.

One final remark.
The lower bound of $h$ varies in a discontinuous way as we change the length of an interval and undergo some phase transitions\footnote{We do not consider phase transition by removing a gap.}.
But this does not mean the Markov gap $h$ varies always discontinuously.
For example, as we vary the length of $[b_1,b_2]$ in phase-D2 in fig.\ref{Fig:D2}, we will encounter a phase transition to phase-D3 in fig.\ref{fig:D3}.
The lower bound changes immediately from $\frac{2c}{3}\log 2$ to $\frac{c}{3}\log 2$.
Though the EWCS undergoes a discontinuous change, its area is continuous (so does the reflected entropy $S_R$), as the phase transition happens when the two possible areas of EWCS coincide.
Therefore, $h$ is continuous.

\section{Conclusion}
In this paper, we studied the Markov gap $h\equiv S_R-I$, especially its lower bound, in the DES, JT gravity models, and generic 2d extremal black holes.
Phases with different island configurations are considered.
To get reasonable results, we correct some formulae in the literature.
Explicitly, we show how the lower bound of the Markov gap stems from the OPE coefficient.
This may shed light on general proof of \eqref{main_clain_boundary}.
Our results support the HPS inequality \eqref{Hayden_ineq}, with a specification that the lower bound only counts the boundaries of EWCS on minimal surfaces.
So \eqref{Hayden_ineq} could be a more general statement for holographic CFT.
However, the general geometric proof for DES model or for island dominance requires further study.

We proposed a boundary statement \eqref{main_clain_boundary}, that the lower bound of the Markov gap $h(A:B)$ is given by $\frac{c}{3}\log 2$ times the number of gaps between $\mathcal I_{R,A}\cup A$ and $\mathcal I_{R,B}\cup B$.
This statement is justified in all the phases we considered.
An analysis of the relation between a gap and $\frac{c}{3}\log 2$ is made in Appendix.\ref{app.adjacent}, where we find that the different cutoffs for the gap in mutual information and reflected entropy give rise to $\frac{c}{3}\log 2$. 
However, \eqref{main_clain_boundary} breaks down in certain situations where the boundary regions contain multi-intervals and  EWCS is disconnected, as only topological information is included in \eqref{main_clain_boundary}.
For multi-interval regions, we provide a possible generalization in Sec.\ref{sec:disc}, and \eqref{main_clain_boundary} is a trivial case.
On the other hand, this statement does not work for states other than vacuum states.
The entanglement entropy of vacuum states is characterized by the length of a region, which is not true for generic states where other parameters appear.
A more generic proof and a physical interpretation of \eqref{main_clain_boundary} from boundary theory are desired, potentially belonging to future exploration.

Apart from reflected entropy, the Markov gap can also be defined {by other} mixed-state measures claimed to be dual to EWCS.
It is interesting to see if there are similar inequalities for other ``Markov gaps''.
For example, in a generic purification instead of the canonical one that corresponds to the definition of reflected entropy, the authors of \cite{Camargo:2022mme} proposed a generalized Markov gap based on partial entanglement entropy.
The holographic entanglement negativity $\mathcal E$ may also admit a ``Markov gap'' with a similar HPS inequality.
But the prefactor should be $\frac{c}{4}\log 2$.
In some sense, this problem reduces to checking the dualities between EWCS and these quantities.
Nevertheless, they may provide further insights and perspectives, as they have different physical origins.

\begin{acknowledgments}
We thank Clément Berthiere for useful communication and information, and Wuzhong Guo and Yang Zhou for their insightful discussions.
Y.L is supported by the China Postdoctoral Science Foundation under Grant No. 2022TQ0140, and the National Natural Science Foundation of China under Grant No.12247161.
J.L is supported by the National Natural Science Foundation of China under Grant No.12047502, No.12247103 and No.12247117.
\end{acknowledgments}

\appendix

\section{The distance between two geodesics}\label{app1}
\begin{figure}
    \centering
    \includegraphics[width=0.7\textwidth]{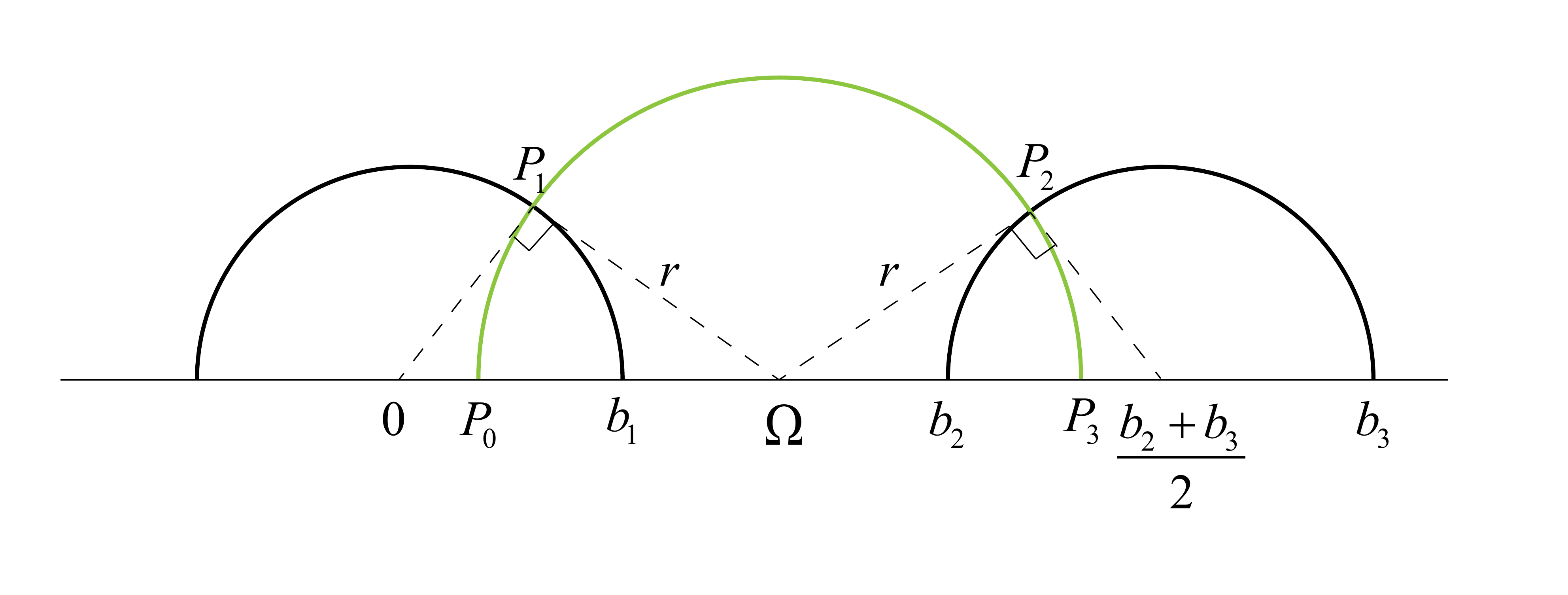}
    \caption{The distance between two geodesics, denoted by black circles, is the geodesic length between $P_1$ and $P_2$ on the green circle. }
    \label{fig:app1}
\end{figure}
As shown in Fig.\ref{fig:app1},
the distance between two parallel geodesics is the distance between $P_1$ and $P_2$.
A unique geodesic, drawn in green, is determined by these two points.
For it to be the shortest, the green geodesic must be perpendicular to the others.
We can obtain two equations by the Euclidean Pythagorean theorem:
\begin{align}
    &b_1^2+r^2=\Omega^2,\\
    &r^2+\sbra{\frac{b_3-b_2}{2}}^2=\sbra{\frac{b_3+b_2}{2}-\Omega}^2,
\end{align}
where $r$ is the radius of green geodesic and $\Omega$ is the $x$-coordinate of its center.
Solve these equations, and we arrive at 
\begin{align}
    \Omega&=\frac{b_1^2+b_2b_3}{b_2+b_3}\\
    r&=\frac{1}{b_2+b_3}\sqrt{
    (b_2^2-b_1^2)(b_3^2-b_1^2)
    }.
\end{align}
The intersections of the green geodesic with $x$ axis are denoted as $P_0$ and $P_3$.
The coordinates of $P_i$ are given by
\begin{align}
   & y(P_0)=0,\quad x(P_0)=\Omega-r,\label{a3}\\
&    y(P_1)=\frac{rb_1}{\Omega},\quad x(P_1)=\Omega-\frac{r^2}{\Omega},\\
   & y(P_2)=\frac{r(b_3-b_2)}{b_3+b_2-2\Omega},\quad x(P_2)=\frac{2r^2}{b_2+b_3-2\Omega}+\Omega,\\
 &   y(P_3)=0,\quad x(P_3)=\Omega+r.\label{a6}
\end{align}
The distance between $P_1$ and $P_2$ is given by
\begin{align}\label{a7}
    D(P_1,P_2)=\left|
    \log\frac{||P_2-P_0||~||P_1-P_3||}{||P_1-P_0||~||P_2-P_3||}
    \right|,
\end{align}
where $||P_1-P_2||=\sqrt{(x_1-x_2)^2+(y_1-y_2)^2}$ is the Euclidean distance.
Insert \eqref{a3}-\eqref{a6} into \eqref{a7}, and we find the explicit expression for $D(P_1,P_2)$
\begin{align}\label{general_distance}
    \boxed{D(P_1,P_2)=\frac{1}{2}\log\sqbra{
    \frac{b_2b_3-b_1^2+\sqrt{(b_2^2-b_1^2)(b_3^2-b_1^2)}}{b_2b_3-b_1^2-\sqrt{(b_2^2-b_1^2)(b_3^2-b_1^2)}}
    }.}
\end{align}
Notice that we assume that the center of one geodesic is located at origin, so $b_1,b_2,b_3$ here are understood as the $x$ coordinates relative to this center.
This allows one to generalize to any case.

\section{The distance between a boundary point and a geodesic}
\begin{figure}
    \centering
    \includegraphics[width=0.7\textwidth]{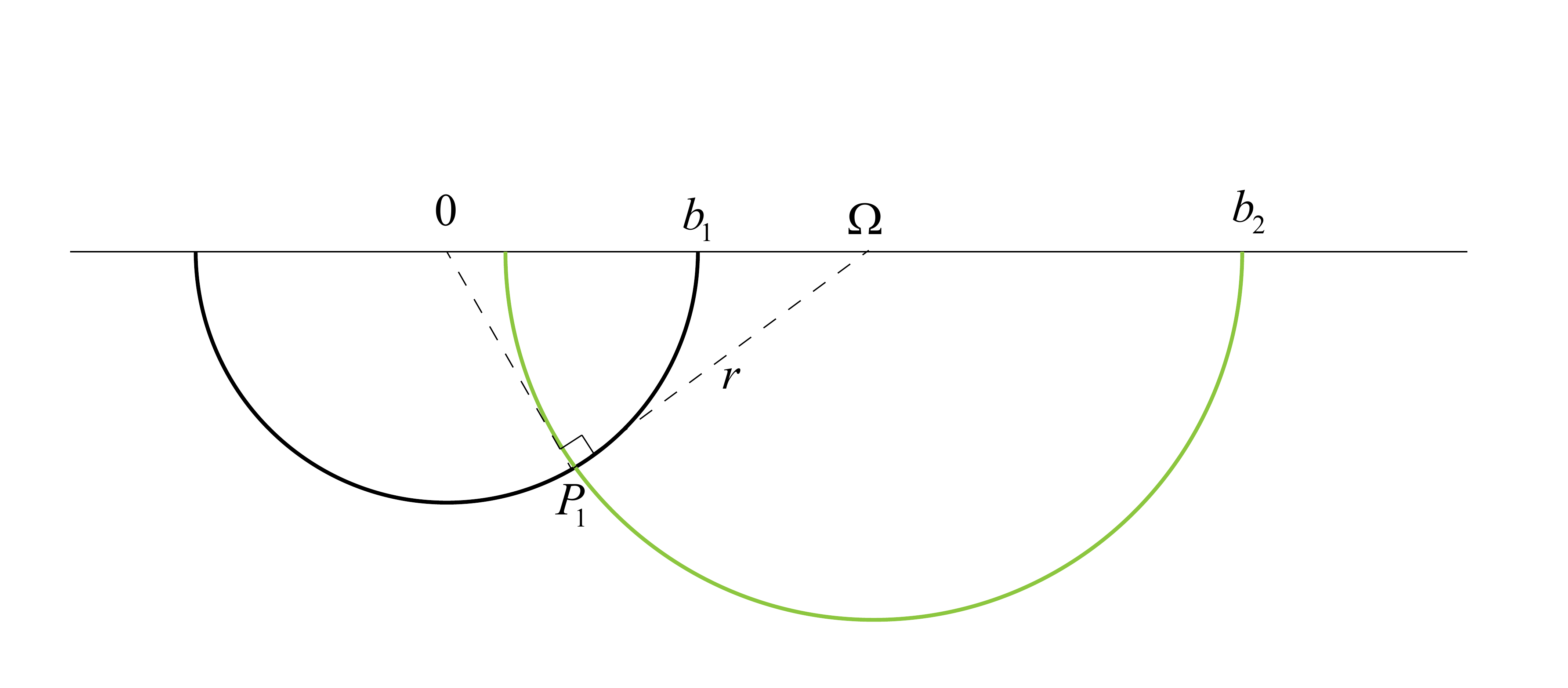}
    \caption{The distance between a boundary point $b_2$ and a geodesic, the black half-circle, is the length of the geodesic between $P_1$ and $b_2$.
    }
    \label{fig:app2}
\end{figure}
We calculate the minimum length of a geodesic that connects a boundary point $b_2$ and a half-circle centered at the origin with radius $b_1$.
As usual, we work in Poincar\'e half-plane with the metric
\begin{align}
    \D s^2=\frac{1}{y^2}(\D x^2+\D y^2).
\end{align}

The target geodesic is shown in green in Fig.\ref{fig:app2}, and it must be perpendicular to the circle with radius $b_1$.
Suppose the green half-circle is centered at $x=\Omega$ with a radius $r=b_2-\Omega$.
We denote its intersection with the geodesic as $P_1$.
By the Euclidean Pythagorean theorem, we have the following equation
\begin{align}
    r^2+b_1^2=\Omega^2.
\end{align}
The solution is 
\begin{align}
    \Omega=\frac{b_1^2+b_2^2}{2b_2},\quad r=\frac{b_2^2-b_1^2}{2b_2}.
\end{align}
Then the length of the geodesic between $b_2$ and $P_1$ is given by
\begin{align}
    D(P_1,b_2)&=\log\frac{2r}{\epsilon}+\operatorname{arctanh}\sbra{\frac{r}{\Omega}}
    \notag\\
    &=\log \frac{2r}{\epsilon}+\log \frac{b_2}{b_1}\notag\\
    &=\log \frac{b_2^2-b_1^2}{b_1\epsilon}.
\end{align}

\section{Adjacent limit}\label{app.adjacent}
We sketched how to obtain the adjacent results from disjoint phases in Sec.\ref{sec.adjacent}.
Here we present a more concrete example on this point.

In Poincar\'e half-plane, the metric is divergent near the boundary CFT $y=0$, corresponding to the IR divergence of the bulk space.
Set the $y_{\rm UV}=\epsilon$, and the entanglement entropy for an interval with length $2l$ is given by the area of the RT surface in unit of $4G_N$\cite{Ryu:2006bv}
\begin{align}\label{RT_formula}
    S(A)=\frac{c}{3}\log \frac{2l}{\epsilon}.
\end{align}
Setting this cutoff means that we only measure the length of geodesics above $y=\epsilon$.
Note that the formula works in the limit $\epsilon\rightarrow 0$.

We would like to get mutual information for the case in which $A$ vanishes from that where $A$ is finite.
Mutual information is just a combination of entanglement entropies, and these entropies are given by the area of their RT surfaces \eqref{RT_formula}.
One can achieve this goal by simply setting $S(A)=0$, which is effectively equivalent to $2l=\epsilon$, even though \eqref{RT_formula} may not work for $2l\sim \epsilon$.

Now we consider the reflected entropy or entanglement wedge cross-section.
Suppose $A$ and $B$ are gapped by two small intervals, as in Fig.\ref{fig:Cutoff_SR}.
Then entanglement wedge cross-section is shown in Fig.\ref{fig:Cutoff_SR} with two ends on the RT surfaces of $AB$.
Holographically, the reflected entropy is given by twice the area of entanglement wedge cross-section.
When evaluating the reflected entropy, we should also set the cutoff as $y_{\rm UV}=\epsilon$ to make sure calculations are consistent.
We let the two gaps to be $[-L-l,-L+l]$ and $[L-l,L+l]$.
The reflected entropy is given by \eqref{takayanagi_SR}
\begin{align}\label{app_C2}
    S_R(A:B)=\frac{c}{3}\log\sbra{
    \frac{2L^2-l^2+2L\sqrt{L^2-l^2}}{l^2}
    }.
\end{align}
We would like to see the vanishing limit of the two gaps.
This cannot be obtained from letting the length of the gap to be $2l=\epsilon$, as the corresponding $y$-cutoff becomes $y_{\rm UV}=\epsilon/2$, see Fig.\ref{fig:Cutoff_SR}.
This is not consistent with $y_{\rm UV}=\epsilon$ we set for entanglement entropy.
In this sense, we have to set $l=\epsilon$, that is $b_3-b_2=2\epsilon$ in Sec.\ref{sec.adjacent}, to get the adjacent limit.
For there is no gap between $A$ and $B$, we have 
\begin{align}\label{app_C3}
    S_{R}(A:B)=2S(A)=\frac{2c}{3}\log \frac{2L}{\epsilon}.
\end{align}
We can see exact agreement between \eqref{app_C2} and \eqref{app_C3} if $l=\epsilon$.

In a word, we can effectively take $2l=\epsilon$ in mutual information and $l=\epsilon$ in reflected entropy to get the results in corresponding adjacent phases.
It is manifest that this procedure results in an additional term $-\frac{c}{3}\log 2$ in the Markov gap, as the cutoff $\epsilon$ is doomed to be canceled there.
This partially explains why the lower bound of the Markov gap is related to the number of gaps between $A$ and $B$.

\begin{figure}
    \centering
    \includegraphics[width=0.7\textwidth]{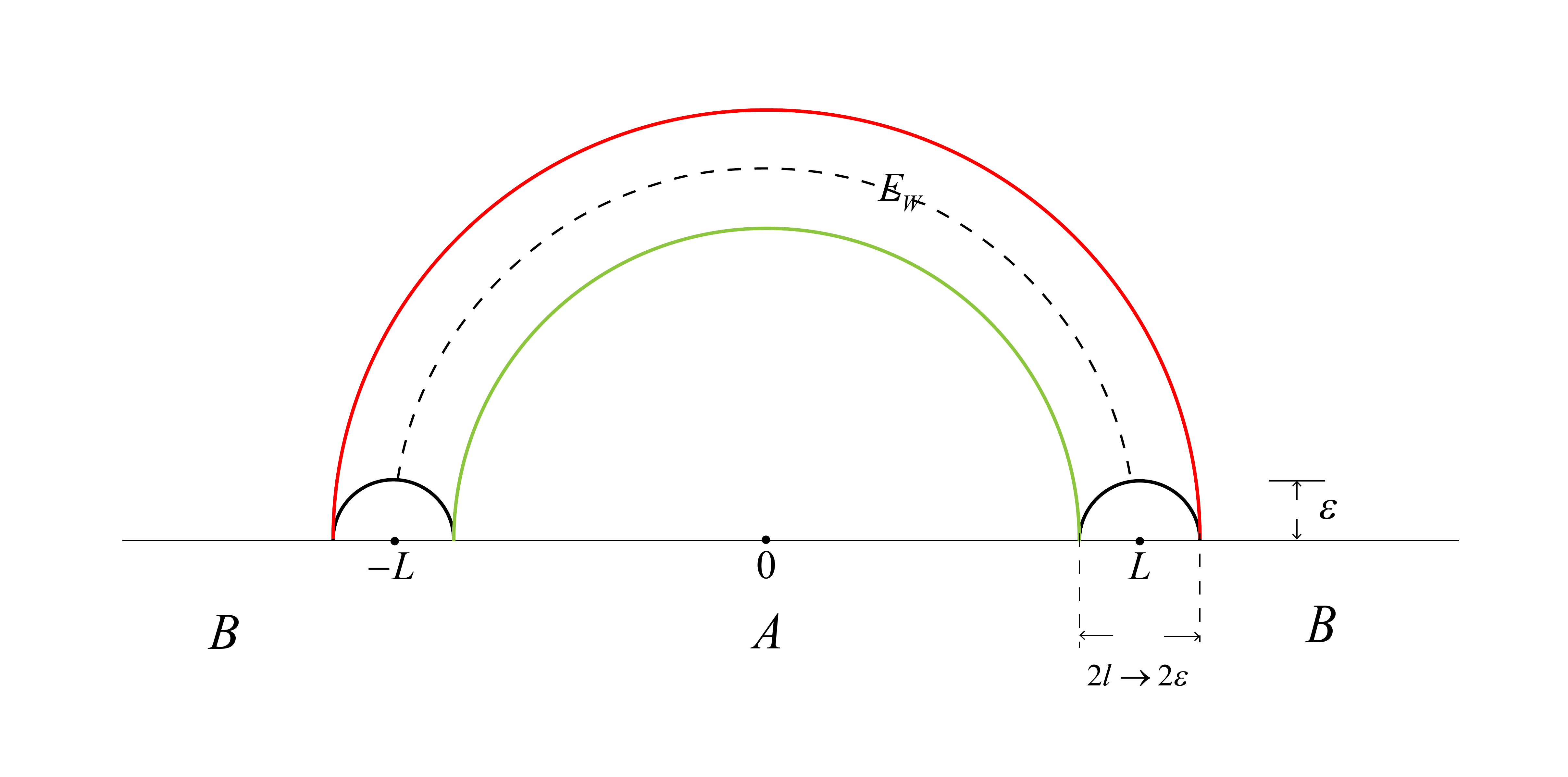}
    \caption{$A$ and $B$ are gapped by two small intervals. 
    The green line and red line denote RT surfaces for $A$ and $B$, respectively.
    The black dashed line denotes the entanglement wedge cross-section $E_W$ between $A$ and $B$.
    The cutoff is $y_{\rm UV}=\epsilon$, which requires a gap of length $2\epsilon$.}
    \label{fig:Cutoff_SR}
\end{figure}

\section{Geometric interpretation of the lower bound with no brane tension}\label{app:geointerpretation}

\begin{figure}
\centering
\includegraphics[width=0.5\textwidth]{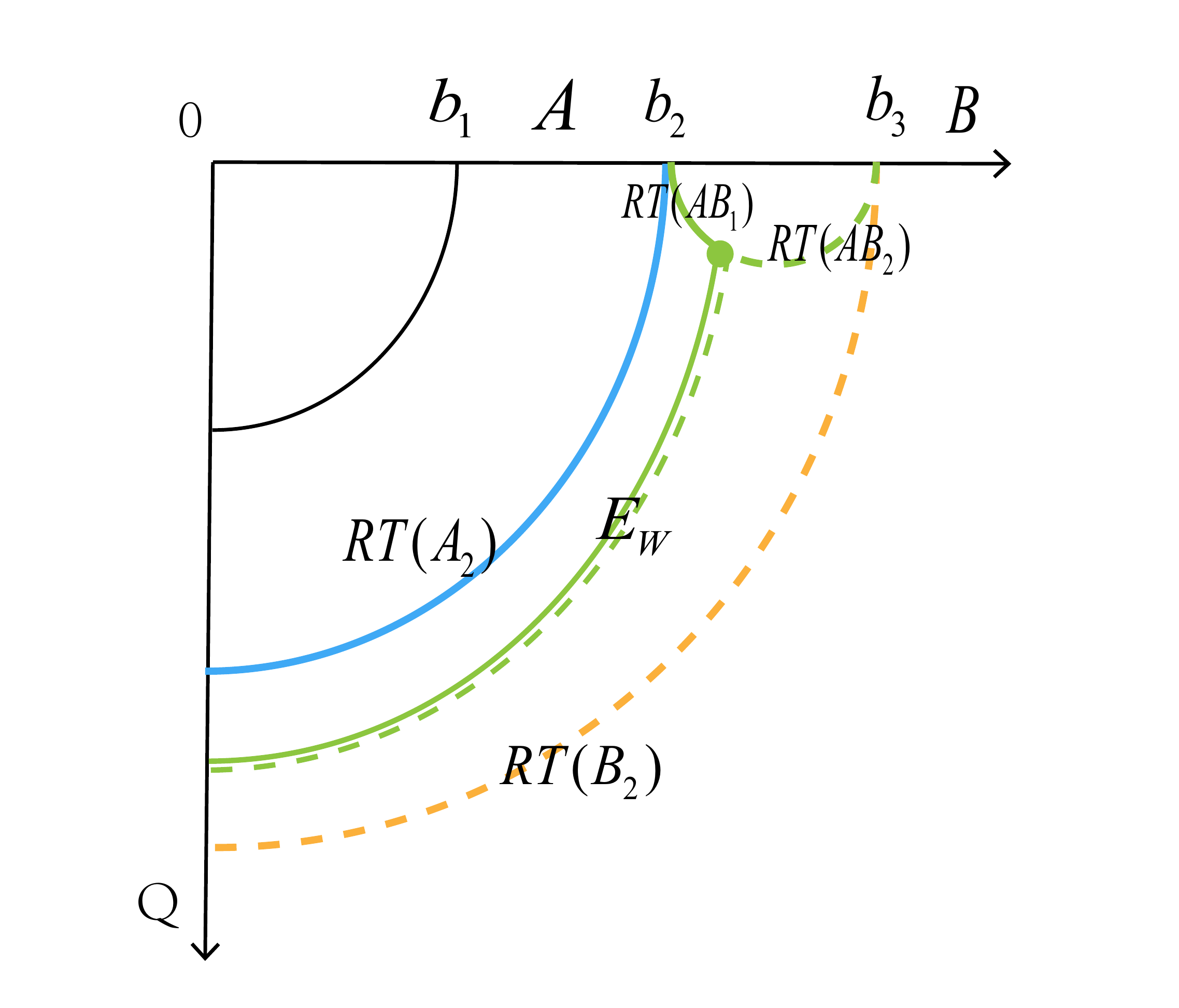}
\includegraphics[width=0.45\textwidth]{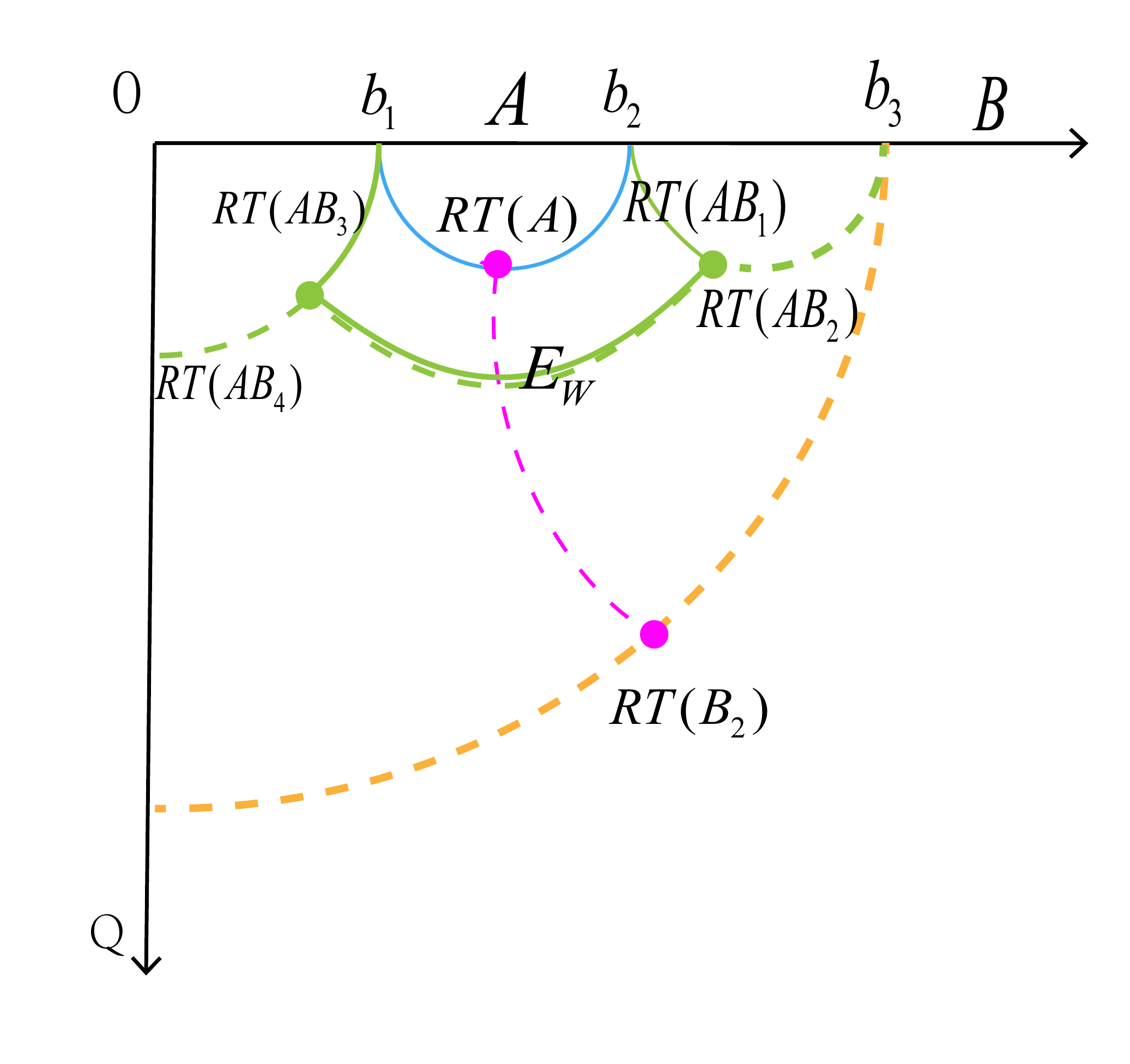}
\caption{Left:  Two right-angled pentagons and thus the lower bound  $\frac{c}{3}\log 2$ for phase-D4.
    Right: Four right-angled pentagons and thus  the lower bound $\frac{2c}{3}\log 2$ for phase-D2.
}
\label{Fig:geo}
\end{figure}

In this section, we will give a geometric interpretation of the lower bound of $S_R-I$ when the brane is tension free.
Without loss of generality, we will only consider phase-D2 and phase-D4.
As shown in Fig.\ref{Fig:geo}, the Markov gap for phase-D4 is 
\begin{equation}
\begin{split}
    h=S_R-I=&\frac{\text{Area}(E_W)+\text{Area}(RT(AB_1))-\text{Area}(RT(A_2))}{4G_N}\\
    &+\frac{\text{Area}(E_W)+\text{Area}(RT(AB_2))-\text{Area}(RT(B_2))}{4G_N}.
    \end{split}
\end{equation}
Since the brane is orthogonal to $x$-axis, it is along a geodesic, and thus the RT surfaces $E_W,A_2,B_2,AB_1$ together with the brane form a right-angled pentagon
\footnote{Different from the proof of the lower bound in pure AdS$_3$ where all sides of the pentagon are made up of RT surfaces or asymptotic degenerate sides, here one side of the pentagon comes from the brane.
Note that if the brane has a non-zero tension, our geometric interpretation using the inequality of the right-angled pentagon does not hold at all, because the brane is not along a geodesic now.  }
with a degenerate side at the asymptotic boundary of AdS$_3$. 
There are two right-angled pentagons for phase-D4.
The key point is that for two adjacent sides $\alpha$ and $\beta$ and a non-adjacent side $\sigma$ of a right-angled hyperbolic pentagon, they satisfy \cite{Hayden:2021gno}
\begin{align}
    \alpha+\beta-\sigma\geq \log2. 
\end{align}
Using this inequality, then we have
\begin{equation}
\begin{split}
    \text{Area}(E_W)+\text{Area}(RT(AB_1))-\text{Area}(RT(A_2))\geq\log 2,\\
   \text{Area}(E_W)+\text{Area}(RT(AB_2))-\text{Area}(RT(B_2)) \geq\log 2,
     \end{split}
\end{equation}
thus
\begin{equation}
    h\geq\frac{c}{3}\log 2,
\end{equation}
where we have used the relation $c=3\ell/2G_N$.
For phase-D2, the Markov gap is given by
\begin{equation}
\begin{split}
    h=S_R-I=&\frac{\text{Area}(E_W)+\text{Area}(RT(AB_1))+\text{Area}(RT(AB_3))-\text{Area}(RT(A))}{4G_N}\\
    &+\frac{\text{Area}(E_W)+\text{Area}(RT(AB_2))+\text{Area}(RT(AB_4))-\text{Area}(RT(B_2))}{4G_N}.
    \end{split}
\end{equation}
Unlike phase-D4, the RT surfaces $E_W,A,AB_1,AB_3$ form a right-angled hexagon.
We can draw a geodesic (pink dashed line in Fig.\ref{Fig:geo}) to decompose a hexagon into two  right-angled pentagons.
Then we have four right-angled pentagons for phase-D2. 
Using the inequality for each pentagon, the Markov gap
\begin{equation}
    h>\frac{4\log2}{4G_N}=\frac{2c}{3}\log2.
\end{equation}
{In fact, similar to pure AdS$_3$, here one may also obtain the lower bound by counting the number of the boundaries of EWCS.
However, for AdS/BCFT, only the boundary on the minimal surfaces of $A\cup B$ contributes to the lower bound while 
the boundary on the brane does not, as we can see from phase-D4. 
This is why we generalize the original HPS inequality to \eqref{main_clain_bulk}.
}



\providecommand{\href}[2]{#2}\begingroup\raggedright\endgroup







\end{document}